\documentclass[sigconf, nonacm, screen]{acmart}
\usepackage{graphicx}
\usepackage{wrapfig}
\usepackage{ifthen}
\newboolean{techreport}
\usepackage{caption}
\usepackage{multirow}
\usepackage{subcaption}
\PassOptionsToPackage{hyphens}{url}
\usepackage{hyperref}
\usepackage{amsmath}
\usepackage{tikz}
\usepackage[]{footmisc}
\usepackage{graphicx}% http://ctan.org/pkg/graphicx
\usepackage{pifont}% http://ctan.org/pkg/pifont
\usepackage{amsmath,amstext}
\let\oldnl\nl% Store \nl in \oldnl
\newcommand{\nonl}{\renewcommand{\nl}{\let\nl\oldnl}}% Remove line number for one line
\usepackage{tikz}
\usepackage{algorithm}
\usepackage{algpseudocode}
\usepackage{amsmath}
\usepackage{xspace}

\usepackage{arydshln}
\usepackage{threeparttable}

\definecolor{codegreen}{rgb}{0,0.6,0}
\definecolor{codegray}{rgb}{0.5,0.5,0.5}
\definecolor{codepurple}{HTML}{C42043}
\definecolor{backcolour}{rgb}{1,1,1}

\newcommand\numberstyle[1]{%
    \footnotesize
    \color{codegray}%
    \ttfamily
    \ifnum#1<10 0\fi#1 |%
}

\newcommand{\lb}{LightBender\xspace}
\newcommand{\lbs}{LightBenders\xspace}
\newcommand{\ov}{overlap\xspace}
\newcommand{\ovs}{overlaps\xspace}
\newcommand{\Ov}{Overlap\xspace}

\newtheorem{example}{Example}[section]

\newtheorem{definition}{Definition}[section]

\usepackage{enumitem}
\renewcommand\footnotetextcopyrightpermission[1]{} % removes footnote with conference information in first column

\begin{document}

\title{Line Drawings using \lbs:  Authoring and Illuminating}
% \author{Haoyu Huang}
% \affiliation{%
%   \institution{University of Southern California}
% }
% \email{haoyuhua@usc.edu}

% \author{Shahram Ghandeharizadeh}
% \affiliation{%
%   \institution{University of Southern California}
% }
% \email{shahram@usc.edu}

\author{Hamed Alimohammadzadeh}
\email{halimoha@usc.edu}
\orcid{0000-0003-2613-5010}
\affiliation{%
  \institution{University of Southern California}
  \city{Los Angeles}
  \state{CA}
  \country{USA}
}

\author{Shahram Ghandeharizadeh}
\email{shahram@usc.edu}
\orcid{0000-0002-1792-7879}
\affiliation{%
  \institution{University of Southern California}
  \city{Los Angeles}
  \state{CA}
  \country{USA}
}

\begin{abstract}

This study presents the hardware and software architecture of a transformative system for illuminating line drawings and letterforms.
These mid-air illuminations are indoors and might be animated. The hardware contribution is a drone equipped with servo-actuated rod joints and a dense, addressable LED strip that enables arbitrary orientation, a \lb.
The software contributions are threefold. First, the system implements algorithms and heuristics to estimate the minimum number of \lbs required to render a line drawing or letterform, stagger swarm formations to mitigate \lb downwash, generate Swarm Flight and Lighting (SFL) files, and execute these files using a swarm of \lbs to illuminate line drawings and letterforms. Second, a Blender add-on enables users to register \lbs, author graphics and animations represented by swarms of \lbs, and deploy the swarm for illumination through one-click functions. Third, users may import SVG files into either the Blender add-on or a standalone LB-Author tool to illuminate line drawings directly from vector graphics.

We present results from an IRB-approved human subject study ($n=21$) to evaluate the impact of \lb misalignment on the perceived illuminations. 
Obtained results demonstrate that the system's 10.1 mm maximum misalignment is perceptually acceptable across tested illuminations, with a median quality rating of 8 on a 0-10 scale.

\end{abstract}

\begin{teaserfigure}
\centering
  \includegraphics[width=0.8\textwidth]{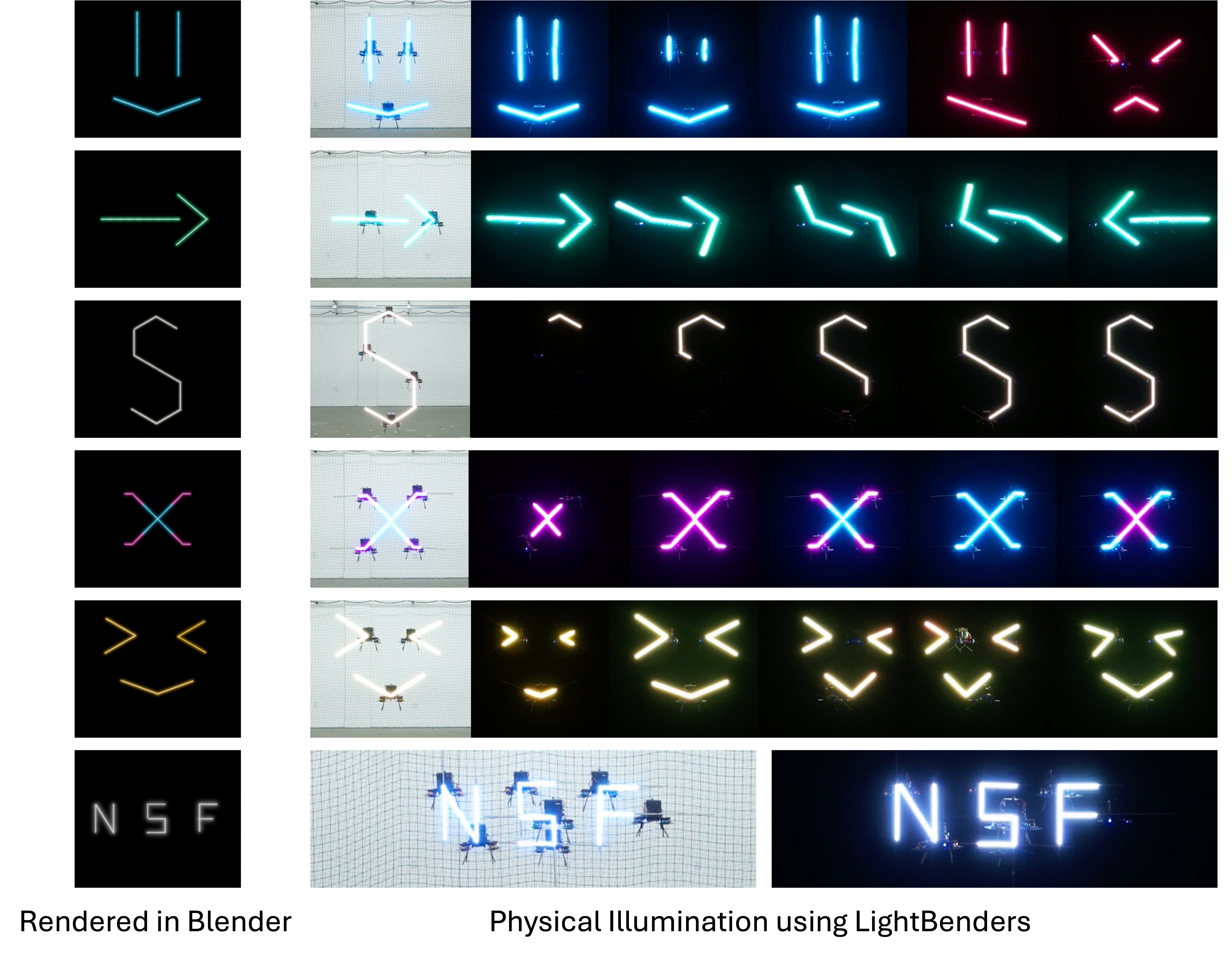}
  \caption{Supplementary video shows the above illuminations using a swarm of LightBenders with the lights on and off. Video also available at \href{https://youtu.be/j4nlgD3iAyM}{https://youtu.be/j4nlgD3iAyM} and \href{https://youtu.be/VzvMgBislcA }{https://youtu.be/VzvMgBislcA }.} 
  \label{fig:teaser}
\end{teaserfigure}
% \received{20 February 2007}
% \received[revised]{12 March 2009}
% \received[accepted]{5 June 2009}

%%
%% This command processes the author and affiliation and title
%% information and builds the first part of the formatted document.
\maketitle

\section{Introduction}
A \lb is a type of Flying Light Speck~\cite{shahram2021,shahram2022}, FLS.  
%It is a drone equipped with a lighting primitive in the form of an actuated rod, local storage, processing and networking capability to implement decentralized algorithms, see Figure~\ref{fig:teaser}a.
%A \lb is a drone with a lighting primitive in the form of an actuated rod.
It is a drone with a lighting primitive in the form of an actuated rod.
Each segment of the rod consists of an array of LEDs.
The brightness and color is adjustable at the granularity of each LED.
A swarm of \lbs illuminate complex shapes and letterforms such as those in Figure~\ref{fig:teaser}.

\begin{figure*}[h]
    \centering
    \includegraphics[width=\linewidth]{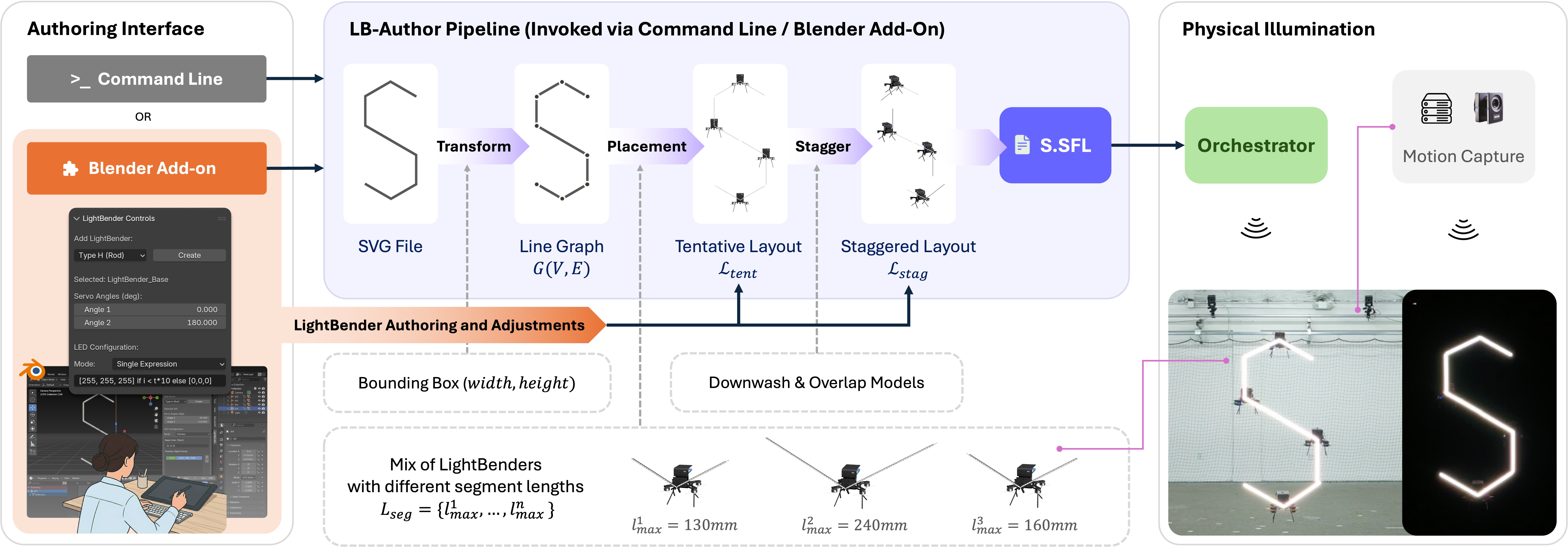}
    \caption{
    The system architecture consists of two authoring interfaces shown on the left: a Command Line Interface (CLI, \S~\ref{sec:lbauthor}) and a graphical Blender add-on (\S~\ref{sec:authoring}).
    When the CLI is executed with a SVG-formatted line drawing, it transforms (\S~\ref{sec:transform}) the drawing into a line graph $G$.
    A placement technique (\S~\ref{sec:sc}) processes this graph to compute the number of required \lbs and their layout. The resulting \lbs are subsequently staggered (\S~\ref{sec:stagger}) to mitigate flickering caused by downwash effects (\S~\ref{sec:downwash}), producing an SFL file. A swarm of \lbs executes this file to illuminate the drawing, as shown on the right side of the figure (\S~\ref{sec:illuminate}). The current system supports a heterogeneous mix of \lbs (\S~\ref{sec:lightbender}) with different segment lengths, illustrated at the bottom center of the figure.  The Blender add-on subsumes the functionality of the CLI, allowing users to import SVG drawings, inspect the computed \lb layout, and make manual adjustments to it. An artist may use this add-on independent of an SVG file by placing \lbs directly in Blender's 3D viewport and author lighting, flight paths, and actuation from scratch.  They may stagger their authored \lbs using a click of a button in the Blender add-on.
    }
    \Description{This figure shows the building blocks of a software system consisting of an authoring interface, a mix of \lbs, and physical illumination.}
    \label{fig:pipeline}
\end{figure*}

%\begin{enumerate}
%    \item What are you trying to do? Articulate your objectives using absolutely no jargon. Build a 3D authoring, monitoring, and visualization tool for Flying Light Specks to illuminate line drawings.  
%    \item What is new in your approach?  Conversion of Blender's line drawings to LightBender's illuminations of lines and circles.  And, their subsequent monitoring to verify the illumination was consistent with the authored line art.
%    \item Why do you think it will be successful?
%    \item Who cares? If you are successful, what difference will it make? Line drawings are the first step in a comprehensive suite of tools for 3D graphics and animations using FLSs.  A room-sized FLS display will resemble the holodeck from the science fiction show Star Trek.  A holodeck with revolutionize how 
%    \item What are the risks? Noise and power are main challenges.
%\end{enumerate}

It is non-trivial to realize a display using \lbs. Two key challenges include (1) generation of content and (2) illumination of the content using LightBenders. The former requires either extension of today’s authoring tools or new ones altogether. The latter requires synchronized flight and coordinated lighting of a LightBender swarm to produce complex graphics. To illustrate, consider the illumination of a simple arrow that must fly in a circular manner. In its simplest form, an authoring tool should incorporate the specification of one or more LightBenders and empower an artist to author a swarm of LightBenders that represents an arrow.
It may produce a {\em Swarm Flight and Lighting} (SFL) specification, fusing the LightBender motion choreography and lighting design into a single, synchronized representation for storage and retrieval.

Displaying an SFL file requires coordinated flight and lighting of a
swarm of LightBenders. A challenge is the design and implementation
of a LightBender that is able to take flight from the ground, fly to
one or more target locations, render its lighting either mid-flight or
once at the destination, and finally land successfully. This challenge
is further exacerbated by the requirement that, once the display of
an illumination is initiated, each LightBender remains synchronized
with the swarm’s flight patterns and lighting. This may require precise
actuation of the LightBenders that is tightly controlled in both time
and space, i.e., temporal and spatial synchronization. A solution must
be cognizant of two physical properties of LightBenders: drift~\cite{drift2019} and downwash~\cite{preiss2017whitewash,downwash2022}.
Drift is the gradual unintended change in a \lb's position, orientation, and altitude in absence of a command, resulting in pixel misalignment in a display.
Downwash is a region of instability below a \lb, impacting the flight of another \lb entering this region.
At its best, it may result in display flickering in the form of unstable \lb flight.
At its worst, it may result in \lb failures and dead regions in the display.
%It may result in display flickering in the form of unstable \lb flight at its best and dead regions in the form of \lb failures at its worst.

We present the architecture of Figure~\ref{fig:pipeline} to address these challenges.
%Figure~\ref{fig:pipeline} shows the overall architecture of the system.  We provide 
It provides an artist with two alternatives to generate illuminations using \lbs.
First, a Blender add-on enables the artist to author graphics using one or more \lbs.
They may use the interface to control each \lb's lit LED and color, and the angle of its rod segments.
Moreover, they may animate these parameters as detailed in Section~\ref{sec:authoring}.
Second, we introduce a new authoring tool, LB-Author.
It enables the artist to provide a drawing in Scalable Vector Graphics (SVG) format.
LB-Author computes the \lbs and their position to illuminate the drawing.
The Blender add-on subsumes LB-Author.
Its graphical user interface (UI) enables the artist to register a SVG drawing for line graph transformation, \lb placement, staggering, and illumination steps of Figure~\ref{fig:pipeline}

While the Blender add-on empowers the artist to author both drawings and animations, LB-Author supports only drawings.
LB-Author is a {\em what} oriented tool.
The artist uses a command line interface (CLI) or the Blender add-on to execute LB-Author with an SVG formatted line drawing.
%The artist draws what they want to illuminate.
It is the responsibility of LB-Author to compute the different types of \lbs and their placement.
We present algorithms to minimize the number of placed \lbs and maximize the utilization of their LEDs.
The Blender add-on enables a {\em how} oriented authoring approach.
It requires the artist to be aware of the different types of \lbs and employ the add-on's UI to describe how a swarm should illuminate a drawing or animation.
Extensions of LB-Author to support animations are future research.

We present Spatial Staggering, labeled Stagger in Figure~\ref{fig:pipeline}, a technique for preventing downwash-induced flickering.
Both authoring tools incorporate this technique, allowing artists to focus on content creation rather than downwash-related constraints.

\noindent{\bf Contributions} of this study are as follows:
\begin{itemize}
    \item Design and implementation of \lbs with actuated rods.  A rod segment consists of an array of LEDs.
    Section~\ref{sec:lightbender}.
    \item A transformative end-to-end architecture, from authoring tools through to illuminations using a swarm of \lbs.
    Figure~\ref{fig:pipeline}.
    \item Two novel \lb authoring tools, a Blender add-on and LB-Author.  
    %LB-Author processes line drawings in Scalable Vector Graphics (SVG) format.
    Sections~\ref{sec:authoring} and~\ref{sec:lbauthor}.
    \item Design, implementation, and evaluation of algorithms and heuristics for (a) placement of \lbs on lines of a drawing, (b) spatial staggering of \lbs to prevent flickering attributed to downwash, and (c) rendering of illuminations using a swarm of \lbs.
    Sections~\ref{sec:placement},~\ref{sec:stagger}, and~\ref{sec:illuminate}.
    \item A quantitative analysis of downwash and drift and their introduced misalignment in illuminations.
    Section~\ref{sec:eval}
    \item A human subject study to evaluate the user perception of the quality of illuminations with \lb misalignment.
% attributed to drift.
    Subjects found the maximum misalignment of 10.1 mm acceptable, rating its quality with different illuminations at an 8 on a scale of 0 to 10.
    Section~\ref{sec:human}.
    \item We open source all our software and hardware designs at \url{https://github.com/flslab/lightbender}.
\end{itemize}
The rest of this study is organized as follows.
Next section introduces related work.
Sections~\ref{sec:lightbender}-\ref{sec:human} present the above contributions in turn.
Brief future research directions are presented in Section~\ref{sec:future}.

%\section{Overall Architecture}
%\input{arch}

\begin{figure*}[htbp]
    \centering
    % ==========================================
    % LEFT COLUMN: Tall Image (Spans rows 1 & 2)
    % ==========================================
    \begin{minipage}[b]{0.39\textwidth}
        \centering
        \begin{subfigure}{\textwidth}
            \centering
            % Adjust the height to match the combined height of the right side
            \includegraphics[width=\textwidth]{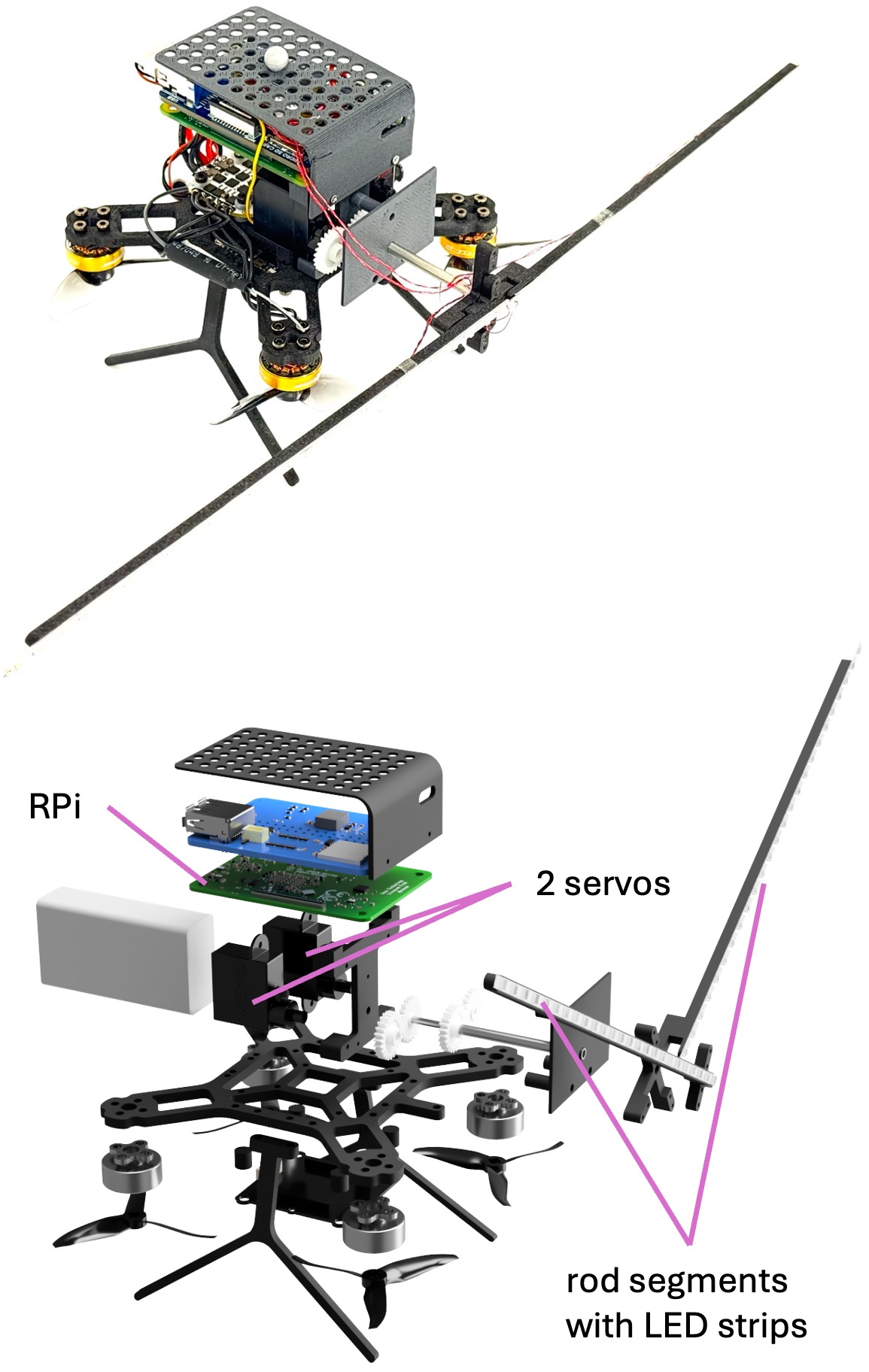}
            \caption{The LightBender. Each rod segment is actuated by a servo.}
            \label{fig:lb}
        \end{subfigure}
    \end{minipage}
    \hfill % Fill space between the two main columns
    % ==========================================
    % RIGHT COLUMN: Two stacked images
    % ==========================================
    \begin{minipage}[b]{0.60\textwidth}
        \centering
        % Row 1, Column 2
        \begin{subfigure}{\textwidth}
            \centering
            \includegraphics[width=\textwidth]{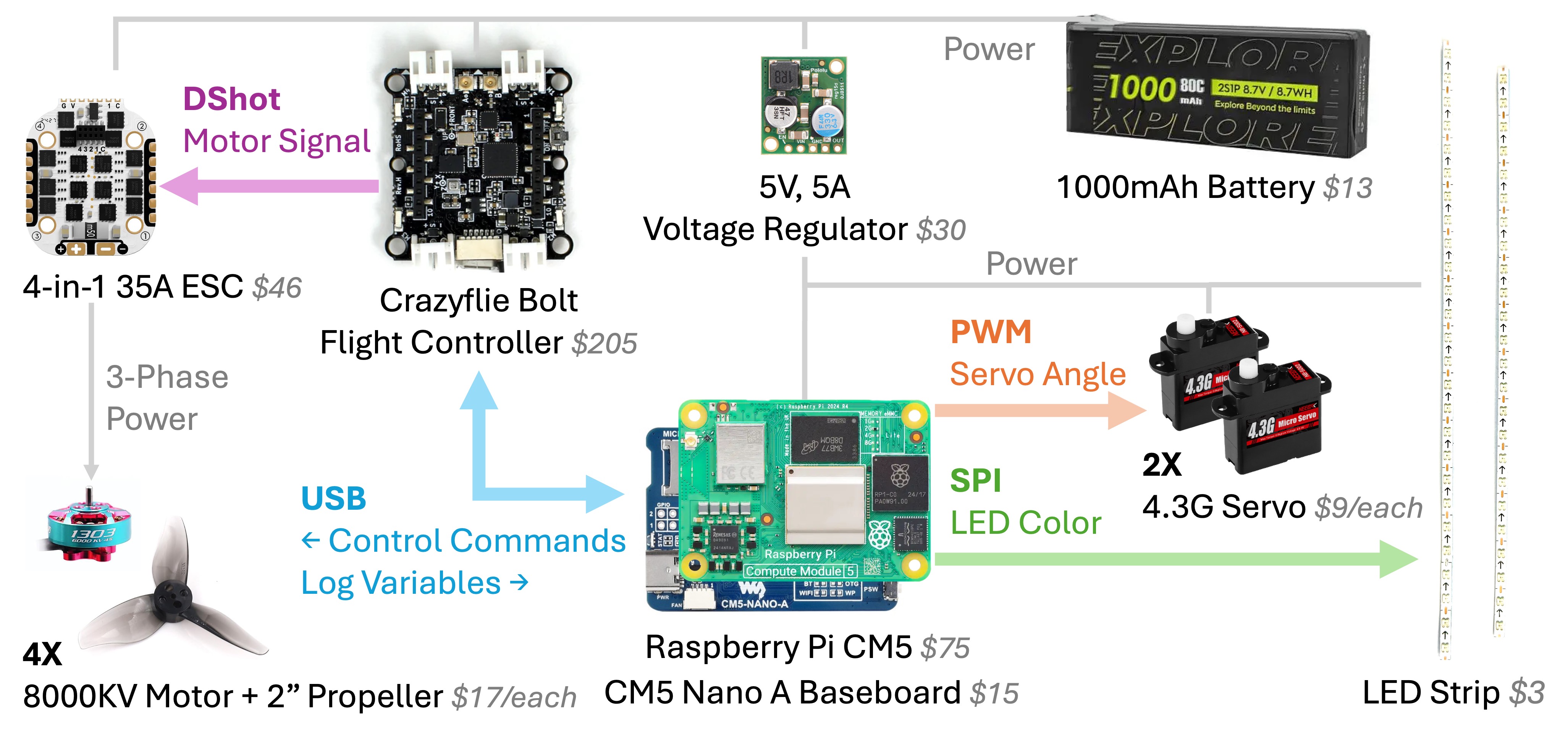}
            \caption{Off-the-shelf electronic components.}
            \label{fig:electrical_parts}
        \end{subfigure}
        
        \vspace{0.4cm} % Adjust the vertical gap between the two right images
        
        % Row 2, Column 2
        \begin{subfigure}{\textwidth}
            \centering
            \includegraphics[width=\textwidth]{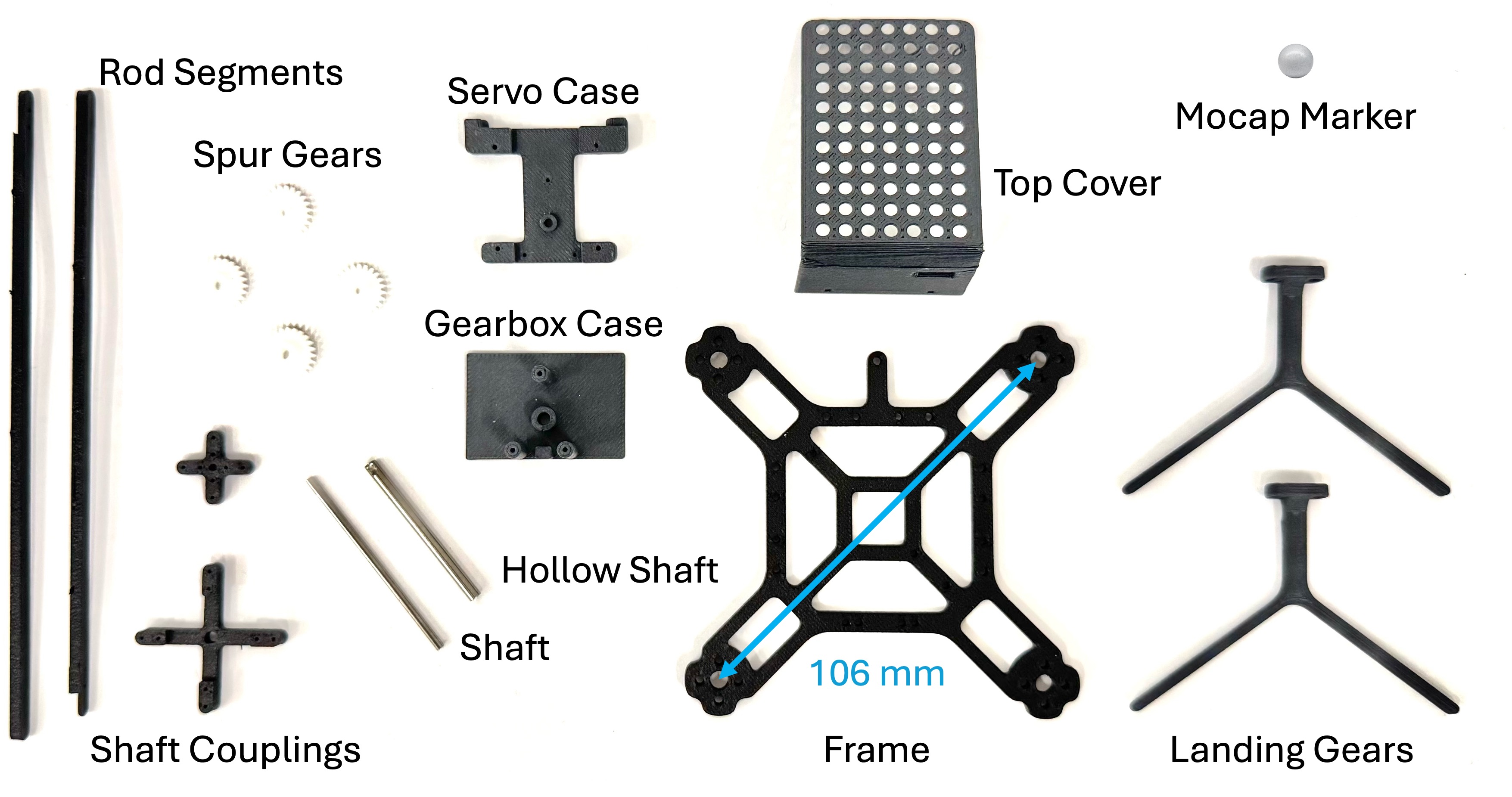}
            \caption{Mechanical and 3D printed parts.}
            \label{fig:mechanical_parts}
        \end{subfigure}
    \end{minipage}
    
    \caption{\lb design and its components. 3D printed parts available from \url{https://github.com/flslab/lightbender}.}
    \label{fig:parts}
\end{figure*}

\section{Related Work}
To the best of our knowledge, \lbs and the end-to-end architecture of Figure~\ref{fig:pipeline} are novel and not described elsewhere.  This includes the authoring tools, placement and staggering algorithms, and the human subject study.
Prior work can be categorized as follows.
%Prior work on aerial displays and drone-based graphics can be organized into six categories.

{\bf Point-Light Drone Displays.} \lbs are inspired by the indoor~\cite{sparked2016,droneperformance2023,bitdrones2016, isphere2017} and outdoor~\cite{gizmodo2018,inteldroneshow,skyelts2023,lumasky2023,sph2024} light-shows with drones as points of light.
%~\cite{dronelight2020,rotatingled2021}.
A key difference is that the LED rod of each \lb supports continuous lines at varying orientations in addition to points of light.
This capability also distinguishes our Blender add-on from existing software~\cite{sph, vergo, skybrush} and hand-gesture~\cite{dronelight2020,dronepaint2021} authoring tools for drones.
%This differentiates our Blender add-on from existing software tools~\cite{} and hand gesture authoring tools~\cite{dronepaint2021} that choreograph today's points of lights.  
%The \lb enables actuated lit lines as a function of time and space, e.g., the arrow moving back and forth with its head changing to become its body.
%Drones of areal night light-shows are points of lights.
\lbs are a class of Flying Light Specks, FLSs~\cite{shahram2021,shahram2022}.  
An FLS as a moving point of light is used to illuminate English letterforms in~\cite{uavmm2025}.
It showed human subjects were challenged to recognize the resulting letters.
In contrast, our human subject study of Section~\ref{sec:human} demonstrates that \lb illuminations enable reliable detection of letters and shapes.
We attribute this to \lb's rod that illuminates continuous lines.
% as a function of time and space.

{\bf Screen-on-Drone Displays.} Several studies mount 2D displays on drones to render graphics~\cite{midairdisplayschi2014,policedrone2019,midairdisplays2014,bitdrones2016}. 
A \lb display differs fundamentally in that it is realized using smaller components, namely, individual \lbs. A swarm of \lbs provides the illusion of self-assembly to render complex graphics via illuminations. Once the illumination is turned off, the \lbs go dark, stop participating as a swarm, and fly back to their hangar. Alternatively, they may transition to a different swarm to illuminate a different shape or animation. This powerful concept of on-demand illuminations is not supported by wall-mounted displays or displays carried by drones. It is particularly useful for environments where an illumination is required on demand and for short intervals of time. An example application is indoor performance, where flying a large screen onto the stage is either infeasible or disruptive.
%to the performance. 
Instead, a \lb display deploys individual \lbs, coordinating their flight patterns and lighting as a swarm to illuminate the desired graphic. After the performance, the \lbs go dark and return to their hangars individually.

{\bf Persistence-of-Vision Displays.} A drone configured with a rotating structure uses the principle of persistence of vision to provide graphics~\cite{rotatingled2021, isphere2017}. This approach differs from \lbs in that the visual effect depends on rapid physical rotation to create the illusion of a continuous surface, whereas \lbs illuminate static or actuated line segments that are directly visible without relying on motion blur or temporal integration by the human visual system.

{\bf Long-Exposure Light Painting.}
DroneLight~\cite{dronelight2020} and DronePaint~\cite{dronepaint2021} use drones to draw in the air using long-exposure photography and gesture-based control, respectively. These systems produce visuals that exist only as photographic artifacts captured over an extended exposure window. In contrast, \lbs illuminations are perceived in real time by a user without requiring any camera mediation.

{\bf Projection-Based Aerial Displays.} A swarm of LightBenders may generate illuminations that complement existing aerial displays using projection systems that may include the use of drones.
An example is dynamic projection mapping technology that uses projectors and specialized software to cast visuals onto any surface to create 3D illusions that blend digital content with the real world~\cite{fibar2020,screen2021,dynamic2022,casper2024,dynamic2024}.
This may include systems that mount a 2D display on or below a drone~\cite{midairdisplayschi2014,policedrone2019,midairdisplays2014,bitdrones2016}, mount a screen on a drone and use a ground projection system~\cite{iuchi2023,projection2025}, a drone carrying a screen followed by a second carrying a projector~\cite{flyingdisplay2014}, a blimp with a tracking system and an image projection system~\cite{blimp2006, floatingAvator2011}.

{\bf Authoring and Software Toolkits.} A variety of toolkits support programming and deployment of aerial display systems across the above categories:  PuReWidgets \cite{purewidgets2012}, SenScreen \cite{senscreen2014}, DisplayDrone \cite{displaydrone2013}, DroneCast \cite{dronecast2017}, Skybrush~\cite{skybrush}, Drone Show Software~\cite{sph}, and Verge Aero Design Studio~\cite{vergo}.
Their applications include indoor and outdoor law enforcement~\cite{policedrone2019} or entertainment~\cite{projection2025,droneperformance2023}.
These support point-light choreography, but do not support the continuous line primitives of \lbs. 
%Hand-gesture based authoring has also been explored~\cite{dronelight2020,dronepaint2021}. 
Our Blender add-on and LB-Author address this gap, providing the first authoring tools designed specifically for actuated-rod lighting primitives.
%The applications of the above systems include indoor and outdoor law enforcement~\cite{policedrone2019} or
%, indoor and outdoor 
%entertainment~\cite{projection2025,droneperformance2023}.
%(references 10-16 of~\cite{projection2025} and stage performance~\cite{droneperformance2023},  

\section{LightBender}\label{sec:lightbender}

%\section{LightBender}
This section describes our implementation of a LightBender that costs\footnote{Purchase prices are as of March 2026.} approximately \$470 and weighs 160 grams. It's a drone built with a Crazyflie Bolt flight controller and a Raspberry Pi CM5 (RPi) as the onboard computer. It carries an actuated lighting module built with two servos with a precision of 0.1 degrees.
Figure~\ref{fig:parts} shows the overall design of a \lb, its electronic components, mechanical, and 3D printed parts.
It is powered by a 2S 1000mAh LiHV battery. At full charge (8.7 V), it sustains hovering flight for up to 362 seconds with the LEDs operating at maximum brightness, until the battery voltage falls below 7.0 V. Recharging the battery from this cutoff to full charge requires approximately 30 minutes.

The LightBender's lighting module is constructed as a jointed rod comprising two independently actuated 160 mm segments. We split and mount an LED strip with 50 LEDs on the rod such that the 26th LED is centered in the middle of the rod. The segments of this rod pivot around a common central rotation axis located at the rod’s midpoint. Due to the mechanical constraints of the servos, each segment is limited to a 180-degree rotational range. To achieve full 360-degree coverage, the segments are calibrated so that their initial positions align to form a straight rod.

\begin{figure}[ht]
    \centering
    \includegraphics[width=1\linewidth]{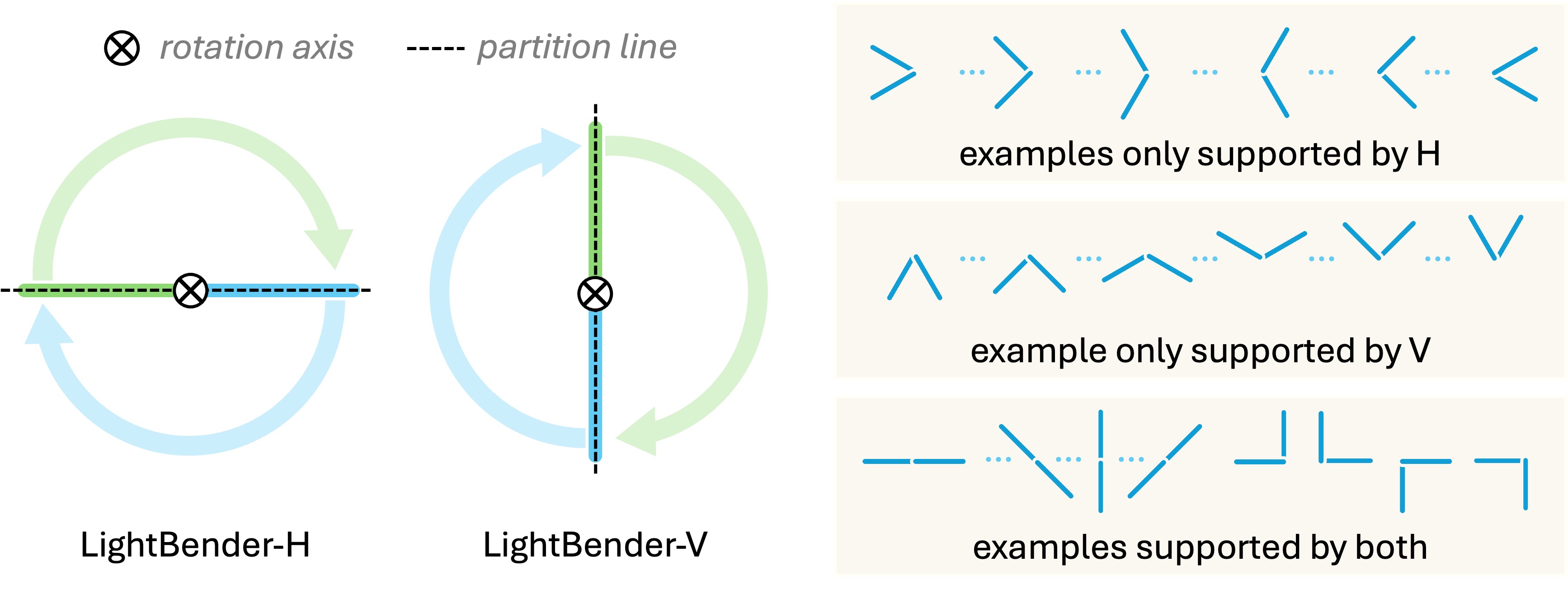}
    \caption{\lb types and their configurations.}
    \label{fig:lb_types}
\end{figure}

As shown in Figure~\ref{fig:lb_types}, this initial alignment defines a partition line, which is a geometric boundary that neither segment can mechanically cross. Consequently, the segments are bound to mutually exclusive semi-circles; it is impossible for both segments to occupy the same side of the partition line simultaneously. This constraint limits the set of shapes a single unit can represent. 
% \begin{wrapfigure}{r}{0.5\linewidth}
% \includegraphics[width=\linewidth]{fig/support_by_type.png} 
% % \caption{Caption1}
% % \label{fig:wrapfig}
% \end{wrapfigure}
To overcome this limitation and ensure a swarm of LightBenders illuminates arbitrary geometries, we introduce two complementary variants: LightBender-H (Horizontal) and LightBender-V (Vertical). In LightBender-H, the partition line is oriented horizontally, and the segments sweep the upper and lower semicircles, respectively. In LightBender-V, the partition line is oriented vertically, and the segments sweep the left and right semicircles, respectively.
% 1) LightBender-H (Horizontal): The partition line is oriented horizontally. In this configuration, the segments sweep the Upper and Lower semicircles, respectively.
% 2) LightBender-V (Vertical): The partition line is oriented vertically. In this configuration, the segments sweep the Left and Right semicircles, respectively.
For example, a LightBender-H cannot generate the lower curve of the letter ``S'' or the smiling mouth of the emoji shown in Figure~\ref{fig:teaser}. These features are instead generated using LightBender-Vs. By interleaving both variants within a swarm, the directional blind spots of individual units are eliminated, enabling the swarm to form lines and angles of arbitrary orientation.

\begin{figure}
\includegraphics[width=.5\linewidth]{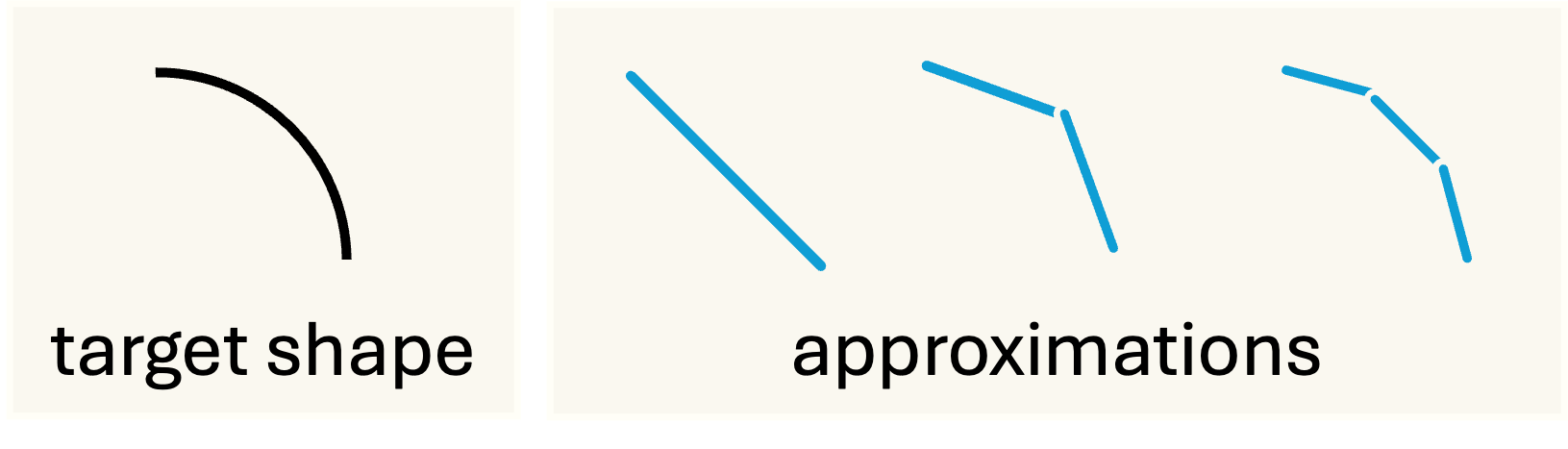} 
\caption{Number and length of segments tradeoff.}
\label{fig:segment_tradeoff}
\end{figure}

The number and length of segments involve a tradeoff between mechanical complexity, flight stability, and stroke fidelity. A longer rod reduces drone count and improves robustness to positioning error compared to two or more \lbs with shorter rods, since a single rod can represent a long stroke without precise relative positioning between multiple \lbs. However, longer rods are less stable in flight due to increased rotational inertia. Additional segments per rod improve the accuracy with which it approximates curved or angled strokes, but each servo increases mechanical complexity and weight. Figure~\ref{fig:segment_tradeoff} illustrates this tradeoff for an example curved stroke. One may approximate it with one rod, or a jointed rod with two or three segments.
The latter approximates the target shape more accurately.
It substitutes for three \lbs with short rigid rods.
We adopt two actuated segments as the design point that best balances stroke fidelity, mechanical simplicity, and flight stability.

\begin{figure*}
    \centering
    \includegraphics[width=\linewidth]{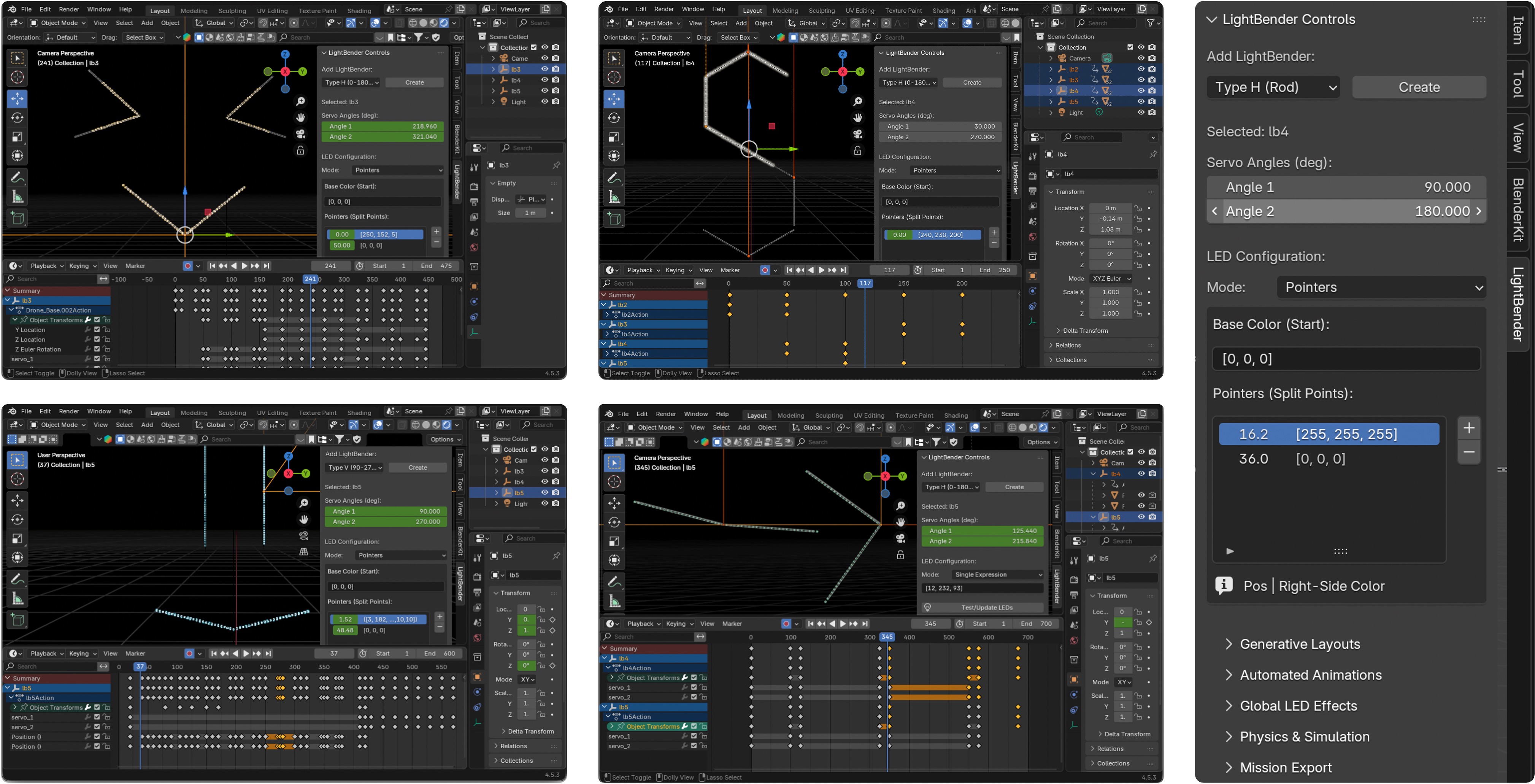}
    \caption{Four screenshots of Blender for different illuminations.  The figure on the right shows the Blender add-on interface for adjusting segment angles, lighting expressions, and pointers for the blue emoji. }
    \label{fig:blender}
\end{figure*}

\subsection{Drift}\label{sec:drift}
Drift is the gradual unintended change in a LightBender's position, orientation, or altitude.
Our first prototype used an inexpensive\footnote{\$70 in March 2026.} flight controller running ArduPilot firmware.
It is originally designed for outdoor, GPS-based operation.
For indoor flight, we used the onboard RPi to overwrite its GPS and vision position estimates using Vicon motion capture data. We tried various update rates: GPS from 5 to 50 Hz and vision position estimate from 25 to 100 Hz.
In addition, we tuned its PID controller.
Its drift remained high along the different axes:  more than 110 mm along the x-axis, 36 mm along the y-axis, and 15 mm along the z-axis. 

We switched to the more expensive\footnote{\$205 in March 2026.} Crazyflie Bolt flight controller, designed for indoor use and supporting positioning data from a motion capture system at 100 Hz.
Prior to tuning its PID controller, its drift was in excess of 40 mm along the x and y axes, and was constantly increasing along the z axis.
While tuning the PID controller stabilized the drift along the z axis to 20 mm, it improved the drift for the x and y axes modestly %(
%These improved modestly 
($< 10\%$). 
We realized our LightBender was suffering from vibration of the flight controller.
We minimized this vibration by using rubber between the flight controller and the LightBender frame.
This reduced the drift along the x-axis to 6~mm, the y-axis to 14~mm, and the z-axis to 12~mm.

\subsection{Downwash}\label{sec:downwash}

\begin{figure}[ht]
    \centering
    \includegraphics[width=0.9\columnwidth]{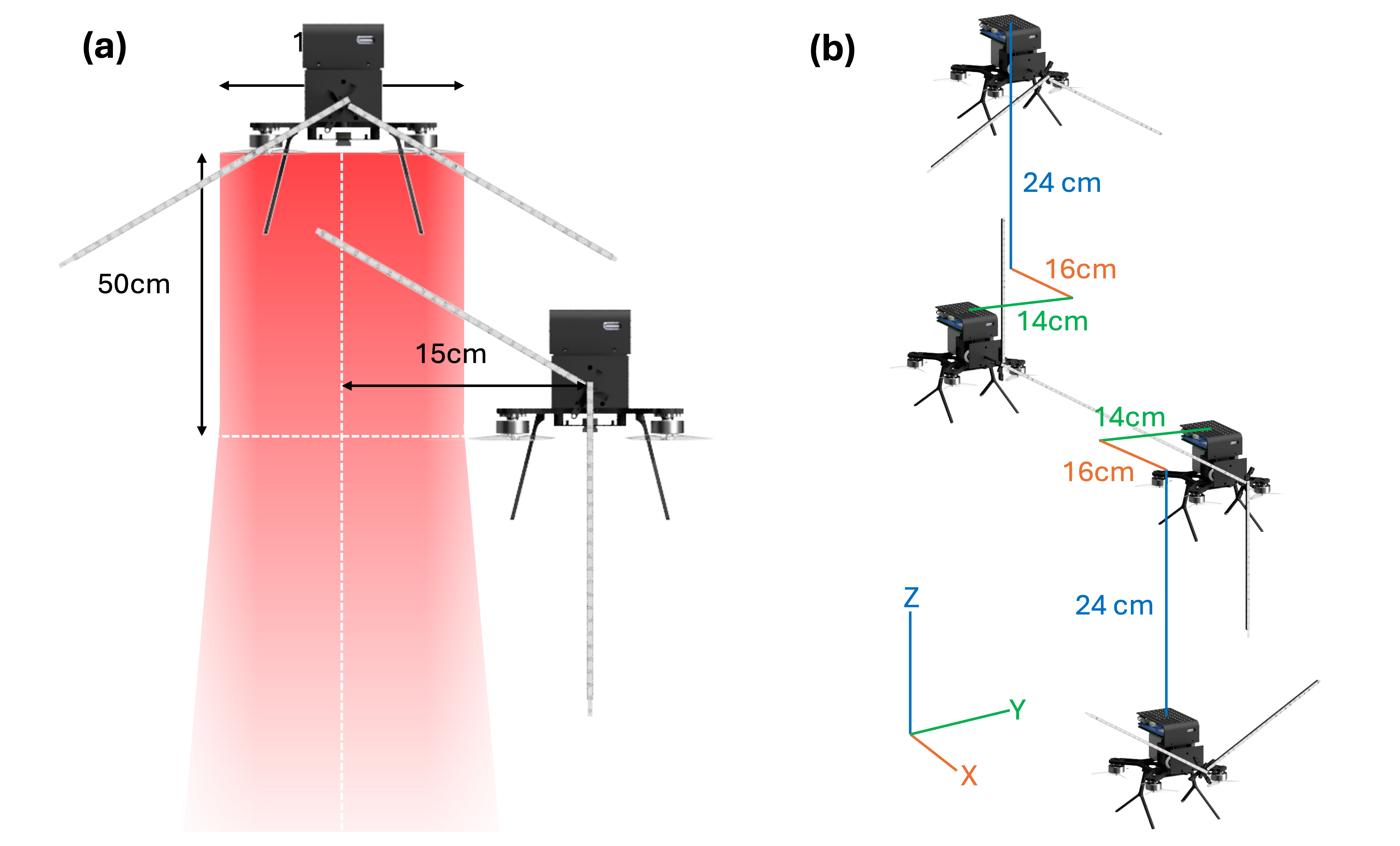}
    \caption{Downwash of a LightBender.  (a) Brighter red denotes stronger downward airflow.
    (b) Physical placement of four LightBenders to illuminate letter S with minimal flickering due to downwash (see discussion of Figures~\ref{fig:rmse_s} and~\ref{fig:rmse_arrow}).}
    \label{fig:downwash}
\end{figure}

The flight of a LightBender generates a downward flow of air produced by its rotors as it generates lift.  It is a region of instability that impacts the flight of another LightBender entering it.
We conducted extensive experiments to understand the downwash of LightBenders.
Figure~\ref{fig:downwash} shows this area with the higher intensity of instability identified in darker red.
We did not observe a LightBender to crash when it entered the downwash (red) region.
However, its flight becomes unpredictable and deviates significantly from the ground truth.
We term this {\em flickering}.

%A LightBender may be positioned below another as long as its center is shifted by 15 centimeters from the center of the LightBender above it.
Two LightBenders may be arranged vertically provided that their centers are offset by 150 mm in the x–y plane.
See Figure~\ref{fig:downwash}.
This realizes illuminations free of flickering.
%This is essential to realizing graphics free from flickering.%, i.e., unpredictable flight of a LightBender that deviates from ground truth.  
%This is essential to evaluate what graphics are feasible without display flickering, i.e., unpredictable flight of a LightBender that deviates from ground truth.
%While we did not observe a LightBender to crash due to downwash, we did observe its flight to deviate from the ground truth significantly.
The spatial staggering technique of Section~\ref{sec:stagger} uses this technique
%We use this technique 
to choreograph \lbs to illuminate letterforms such as S.

%{\bf What does the downwash look like when a LightBender is stationary? }

\section{
%LightBender Authoring Tool:  An add-on for Blender
Blender Add-On}\label{sec:authoring}

We developed a custom add-on for Blender 3.0+ that serves as a unified authoring environment and visualization tool for creating animations using LightBenders.
See Figure~\ref{fig:pipeline}.
The tool parametrically instantiates the LightBender light primitives based on specific hardware configurations (LightBender-H or LightBender-V). It generates a kinematic model of a jointed rod consisting of two rotatable segments, each 160 mm long. It procedurally places 2$\times$2~mm LED indicators at 6~mm intervals along the rod to model the LED strip of the LightBender. Actuation of the rods is abstracted through Blender Drivers, which map custom properties (servo angles in degrees) to the Euler rotations of the rods. This allows the artist to animate the angle of rods using Blender's animation tools.
See Figure~\ref{fig:blender}.

The add-on includes a Python-based expression evaluator to simulate LED lighting behavior in Blender. An animator defines LED patterns using mathematical formulas that depend on time, the LED index, and the total number of LEDs. A frame-change handler evaluates these expressions per frame, updating the emission shader of individual LED instances in the viewport. This enables simultaneous refinement of light patterns, LightBender trajectory, and rod angle adjustments. 

To enable richer animation design through light control, the add-on allows artists to define \textit{pointers} along the LED strip. Pointers act as dynamic delimiters, partitioning the strip into divisions and allowing the assignment of distinct color expressions to each division. The position of each pointer can be animated over time to create wiping or transition effects. See Section 3 of~\cite{siggraph2026}. This feature was essential in choreographing sequences, such as blinking of the emoji, color shifting in 'X', and the drawing of the letter 'S'. See Appendix~\ref{sec:pointer} for a detailed description of LED expression types.

Given a \lb drawing, the add-on uses the spatial staggering technique of Section~\ref{sec:stagger} to detect \ovs and downwash.
It prevents them by adjusting the position of the \lb relative to the user's viewpoint. The user's viewpoint is modeled via a camera in the Blender scene.

%Our add-on detects vertical alignments of LightBenders that may suffer from downwash. It resolves them by automatically staggering LightBenders along the camera's (user's) depth of view.
%To avoid downwash, this Blender add-on generates SFL files that automatically stagger LightBenders along the camera's depth of view.
%Its algorithm first clusters interacting LightBenders into downwash groups. If a group contains $D$ LightBenders, the system generates $D$ depth layers. A depth layer is a plane perpendicular to the view axis, separated by a fixed threshold. This threshold is determined by the size of the LightBender’s downwash region, i.e., 15x50 cm as shown in Figure~\ref{fig:downwash}. The LightBenders are then distributed across these layers to prevent downwash. The layers may be positioned in a direction along the x-y plane, towards and away from the camera (user), see Figure~\ref{fig:downwash}b. The staggering order within the layers is arbitrary. In Figure~\ref{fig:downwash}b, the LightBenders are facing the camera's (user's) view along the x-axis.

The final stage is the generation of the Swarm Flight and Lighting (SFL) specification.
See Figure~\ref{fig:pipeline}.
The add-on discretizes LightBender's position (x, y, z), orientation (yaw), and rod segment angles, along with the user-defined LED expression(s) and pointer(s). The tool extracts control points only at animator-defined keyframes. The resulting SFL data is serialized into a structured YAML format that contains flight path waypoints, servo angles, and LED expressions and pointers.
This SFL file is used to deploy a swarm of LightBenders to illuminate the authored animation.
See Section~\ref{sec:illuminate} for additional details.

%The final stage of the pipeline is the generation of flight plans. The tool discretizes the LightBender's position (x,y,z) and orientation (yaw) rod angles, alongside the user-defined LED expression. The tool extracts control points only at animator-defined keyframes. The resulting data is serialized into a structured YAML format that contains flight path waypoints, servo angles, and LED expressions for deployment using LightBenders. 

% \begin{figure}
%     \centering
%     \includegraphics[width=\linewidth]{fig/blender_addon.png}
%     \caption{Creating an emoji animation using the Blender add-on.}
%     \label{fig:placeholder}
% \end{figure}

\section{LB-Author: Processing of SVG Line Drawings}\label{sec:lbauthor}
LB-Author processes line drawings provided by an artist.  
%The input to LB-Author is an SVG file.
Its input is an SVG file and its output is an SFL file, see Figure~\ref{fig:pipeline}.
LB-Author processes its input in three sequential steps: Transform, Placement, and Spatial Staggering.
%First, it transforms the input SVG line drawing into a Line Graph $G=(V, E)$.
Transform converts the input SVG into a Line Graph $G=(V, E)$.
Subsequently, Placement maps $G$ to \lbs, ensuring every edge of the graph (i.e., SVG line) is covered by a \lb segment. 
%This step calculates the physical state for each \lb by considering a mix of \lbs with different segment lengths. 
%The output of placement may include conflicts in the form of collisions and \lbs in the downwash region of one another. 
Its output may include \ovs and flickering due to downwash.
Finally, Stagger detects these conflicts and resolves them by staggering the participating \lbs relative to the user's viewpoint.

% Figure~\ref{fig:pipeline} shows the overall architecture of the system.
% First, it reduces the authored Blender objects to scene geometry consisting of lines.
% Subsequently, it converts this representation to a graph
% consisting of vertices and edges, G=(V,E).
% Each vertex $v_i$ represents a 3D position.
% An edge $e_{i,j}$ from vertex $v_i$ to vertex $v_j$ represents a line with of $v_i$ and $v_j$ as its two ends.
% We introduce two algorithms to process $G$ and assign LightBenders to edges of the graph, i.e., the lines.
% Subsequently, we detect conflicts in the form of collisions and LightBenders in downwash regions of others LightBenders.
% We resolve these conflicts by staggering the participating LightBenders relative to the user's Field of View, FoV.

% There are two possible interpretations of the first step.
% One may compute a wireframe~\cite{cg1990} or a skeleton~\cite{schaefer2007example,AuSkeleton2008,bucksch2010skeltre,YuSkeleton2014,MA202256} representation of a volume.
% Both compute lines that may be converted into our graph $G$ for LightBender assignment.
% There is a substantial body of literature on both representations.
% In this paper, we assume volumes consisting of cubes and use a simple skeleton technique that converts them to lines.

\subsection{Transform}\label{sec:transform}

The \textit{Transform} step converts a 2D SVG line drawing into a Line Graph $G=(V, E)$. The vertex set $V$ represents 3D points and the edge set $E$ represents the straight lines that connect these vertices. The output graph is scaled and translated to fit within a bounding box, defined by the user via input parameters.

SVG is an XML-based file format for 2D vector graphics designed for the web. We support the <path> element in SVG. A path can be used to create complex shapes composed of multiple straight or curved lines that may overlap. This may include Bezier curves. Transform parses paths in the input SVG to construct vertices and edges of the graph $G$. A straight line is represented as an edge with two vertices at its endpoints. A curved line is first sampled into linear segments at a resolution determined by the \lb segment size.  Each resulting straight line becomes an edge in $G$. Closed shapes are handled by detecting and breaking cycles in the resulting graph. SVG scaling is applied uniformly to fit the bounding box shown in Figure 3, preserving aspect ratio. Overlapping strokes are represented as overlapping edges and merged where collinear. Finally, scaling is applied to fit the position of vertices into the input bounding box defined by width and height, shown in Figure~\ref{fig:pipeline}. The time complexity of transform is $\mathcal{O}(\lambda)$ where $\lambda$ is the total number of lines included in all <path> elements in the SVG file\footnote{This number equals the number of edges in $G$ for the letterforms of Figure~\ref{fig:a-z}.}.

% {\bf What is the complexity of transform?}

\subsection{Placement}\label{sec:placement}
The \textit{Placement} stage maps the graph $G = (V, E)$ to a set of \lbs,
producing a tentative layout $\mathcal{L}_{tent}$.
Its objective is to fully illuminate each edge in $E$ while minimizing the number of \lbs and minimizing the total number of dark LEDs.
Each \lb is modeled by its physical state: 3D position $(x, y, z)$, global rotation about the z-axis (yaw), and the angles of its segments $(\theta_1, \theta_2)$. Each segment can illuminate an edge up to its maximum length, $l_{max}$. The placement supports a heterogeneous mix of \lbs with different segment lengths, $L_{seg}=\{l_{max}^{1}, ..., l_{max}^{n}\}$.
We use the following definitions throughout.

\begin{definition}\label{def:bestfit}
Best-fit LightBender:
Given a line and multiple types of \lbs with different segment lengths, the best-fit \lb is the one with the smallest length that equals or exceeds the length of the line.
\end{definition}

\begin{definition}
Valid edge pair: A pair of incident edges that share a common vertex $v$ is valid if their vectors originating from $v$ are collinear in the XY plane.
\end{definition}

\begin{definition}\label{def:util}
Utilization of a placement: the ratio of total lit rod length of all \lbs in a placement to their total rod length. $$U = \left( \frac{\sum_{i=1}^{|X|} (l^{i}_{1} + l^{i}_{2})}{\sum_{i=1}^{|X|} 2 \cdot l^{i}_{max}} \right) \times 100$$
Where $i$ is a \lb in placement $X$ with segment length $l^{i}_{max}$~mm required to lit $l^{i}_{1}$~mm of its first segment and $l^{i}_{2}$~mm of its second segment.
\end{definition}

We present two complementary algorithmic solutions for placement: \textit{Set Cover (SC)} in Section~\ref{sec:sc} and \textit{Vertex-First Greedy (VFG)} in Section~\ref{sec:vfg}. Both are robust to arbitrary graph representations, making no assumptions about the edge lengths or vertex degree.
They may be combined into a hybrid, as discussed in Section~\ref{sec:discuss}.

\subsection{Solution 1: Set Cover (SC)}\label{sec:sc}
The Set Cover formulation identifies the optimal placement of LightBenders by treating the problem as a multi-objective exact cover over a discretized geometry. It operates in three sequential steps: candidate generation, chunk discretization, and branch-and-bound search. 
Below, we provide a formal problem definition and its Integer Linear Programming (ILP) formulation, followed by a description of the SC algorithm.

\subsubsection{Problem Definition}\label{sec:sc-definition}

Assume the geometry of $E$ is discretized into a finite set of segments, or \textit{chunks}, denoted as $M = \{m_1,m_2,...,m_k\}$. 
A chunk is either an edge or a segment of an edge.
We define a finite universe of feasible \lb candidates $C = \{c_1, c_2, \dots, c_n\}$ that can be placed on geometry $E$.
Each candidate $c_j \in C$ is a \lb that covers one or more chunks. $S(c_j) \subseteq M$ maps $c_j$ to the set of the chunks it covers.

The goal is to find a placement $X$, a subset of candidates, $X \subseteq C$, subject to the following objectives in order.
First, minimize the number of \lbs.
Second, minimize the overlap penalty.
Third, with multiple types of \lbs, minimize the total length of selected \lbs.
A placement is valid when every chunk in $M$ is covered by at least one \lb. 
We formulate these in the next section.

\subsubsection{Integer Linear Programming (ILP) Formulation}\label{sec:sc-formulation}
The problem of Section~\ref{sec:sc-definition} can be expressed as a binary Integer Linear Program.
For each candidate $c_j$, a decision variable $x_j$ is defined, where $x_j$=1 if $c_j$ is included in $X$, and $x_j$=0 otherwise.

Then we define three linear objectives combined via a weighted-sum with the value of weights $W_1$, $W_2$, and $W_3$, enforcing a priority:
\begin{equation}
Minimize \quad f(x) = W_1 \times f_1(X_k) + W_2 \times f_2(X_k) + W_3 \times f_3(X_k)
\end{equation}
subject to
\begin{comment}
\begin{equation}\label{eq:cover}
\sum_{\{j \mid m_i \in S(c_j)\}} x_j \geq 1,
\qquad \forall m_i \in M
\end{equation}
\begin{equation}
 x_j \in \{0,1\},
j=1, 2, ..., n
\end{equation}
\begin{equation}
\sum_{j=1}^{n} x_j \leq n
\end{equation}
\end{comment}
\begin{alignat}{2}
    &\sum_{\{j \mid m_i \in S(c_j)\}} x_j \geq 1 \qquad &&\forall m_i \in M \label{eq:cover}\\
    &x_j \in \{0,1\}                              \qquad &&j=1,2,\dots,n \\
    &\sum_{j=1}^{n} x_j \leq n
\end{alignat}
\noindent where
\begin{alignat}{2}
    f_1(x) &= \sum_{j=1}^{n} x_j              \qquad &&\text{Number of LightBenders} \label{eq:minlb}\\
%f_2(x) = \sum_{i \in M} max \left( 0, \sum_{\{j \mid m_i \in S(c_j)\}} x_j - 1 \right)
    f_2(x) &= \sum_{j=1}^n |S(c_j)|\, x_j    \qquad &&\text{Segment overlap penalty} \\
    f_3(x) &= \sum_{j=1}^n l(c_j)\, x_j      \qquad &&\text{Total length, length penalty}
\end{alignat}
The first constraint, Equation~\ref{eq:cover}, is a set cover condition, guaranteeing that every chunk is covered by at least one \lb. The second constraint imposes binary integrality on the decision variables. The third enforces a budget constraint, ensuring the total number of placed \lbs does not exceed $n$.

Objective 1, $f_1(x)$ in Equation~\ref{eq:minlb}, minimizes the number of participating \lbs.
Objective 2, $f_2(x)$, minimizes the amount of overlap between segments of \lbs, i.e., overlap penalty.
It is equivalent to minimizing the number of covered chunks.
%Since redundant coverage is equivalent to minimizing the number of covered chunks, $f_2$ can alternatively be expressed as $f_2(x) = \sum_{j=1}^n |S(c_j)| x_j$.
Objective 3, $f_3(x)$, applies only when \lbs are heterogeneous in segment length $l$.
It minimizes the total length of selected LightBenders, i.e., length penalty, preferring those whose length most closely fits the chunks they cover.

%Objective 2 is realized when its number of covered chunks is minimized.  Hence, it is possible to formulate it as: $f_2(x) = \sum_{j=1}^n |S(c_j)| x_j$.

Minimizing Objective 1 does not minimize Objective 2.
To illustrate, consider the letter Z which has 4 vertices and 3 edges.
Two vertices have degree 2 and two have degree 1.
Assume each edge is the same length as a \lb segment.
Placing \lbs at both degree-2 vertices causes one segment to overlap with another.
Alternatively, placing one \lb at a degree-2 vertex and the other at an uncovered degree-1 vertex eliminates the overlap.
  Both placements use two \lbs, satisfying $f_1$ equally.
First placement covers a total of 4 chunks while the second covers 3 chunks.
Objective 2 strictly prefers the second because it minimizes the amount of overlap between segments of the two \lbs.

It is possible to formulate this ILP as an incremental optimization.  It solves the objectives sequentially and each optimal value is fixed before proceeding to the next. 
Specifically, $F1=f_1(x)$ is minimized first.
Next, $f_2(x)$ is minimized subject to its solution having $f_1(x)=F1$, yielding $F2$.
Finally, it minimizes $f_3(x)$ subject to $f_1(x)=F1$ and $f_2(x)=F2$.
The next section implements this incremental optimization.

\subsubsection{Set Cover, SC, Algorithm}
%There are multiple ways of solving this optimization problem.
%This section describes a Set Cover solution, SC.
%A Set Cover, SC, solution is as follows.
%It identifies all collinear incident edges and merges them into one edge, producing $G' = (V', E')$.
%A. graph simplification: $G' = (V', E')$, where all collinear adjacent edges have been merged into one edge. Set Cover, SC, 
SC identifies all collinear incident edges and merges them into one edge, producing $G' = (V', E')$.
It generates candidates and discretizes the edges.
Finally, it uses a branch-and-bound technique to enumerate the search space of all subsets of the candidates while pruning this space effectively. 
%Finally, it selects a subset of these candidates using a branch-and-bound technique.
Below, we describe these two steps in turn.
This section concludes with an example to illustrate these steps.

%B. Candidate generation ($C$): The universe of feasible candidates $C$ is generated using two methods:
SC generates the universe of feasible candidates $C$ by placing \lbs at every vertex $v \in V'$ and each edge $e \in E'$.
For each vertex with a degree higher than two, it generates a candidate for each valid edge pair.
%1. Vertex Spanning: For every vertex $v \in V$, a \lb is placed at $v$ for all valid pairs of incident edges.
%2. Edge spanning: For each edge $e \in E$ with length $L$, the theoretical minimum number of \lbs required is computed as $N = \lceil \frac{L}{2  \times l_{max}} \rceil$. Two subsets of candidates are generated:
% With an edge $e$, SC considers its length, $\left\| e \right\|$, and the maximum length of a LightBender segment, $l_{max}$.
% The theoretical minimum number of \lbs is $N = \lceil \frac{\left\| e \right\|}{2  \times l_{max}} \rceil$.
With an edge $e$, SC considers its length, $\left\| e \right\|$.
For this edge and a type of \lb with segment $l_{max}$, SC generates candidates by considering the edge's two vertices. 
Starting with one vertex, it places a \lb with both segments until the entire edge is covered.
The \lbs are placed one segment, $l_{max}$ apart, ensuring two consecutive \lbs are overlapped by one segment.
It generates additional candidates by repeating this process using the other vertex.

To transform the continuous edges into the discrete set $M$, SC uses two consecutive segment tips\footnote{A tip is the end of a segment farthest from the center of its rod.} to define non-overlapping chunks.
Chunks smaller than a threshold ($\epsilon = 1~mm$) are merged.
Thus, $\epsilon$ controls the gap between \lbs.

%D. Branch and Bound

SC uses branch and bound to explore the space of all possible subsets of
$C$.
% It seeds the solution $X$ with a candidate $c_i$.
It uses a binary decision tree that includes or excludes a candidate $c_i$ in $X$.
This search space is large and SC bounds (prunes) it in several ways.
First, prior to adding $c_i$ to $X$, it evaluates the feasibility of $X$.
%3. Feasibility: At any depth in the search tree, 
If the union of the current coverage of $X$ and all remaining candidates fails to cover $M$, the branch is infeasible and is pruned.
SC discards $X$ and starts over by seeding it with the next candidate in $C$, $c_{i+1}$.

Second, SC maintains the best solution $X_{best}$ it has found while searching.
It is the one with the minimum number of LightBenders, minimum overlap penalty, and minimum length penalty.
If the inclusion of $c_i$ in $X$ covers $M$ (i.e., satisfies the set cover condition of Equation~\ref{eq:cover}) and results in fewer \lbs than the best solution, then $X$ replaces $X_{best}$.
If the inclusion of $c_i$ in $X$ results in a subset size equal to the identified minimum, the algorithm evaluates its overlap penalty and length penalty.
If $X$ covers $M$ and has a lower overlap penalty than $X_{best}$ then $X$ replaces $X_{best}$.
Finally, if the overlap penalty is equal and $X$ has a lower length penalty then $X$ replaces $X_{best}$.
Otherwise, the branch is pruned, i.e., $X$ is discarded.
SC seeds $X$ with the next candidates in $C$ and starts the search.
In essence, SC implements the incremental optimization of Section~\ref{sec:sc-formulation}.
Finally, user-defined thresholds $\beta$ and $\Theta$ bound the total number of branches explored and the execution time of SC, respectively. If either threshold is reached, SC terminates and returns $X_{best}$.

%If $X$ covers $M$ and has a lower overlap penalty and length penalty than $X_{best}$ then $X$ replaces $X_{best}$.
%If these penalties are equal to or worse than the current solution, 
% Otherwise, the branch is pruned, i.e., $X$ is discarded.
% SC seeds $X$ with the next candidates in $C$ and starts the search.

SC uses two heuristics to expedite pruning of the search space and minimize the number of visited branches.
First, it seeds $X_{best}$ prior to executing the branch and bound using a greedy algorithm.
This greedy algorithm iteratively selects the candidate with the maximum coverage of uncovered chunks and terminates once all the chunks are covered by \lbs.
%It terminates once a \lb is placed on every chunk.

Second, prior to executing branch and bound, SC sorts the candidate list $C$ in descending order of chunk coverage, $|S(c)|$.
Hence, it starts with a candidate (i.e., a LightBender) that covers the maximum number of chunks.
And, expands the binary search using similar candidates (i.e., LightBenders) first.
This is aligned with the objective to minimize the number of LightBenders.
This ordering strives to identify a best solution quickly in order to prune the rest of the search space when seeded with a poor choice of candidates.

SC supports a mix of \lbs by generating candidates for each available type.
%Its runtime depends on the size of candidates set. 
Its time complexity is $\mathcal{O}(2^{| C |})$ and
space complexity is $\mathcal{O}(|C| + |M|)$. Cardinality of candidates set is $|C| = \mathcal{O}(|E'| \cdot k + |V'| \cdot d^2_{max})$
and chunks set is $|M| = \mathcal{O}(|E'| \cdot k)$ where $k = \left\lceil \frac{\left\| e_{max} \right\|}{l_{min}} \right\rceil$,
$e_{max}$ is the longest line (edge), and $l_{min}=min(L_{seg})$ is the shortest segment length in the mix of \lbs.

%SC's time complexity is exponential. With line drawings consisting of tens and hundreds of segments and vertices, even with branch and bound, it may require years of computation. A threshold $\beta$ limits the maximum number of branches visited by SC. Similarly, $\Theta$ limits the maximum amount of time SC is allowed to execute. SC terminates once it either exhausts the search space, reaches $\beta$ or $\Theta$, returning $X_{best}$.

SC has exponential time complexity. For line drawings with tens or hundreds of segments and vertices, the search may become computationally prohibitive even with branch-and-bound pruning.
A threshold $\beta$ bounds the maximum number of branches explored by SC, while $\Theta$ bounds its execution time. SC terminates when it either exhausts the search space or reaches $\beta$ or $\Theta$, and returns $X_{best}$.

\begin{figure}[h]
    \centering
    \includegraphics[width=\linewidth]{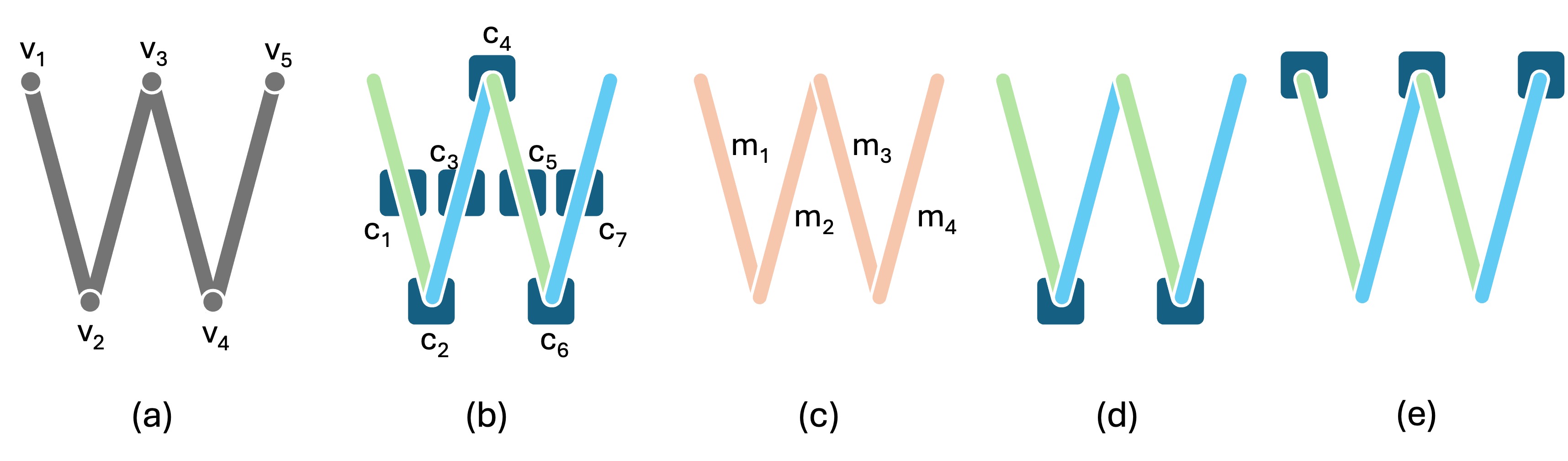}
    \caption{(a) $G(V,E)$, (b) candidates, (c) chunks, and (d, e) alternative placements of \lb for letter W.}
    \label{fig:w}
\end{figure}

%Example block
\begin{example}
To illustrate, consider the geometry $E$ of letter W in Figure \ref{fig:w}a with one type of \lb that has a fixed segment length $l_{max}$.
$E$ consists of 5 vertices and 4 edges.
Assume the length of each edge equals the length of one \lb segment, $\| e_{max} \| = l_{max}$.
First, SC generates $C$ as shown in Figure \ref{fig:w}b.
It consists of seven candidate \lb placements: One per vertices $v_2$, $v_3$, and $v_4$, and one at the midpoint of each edge.
%It consists of seven candidate \lb placements: Three vertices spanning candidates at $v_2$, $v_3$, and $v_4$, and four edges spanning candidates at the midpoint of each edge.
Next\footnote{An astute reader notices that we compute $C$ prior to $M$.  This minimizes the number of chunks based on the \lb segments and reduces SC's complexity.}, SC discretizes $E$ into chunks using the tips of the candidate \lb segments.
Since the candidates align with the edge lengths in this example, the algorithm generates four chunks (set $M$) shown in Figure \ref{fig:w}c.

%SC computes a solution $X$ that is a subset of $C$, $X \subseteq C$, that minimizes both the number of candidate \lbs and the overlap penalty while covering all chunks in $M$. This is the first and second objectives of the stated optimization. With a heterogeneous mix of \lbs, the third objective favors \lb types with smaller segment sizes. 

SC starts seeding the best solution, $X_{best}$, using a greedy algorithm. 
%The algorithm iteratively selects the candidate with the maximum coverage of uncovered chunks. 
With this algorithm, $c_2$, $c_4$, and $c_6$ of Figure \ref{fig:w}b tie for maximum coverage, as each covers two chunks. Assuming the algorithm selects $c_2$, chunks $m_1$ and $m_2$ are marked as covered. In the next iteration, $c_4$ can only cover one uncovered chunk, $m_3$, while $c_6$ covers two, $m_3$ and $m_4$. Hence, $c_6$ is selected. The resulting greedy solution of Figure~\ref{fig:w}d fully covers $M$ using two \lbs, $f_1(x)$=2. The overlap penalty for this solution is zero, $f_2(x)$=0.

%After computing the greedy baseline ($|X_{best}| = 2$, $P(X_{best}) = 0$), 
SC uses the branch and bound to explore the search space for a better solution.
It sorts the candidates in $C$ in descending order of their coverage ($|S(c)|$): $\{c_2,c_4,c_6,c_1,c_3,c_5,c_7\}$. The recursive traversal of the decision tree considers inclusion or exclusion of each candidate in solution $X$.
%It updates $X_{best}$ if it finds a solution $X$ with fewer \lbs or if it finds the solution of the greedy with a lower overlap penalty. The algorithm prunes the search space if adding one more candidate will exceed the best known solution, 
%all candidates in $C$ have been exhausted, remaining candidates in $C$ do not cover the rest of chunks, or if taking a candidate generates the same solution size but worsens the overlap penalty.

%SC terminates once it exhausts all candidates in $C$. When the search space is large, SC terminates after visiting a fixed number of branches $\beta$. 

With W, a single \lb candidate covers a maximum of two chunks.
Hence, it is impossible to cover four chunks with fewer than two \lbs. 
Furthermore, the baseline overlap is already at its minimum of zero. Hence, the branches with more than two candidates selected or the ones with a non-zero overlap penalty are pruned.
The algorithm terminates and shows the greedy baseline of Figure~\ref{fig:w}d as the optimal solution for this example.~$\blacksquare$

%An example for a non-zero overlap penalty is when $c_1$ is selected as the seed for $X$ and $c_2$ is added to $X$. In this case, the overlap penalty is 1 since two \lbs cover a segment of $m_1$.
\end{example}

\subsection{Solution 2: Vertex-First Greedy, VFG}\label{sec:vfg}

Vertex First Greedy, VFG, places \lbs by traversing the graph structure.
It consists of two passes.
In Pass 1, it places \lbs at vertices with degree two and two uncovered edges.
%traverses the graph of vertices and edges to place \lbs.
This pass may produce one or more uncovered edges.
In Pass 2, VFG assigns a \lb to each such uncovered edge.
The following describes the two passes with one \lb type assuming the length of an edge is equal to or shorter than its segment length $l_{max}$.
Subsequently, we expand the discussion to edges longer than $l_{max}$ and \lbs with different segment lengths.

A key requirement is that Pass 1 start with a degree 2 vertex with one edge that has a leaf vertex, i.e., a vertex with degree one.
%Pass 1 starts by selecting a degree two vertex with one edge that has a leaf vertex, i.e., a vertex with degree one.
Examples include $V_2$ and $V_4$ of Figure~\ref{fig:w}a.
%$V_3$ does not qualify because none of its edges have a vertex with degree one.
VFG places a \lb at one such vertex and marks its edges as covered.
Next, it traverses its outgoing edges to identify the next reachable vertex with two uncovered edges, placing a \lb at it.
%Each time it encounters such a vertex, it places a \lb at it.
It repeats this process, traversing the graph in either a breadth-first or a depth-first manner.
Pass 1 terminates when there are no vertices with degree two or higher that have two uncovered edges.
%In Figure~\ref{}, it navigates to $V_3$ from $V_2$. $V_3$ has one uncovered edge. Its other edge was covered by assignment of \lb to $V_2$. Hence, VFG navigates to the next vertex, $V_4$. Since it has two uncovered edges, VFG places a \lb at $V_4$. Once the traversal of the graph ends, 
%How does Pass 1 avoid an infinite loop attributed to cycles in the graph?
It avoids cycles by not visiting a vertex more than once.

Pass 2 places 
%a best-fit \lb at a vertex of each uncovered edge.
a best-fit \lb at an uncovered edge $e$.
%vertex $v$ with only one segment activated to cover $e$. 
We considered alternatives such as placing the \lb at a vertex of $e$ or the midpoint of $e$, i.e., 
%the edge 
at $\frac{u+v}{2}$ where $u$ and $v$ are the positions of the vertices for $e$. The experimental results showed that their computed number of \lbs, overlap penalty, and length penalty to be approximately the same.
%and their LED segment utilization is approximately the same across these alternatives.

%VFG assigns a \lb to the uncovered edges of the graph. In our example, once $V_2$ and $V_4$ are assigned a \lb, no uncovered edges remain. Hence, VFG terminates. 

With edges longer than $l_{max}$,
%More formally, 
VFG constructs G'=(V',E') by merging all collinear incident edges into one edge.
%Edge subdivision: to account for \lb 's physical limitations, the algorithm first normalizes the target graph $G$. 
It subdivides an edge $e \in E'$ with the Euclidean length $\left\| e \right\| > l_{max}$ into $k = \left\lceil \frac{ \left\| e \right\| }{ l_{max} }\right\rceil$ segments of length $l_{max}$ with new intermediate vertices.
%It subdivides an edge $e \in E'$ with the Euclidean length $\left\| e \right\| > l_{max}$ into $k = \lceil \frac{ \left\| e \right\| }{ l_{max} }\rceil$ uniformly sized segments with new intermediate vertices.
%It adds new intermediate vertices to form the normalized graph $G' = (V', E')$ where $\forall e' \in E', \|e'\| \le l_{max}$.
%$l_{max}$ 
%$l$ is the length of a \lb segment.
%With one type of \lb, $l$ is its segment size and a constant,  $l=l_{max}$.

To allow a gap smaller than or equal to $\epsilon$ between \lbs, we adjust the computation of k.
If the fractional part of $\frac{ \left\| e \right\|  }{l_{max}}$ is smaller than $\frac{(k-1) \epsilon}{l_{max}}$, we subdivide the edge into k-1 segments of length $l_{max}$.
This induces a maximum $\epsilon$ between the \lbs.

%To process the adjacent edges sequentially, the vertices $V'$ are sorted using a Breadth-First Search traversal.

%After these preparations, 
%VFG follows a two-pass greedy placement of \lbs over the ordered vertices. Pass A (dual-edge coverage): For each vertex $v \in V'_{ordered}$, the algorithm evaluates all pairs of uncovered incident edges $(e_1, e_2)$. For every valid edge pair, a best-fit \lb for $(e_1, e_2)$ is placed at $v$, and both $e_1$ and $e_2$ are marked as covered.

%Pass B (single-edge coverage): The algorithm iterates through the remaining uncovered edges in $E'$. For each edge $e = (u, v)$, a best-fit \lb is placed at vertex $u$ with only one segment activated to cover $e$. We considered alternatives such as placing the \lb at $v$ or the midpoint of the edge at $m=\frac{u+v}{2}$. Experimental results showed the computed number of \lbs and their LED segment utilization is approximately the same across these alternatives.

With a mix of \lbs, one may set $l_{max}$ to the length of one of the \lb 's segments.
The edges of the resulting graph are the same length or shorter than $l_{max}$.
VFG uses the definition of best-fit \lb (see Definition~\ref{def:bestfit}) when placing a \lb at either a vertex or an edge.
It is computationally fast with complexity $\mathcal{O}(|V'| \cdot d^2_{max})$ where $d_{max}$ is the maximum degree of vertices. Its space complexity is $\mathcal{O}(|V'| + |E'|)$.

\begin{example}
With W in Figure \ref{fig:w}a, both $V_2$ and $V_4$ qualify as a vertex with degree two and an edge that has a leaf vertex.
Pass 1 starts with one of these, say $V_2$, and places a \lb at this vertex.
It traverses the graph to $V_3$.
However, $V_3$ has only one uncovered edge (after placement of a \lb at $V_2$).
Hence, VFG visits the next reachable vertext, $V_4$.
It has two uncovered edges.
VFG places a \lb at vertex $V_4$, covering its edges.
There are no other vertices with two uncovered edges to visit.
Hence, Pass 1 terminates.
There is no Pass 2 because all edges are covered. 
Hence, VFG terminates with two \lbs that cover W as shown in Figure~\ref{fig:w}d.
The stated requirement for the start vertex prevents VFG from using $V_1$ or $V_5$ to place \lbs.  This prevents VFG from computing the solution shown in Figure~\ref{fig:w}e.~$\blacksquare$

%Letter W consists of 5 vertices and 4 edges, see Figure \ref{fig:w}a. Assume every edge is smaller than a \lb segment. VFG processes vertices with 2 or more incident edges first. With W, these are $V_2$, $V_3$, and $V_4$. If Pass A starts with either $V_2$ or $V_4$, it computes two \lbs placed at $V_2$ and $V_4$ as the solution. The heuristic terminates because there are no uncovered edges to process using Pass B. On the other hand, if Pass A starts with $V_3$, it assigns a \lb to $V_3$ and covers its two outgoing edges.  It generates two outer edges of W for processing by Pass B. Pass B places a \lb at each of $V_1$ and $V_5$ to cover these edges. Hence, this alternative uses three \lbs.
%Note if a greedy heuristic started with a degree 1 vertex, it would compute three \lbs.
\end{example}

%Figure \ref{fig:y_vertex} shows the how Vertex solution traverses the graph and places 7 \lbs for letter Y.

%star example

% VFG is a greedy topology-driven heuristic that uses the vertices of $G$ to guide the placement of LightBenders.
% Because each \lb has a central pivot from which its two segments extend, the vertices are an ideal location.
% It starts by identifying the length of the available rod with the highest length, maxLimit.
% Next, it identifies all lines (edges of the graph) with length $L$ greater than maxLimit and represents each with S segments of maxLimit, $S=\frac{L}{maxLimit}$.
% Next, it traverses the vertices with degree two (i.e., two edges) in a breadth first manner, placing a LightBender at the vertex and extending its rod segments along the edges of the vertex.
% Amongst the available LightBenders, it selects the one with the smallest segment length that covers the edges (replace with best-fit as defined in the paragraph before this section). 
% It removes the edges that are covered by a segment of the \lb.

% Once all vertices have been processed, VFG iterates through the remaining uncovered edges.
% For each edge, it places a LightBender at its vertex.
% It selects the LightBender with the segment that is larger than the edge (replace with best-fit LightBender).

\begin{figure}[h]
    \centering
    \begin{subfigure}{\columnwidth}
        \centering
        \includegraphics[width=\linewidth]{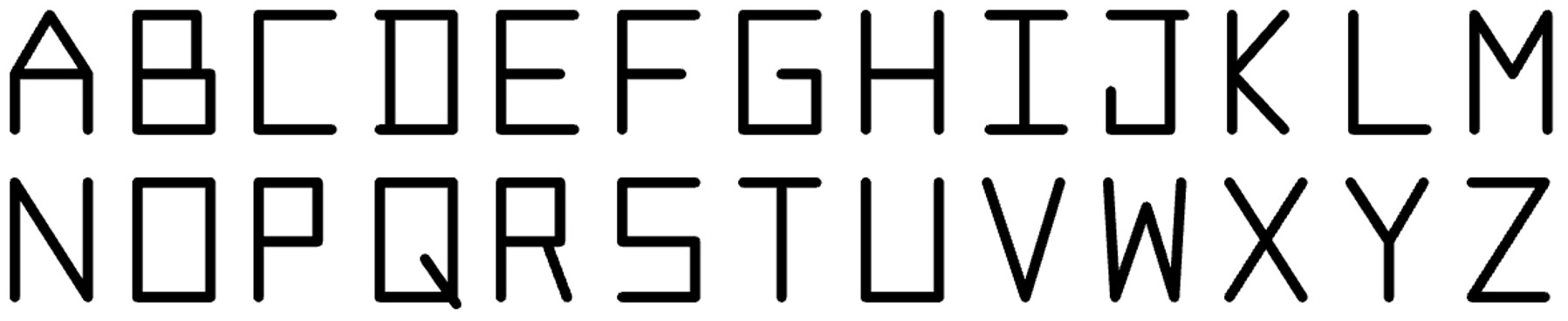}
        \caption{Letterforms A to Z~\cite{jsbin_nacirij} with 3 to 8 vertices and 2 to 6 edges.}
        \label{fig:a-z}
    \end{subfigure}
    \begin{subfigure}{\columnwidth}
        \centering
        \includegraphics[width=\linewidth]{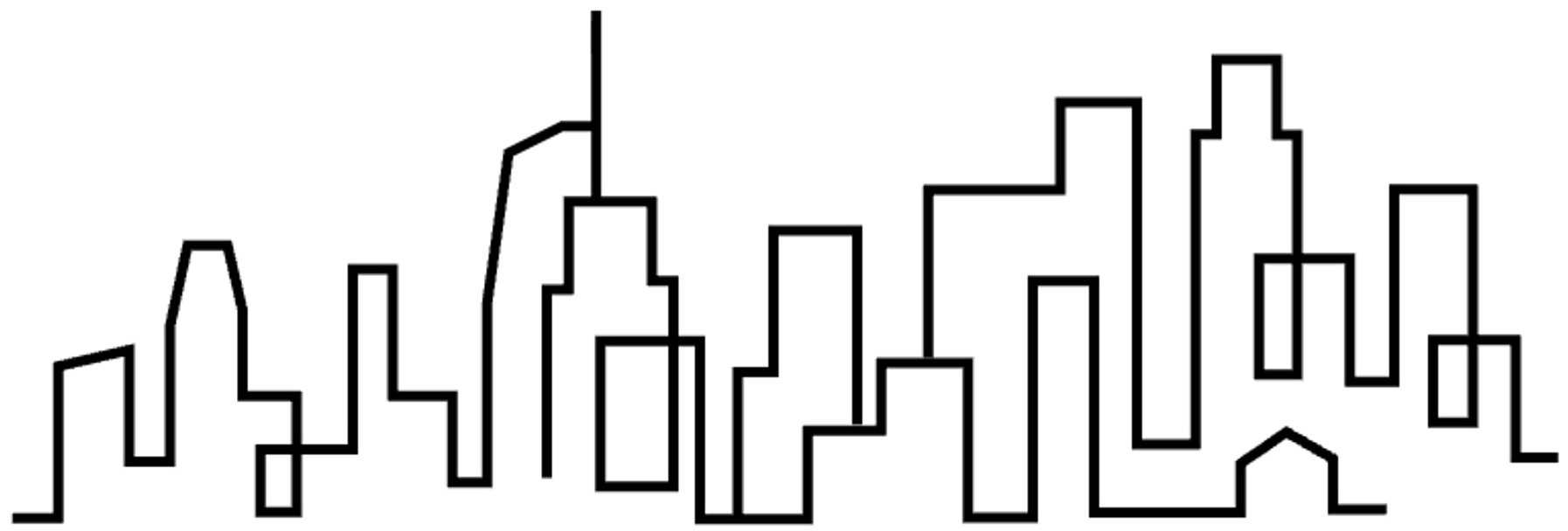}
        \caption{Skyline graphic with 89 vertices and 84 edges.}\label{fig:skyline}
    \end{subfigure}
    \caption{Two line drawings.}\label{fig:twolds}
\end{figure}

\subsection{A Comparison}\label{sec:vfgvssc}
We compared SC with VFG on two sets of data:
each of the English letterforms of Figure~\ref{fig:a-z} and the vector skyline graphic of Figure~\ref{fig:skyline}.
While an English letterform consists of 3 to 8 vertices and 2 to 6 edges, the skyline graphic consists of 89 vertices and 84 edges.
The English letterforms are relatively simple with adjustable heights, requiring 2 to 14 LightBenders with heights ranging from 250 mm to 1000 mm. 
With these, an expert may evaluate the quality of placement provided by SC and VFG.
In contrast, the skyline graphic is complex. It is very difficult for an expert to evaluate a placement.
We use the following criteria to compare the placement techniques:
\begin{itemize}
    \item Number of required \lbs.
    Fewer is better.% and their utilization. A placement technique that minimizes the number of \lbs and maximizes the utilization of their LEDs is superior.
    \item Utilization of the LEDs on rod segments placed on different lines, see Definition~\ref{def:util}.  %With an entire segment consisting of a row of LEDs, 
    This metric quantifies what percentage of placed LEDs are lit. 
    Higher is better.
    \item Execution time defined as the elapsed time to run a placement technique. We restrict SC to explore up to 1 million states in the binary decision tree. If it exhausts the entire solution space prior to reaching this limit, the process terminates early.
    Faster is better.
\end{itemize}
A placement technique that minimizes the number of required \lbs, maximizes their LED utilization\footnote{This is consistent with minimizing segment and length penalties of Section~\ref{sec:sc-formulation}.}, and executes faster is superior.

% \begin{figure}
%     \centering
%     \includegraphics[width=\linewidth]{fig/a_z_dtla.jpg}
%     \caption{Line drawings of English letters A-Z and skyline graphic.}
%     \label{fig:line_drawings}
% \end{figure}

Obtained results highlight the following lessons:
\begin{enumerate}
    \item VFG executes faster than SC.
    \item There is a tradeoff between the number of placed \lbs and LED utilization.
    VFG may be able to place the fewest number of \lbs.
    However, its LED utilization may be lower than SC. 
    \item With the English letterforms, when SC is able to complete an exhaustive search of all possibilities, it uses either fewer or the same number of \lbs as VFG.
    Otherwise, in almost all cases, VFG uses fewer \lbs. 
    %In our experiments, SC placed at most one more \lb than VFG.  There were scenarios where VFG placed 2 more \lbs than SC.  In all our experiments, SC provided a higher utilization than VFG.  This is true even when SC placed a higher number of \lbs - {\bf Hamed, provide the placement with letter a at 1000 mm.}
    \item With the skyline graphic, VFG places fewer \lbs.
    With one type of \lb, Base, its utilization is higher than SC.  With multiple types of \lbs, Mixed, its utilization is lower.  See discussions of Section~\ref{sec:discuss}.
\end{enumerate}
%Below, we compare SC with VFG using the English letters and the skyline graphic in turn.
Below, we present results with the English letterforms and the skyline graphic in turn.

\subsubsection{English Letters}
We consider the English letterforms of Figure~\ref{fig:a-z} at sizes of 250 mm, 500 mm, and 1000 mm.
We considered two system configurations:
1)
Base with a single type of \lb and segment size 160 mm.
2) 
Mixed with heterogeneous \lbs and segment sizes of 130 mm, 160 mm, and 240 mm.
We present results for each in turn.
%We discuss obtained results with each in turn. 

%Figure~\ref{} shows the English letters.
%We assumed one LightBender with a fixed segment size of X mm.
%We scaled the size of each letter from 250 mm to 500 mm and 1000 mm.
\noindent{\bf Base:}
SC and VFG compute the same placement with 250 mm and 500 mm sized letterforms.
SC terminates early with both.
Its execution time is approximately the same as VFG with 250 mm letterforms.
It is slower than VFG with 500 mm letterforms due to a larger number of candidates $|C|$.

%SC is able to search the solution space exhaustively with 250 and 500 mm sized letterforms, computing the same placement as SC.
%With 250 mm, except for K, VFG computes the same placement as SC.  
%With 500 mm, VFG is able to compute the same placement as SC for all letterforms.
With 1000 mm letterforms, SC's placement requires (a) 1 more \lb than VFG for M and N, (b) 1 fewer \lb than VFG for A, B, and G, and (c) same number of \lbs as VFG for other letters.
The execution time of VFG is significantly faster than SC.
SC is unable to search the entire solution space when limited to visit $\beta$=1 million branches for 9 of the letters (34\% of the letters).
This explains why it does not compute the same solution as VFG for M and N.
At the same time, this does not mean SC is inferior when it fails to search the entire solution space.
B is an example.
SC computes its placement requiring 1 less \lb than VFG while terminating early due to $\beta$.
%(With A and G, SC searches the entire solution space by pruning it effectively.)
%For the other letters, VFG and SC place the same number of \lbs.
%Every time a technique computes fewer \lbs for a letter, its utilization is proportionally higher.

%The placement of LightBenders with K is shown in Figure~\ref{}.

\begin{figure}
    \centering
    \includegraphics[width=0.6\linewidth]{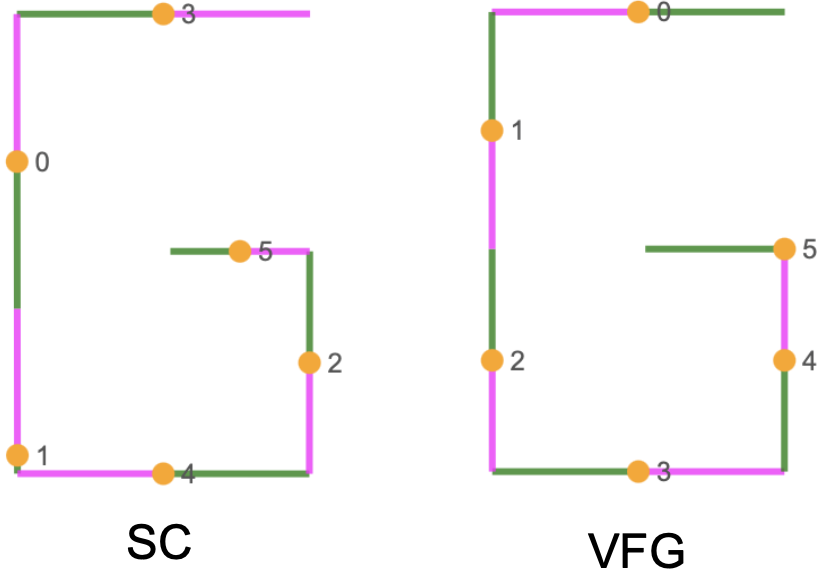}
    \caption{500 mm sized G. VFG and SC use 6 \lbs with the same utilization. VFG uses 1 segment of \lb 5, while SC uses both segments of \lb 5 partially.}
    \label{fig:g}
\end{figure}

With Base, utilization is a function of the number of placed LightBenders. SC and VFG may place the same number of LightBenders for a letter in different ways.
However, their utilization may still be equal.
An example is 500 mm G in 
Figure~\ref{fig:g}.
%While SC uses one more segment than VFG, the lighting of these segments is such that the overall utilization is the same.

These results highlight the tradeoff between the speed of execution and the number of required \lbs.
VFG executes faster and places either the same number of \lbs or one less for most cases. 
There are a few cases where SC places fewer \lbs.
However, its execution time is significantly slower than VFG.

\noindent{\bf Mixed:}
A heterogeneous mix of \lbs enlarges the solution search space.
VFG uses best fit (see Definition~\ref{def:bestfit}) to guide its placement of \lbs while SC considers all possibilities.
With 250 mm and 500 mm letter sizes, SC is able to prune the search space effectively to consider all possibilities without exhausting its 1 million limit.
With 1000 mm letter sizes, SC exhausts the $\beta$=1 million limit with 70\% of (18 out of 26) letters.

While VFG places fewer \lbs than SC, SC's utilization is higher because it uses \lbs with shorter segments more frequently than VFG.
To elaborate, with 250 mm letterforms, SC and VFG place the same number of \lbs.
However, SC's utilization is higher for 88\% of the letters and the same as VFG for the remaining letters. 
With 500 mm letterforms, VFG places the same number of \lbs as SC for all letters except Y, for which it uses one fewer LightBender.
However, SC's utilization is higher for all letters including Y.
With 1000 mm letterforms, VFG uses fewer \lbs than SC for most letters.
However, SC's utilization is also higher for all letters.
For example, SC's utilization is 18.8\% higher with T.  
It uses the following mix of 5 \lbs:
two 130 mm, two 160 mm, and one 240 mm.
VFG also uses 5 \lbs.
However, its mix favors \lbs with longer segments:
one 130 mm, one 160 mm, and three 240 mm.
This is due to its use of best fit \lbs, see Definition~\ref{def:bestfit}.
%While its execution time is slower for these cases, it is difficult to argue its placement is inferior. Even though SC uses more \lbs than VFG, its overall LED utilization of its placement is higher.

%We highlight the tradeoff between fewer \lbs and a higher utilization with the following.

%With 250 mm letter sizes, SC and VFG place the same number of \lbs for all letters.  A key difference is that SC maximizes the utilization of LEDs. To illustrate, consider the placement of \lbs for V. See Figure~\ref{}. Its two sides are 261 and 263 mm in length.  VFG's best fit places two \lbs, each with 240 mm segment, on each side. This results in approximately a 55\% utilization. SC also uses two \lbs:  one with 130 mm segment size and the other with 160 mm segment size.  This results in 90\% utilization.

\subsubsection{Skyline Graphic}
With both Base and Mixed, VFG computes a placement in less than two milliseconds while
SC requires a minimum of 400 milliseconds by exhausting its $\beta$=1 million limit.

\noindent{\bf Base:} VFG places 57 \lbs and utilizes 61.5\% of their LEDs.
SC places 61 \lbs with 57\% utilization.
With SC, we increased $\beta$ to 10 million and 100 million.
This increased execution time without providing a better solution.
Hence, we conclude VFG is superior to SC with these experimental settings.

%With the Mixed system, 
\noindent{\bf Mixed:} VFG uses fewer \lbs than SC, 49 versus 53. 
However, its utilization of LEDs is lower than SC, 60.5\% versus 64.4\%. 
While SC considers a mix of \lbs with short segments, VFG's best fit (Definition~\ref{def:bestfit}) does not consider such possibilities.
VFG's solution uses almost twice as many 240 mm \lbs.
%when compared with SC.
Its placement requires nineteen 130 mm, five 160 mm, and twenty five 240 mm \lbs.
SC's placement requires thirty 130 mm, nine 160 mm, and fourteen 240 mm \lbs.
%VFG's use of best fit (Definition~\ref{def:bestfit}) does not consider fitting several \lbs with shorter segments. It favors \lbs with longer segments, using nineteen 130 mm, five 160 mm, and twenty five 240 mm \lbs.

\subsection{Discussion}\label{sec:discuss}
Results of Section~\ref{sec:vfgvssc} highlight the tradeoff between VFG and SC as two alternative placement techniques.  The choice between them trades execution time for the number of placed \lbs and their LED utilization.
It is possible to combine these techniques into a hybrid, HYB, with a not to exceed execution time $\Theta$.
It is based on the following observations from Section~\ref{sec:vfgvssc}.
First, VFG's placement is typically superior to SC's greedy by minimizing the number of placed \lbs.
Second, SC uses its own greedy heuristic as an $X_{best}$ seed solution to prune the search space effectively.
The key difference between this heuristic and VFG is that it considers a set of candidates in $C$ while VFG traverses a graph of vertices and edges.
%While this technique computes a set of candidates $C$ that it considers in turn, VFG traverses a graph of vertices and edges.
Third, both SC's greedy and VFG are fast with sub-millisecond execution times for the different letterforms.
With the skyline graphic, their execution time is faster than 2 milliseconds.
Based on these observations, HYB may invoke both VFG and SC's greedy techniques to identify $X'_{best}$.
%This is feasible because both greedy techniques are extremely fast with sub millisecond execution times.
%replace SC's greedy with VFG.Alternatively, since both are extremely fast (sub millisecond execution times), another possibility is to require SC to execute both its greedy technique and VFG, selecting the best placement to prune the search space of possibilities.
Subsequently, it may use SC's branch and bound 
%and its pruning 
with $X'_{best}$ for $\Theta$ time units or $\beta$ branches.

\noindent{\bf Base:} 
%With one type of \lb, 
HYB is superior to both SC and VFG because it capitalizes on both their strengths.
To elaborate, for the 1000 mm letterforms, HYB minimizes the number of required \lbs, similar to VFG, while also maximizing segment utilization. The number of placed \lbs is as low as VFG’s (and typically one fewer than SC’s), while the segment utilization matches SC’s and exceeds that of VFG. In this sense, HYB combines the advantages of both approaches. For the skyline graphics, HYB produces the same placement as VFG, which is superior to SC.
%These results can be explained as follows. HYB selects, as the baseline for its search, the better solution produced by either VFG or SC Greedy. Because VFG always has zero overlap, HYB chooses the VFG solution whenever the number of \lbs computed by VFG is less than or equal to that computed by SC Greedy. If the subsequent search identifies a better solution than the baseline, that improved solution corresponds to one that SC could compute. Otherwise, the baseline solution is retained. In effect, HYB always returns the better of the VFG and SC solutions.
%With only one \lb type, any solution using fewer \lbs necessarily achieves higher utilization. Consequently, HYB naturally attains both high utilization and a low number of \lbs.

\noindent{\bf Mixed:} 
HYB minimizes the number of required \lbs. However, minimizing the number of \lbs does not necessarily maximize utilization, as observed for several of the 1000 mm letterforms and the skyline graphics. Utilization is determined by the total rod length used rather than the number of \lbs alone. Consequently, solutions with fewer \lbs may exhibit lower utilization than those produced by SC.

%Moreover, it enables an artist to specify a not to exceed time $\Theta$. The research results of Section~\ref{sec:vfgvssc} are essential to identify HYB.

\begin{comment}
\begin{definition}\label{def:placement_min}
A theoretical lower bound on the number of \lbs:
$$\left\lceil \frac{\sum_{e \in E}\left\| e \right\|}{2 \cdot l_{max}} \right\rceil$$
where the numerator is the total edge length and $l_{max}$ 
is the LightBender segment length. For a heterogeneous mix, $l_{max}$
is substituted with 
$min(L_{seg})$ 
to reflect the shortest available segment.
\end{definition}
\end{comment}

\subsubsection{Theoretical Lower-Bound}
A theoretical lower-bound on the number of required \lbs is to divide the total length of a line drawing by the rod length:
$\left\lceil \frac{\sum_{e \in E}\left\| e \right\|}{2 \cdot l_{max}} \right\rceil$
where the numerator is the total edge length and $l_{max}$ 
is the LightBender segment length\footnote{For a heterogeneous mix, $l_{max}$
is substituted with 
$min(L_{seg})$ 
to reflect the shortest available segment.}.
With letterforms and one type of \lb, HYB realizes this lower bound most of the time, but not always. 
It is infeasible when HYB is able to conduct an exhaustive search.
To elaborate, with 250 mm letterforms and 160 mm rod length, HYB realizes the lower bound for 70\% of the letterforms.
It uses one more \lb for the remaining 30\%.
This is 33\% to 50\% more than the theoretical minimum, e.g., 3 instead of 2.
For these, we know the theoretical lower-bound is infeasible because (a) only a few \lbs are required and it is trivial to show that the lower-bound is not possible and (b) HYB searches the space exhaustively. 
With the skyline graphic, HYB uses 58.3\% more \lbs than the theoretical lower-bound, 57 instead of 36.

\section{Stagger}\label{sec:stagger}
%projection is a 2D image rendered from a 3D scene.

% [x] Must clearly state:  Resolve conflicts by adjusting the positions of one or more of the overlapping LightBender either towards or away from the user's FoV. (Not sure where this should be stated - may require a restart.)
% Knowing the user's FoV, the system may adjust the positions of the conflicting \lbs towards or away from the user to enhance the QoI. 
% The objective of \termss is to minimize flickering due to downwash and to enable the illumination of line drawings that would be infeasible without staggering, e.g., LA in Figure~\ref{}.

% The tentative \lbs layout computed by the Placement phase optimizes the mapping of line segments to \lbs but makes no assumptions about 3D relationships between them. Hence, two types of physical conflicts may arise that degrade the Quality of Illumination (QoI): 

With both authoring tools, a \lb layout may result in two types of conflicts that degrade the Quality of Illumination (QoI):
%With both authoring tools, the resulting tentative \lbs layout, two types of physical conflicts may arise that degrade the Quality of Illumination (QoI): 
downwash and \ovs. 
Downwash results in flickering of Section~\ref{sec:downwash}.
\Ov occurs when two or more \lbs occupy overlapping positions.
This is physically impossible and its forced implementation results in collisions and crashes that may produce dead regions in the display.
%causing them to crash and producing resulting in crashes that produce dead regions in the display.

%To address this, we introduce 
%We address both using Stagger.It resolves conflicts by adjusting the 3D position of conflicting \lbs in regard to the user's viewpoint. We call the resulting layout a staggered layout $\mathcal{L}_{stag}$.
%Its objective is to minimize 
% It prevents \ovs and minimizes flickering, enabling the illumination of line drawings that would be otherwise infeasible. E.g., LA in Figure~\ref{fig:la_example}.
% [ ] add figures that clarify

Spatial staggering, Stagger, addresses both.
It consists of three distinct phases: \textit{Detect}, \textit{Select}, and \textit{Resolve}. 
First, the system processes a tentative layout, $\mathcal{L}_{tent}$, to detect conflicts. Second, it selects the minimum subset of \lbs requiring adjustment. Third, it adjusts the position of these \lbs to resolve the conflicts relative to the user's viewpoint.
Its output is a new layout $\mathcal{L}_{stag}$.
See Figure~\ref{fig:pipeline}.
These steps and their alternative implementations are detailed below.
Their execution time in less than a millisecond. The Blender add-on allows users to revert $\mathcal{L}_{stagger}$ to $\mathcal{L}_{tent}$ and subsequently apply a different heuristic to compute an alternative $\mathcal{L}_{stagger}$ for evaluation.
Section~\ref{sec:ss_eval} evaluates these implementations.
% [ ] we use LA as an example to illustrate the steps.

% The objective of spatial staggering is to minimize flickering due to downwash and to enable line drawings that would not be feasible without staggering, e.g., LA in Figure~\ref{}.
% It consists of three distinct steps: Detect, Identify, and Stagger. The first step processes the graph representation of a line drawing and its monitored data (if any) to detect the requirement to stagger LightBenders spatially.
% In Step 2, it identifies the LightBenders that should be staggered spatially and those whose lighting should be adjusted.
% Finally, in Step 3, it staggers the LightBenders by adjusting their position.
% This step may adjust the lighting of one or more LightBenders in response to staggering.
% To illustrate, if a LightBender is moved closer to the user's Field of View, FoV, its rod may appear longer. 
% In this case, some of the LEDs on its rod may be turned off to restore its intended length.
%Its objective is to minimize flickering due to downwash and to enable line drawings that would not be feasible without staggering. 

\subsection{Detect}
We model \ov and downwash conflicts using a 
%We detect conflicts in the layout where \lbs are either physically overlapping or positioned within the downwash of another \lb. We model these interactions using a 
conflict graph $G_k=(V_k, K)$, where $V_k$ is the set of \lbs that participate in at least one conflict and the edge set $K$ connects the conflicting pairs. Each edge in $K$ encodes the type of conflict, overlap or downwash.
% A conflict is defined as a binary interaction between exactly two \lbs. It occurs when they collide or when one \lb is positioned within the downwash region of another. 
% To model this, we construct a conflict graph $G_k=(V_k, K)$, where $V_k$ represents the set of \lbs and $K$ represents the set of edges connecting conflicting pairs.
%Conflicts 
We detect these using the following geometric approximations.
With \ovs, we model 
%\textbf{\Ovs:} We model 
each \lb as a sphere with diameter $d_s$ determined using \lb's dimension (150 mm). An edge is added to $K$ between two \lbs if their Euclidean distance is less than $d_s$.
%\textbf{Downwash:} We 
With downwash, we model the downwash region as a cylinder with diameter $d_c$ ($d_c$=$d_s$=150 mm in this paper) extending below each \lb along the entire z axis. A downwash conflict exists between two \lbs if their downwash cylinders intersect.
%one intersects the downwash cylinder of another. 
This is geometrically equivalent to projecting both \lbs onto the XY plane and testing if their planar distance is less than $d_c$.

%Constructing $G_k$ 
Figure~\ref{fig:dtla_conflicts} shows the conflicts for the skyline graphic using the above methodology.
Each of the 57 \lbs in this layout conflicts with at least 3 other \lbs,
forming a conflict graph $G_k$ with $V_k$=57 vertices and $K$=250 edges (167 downwash and 83 \ov edges).

\subsection{Select}\label{sec:staggerselect}
The objective of Select is to identify the minimum subset of \lbs whose position must be adjusted to eliminate all conflicts $K$ in $G_k$. Minimizing this subset is important for two reasons. First, each displaced \lb degrades QoI proportional to its displacement magnitude (Section~\ref{sec:dist}). Second, displacing fewer \lbs reduces the total perspective correction required, limiting the possibility of exceeding rod length which requires additional \lbs. 

A Minimum Vertex Cover\cite{Karp1972} (MVC) formulation of this problem is as follows:
Find the smallest subset $S \subseteq V_k$ such that every edge in $K$ is incident to at least one vertex in $S$. $S$ is the cover set.
Since MVC is NP-hard, we employ a hybrid approach depending on the size of $V_k$:

% Since MVC is NP-hard, we employ a brute-force solution for small instances of $V_k$ and evaluate three heuristic approximations for larger swarms: a greedy approach based on the maximum vertex degree, a Top-Z first approach, and a Bottom-Z first approach.

\subsubsection{Brute-Force}
For small conflict graphs such as those for letterforms (typically $|V_k| < 20$), we exhaustively enumerate all $2^{|V_k|}$ subsets of $V_k$ to find the smallest that covers all edges in $K$. This guarantees the absolute minimum number of \lbs at time complexity $O(2^{|V_k|} \cdot K)$.

\subsubsection{Heuristic Approximations}
\label{sec:ss_heuristics}
For larger conflict graphs such as the skyline graphic, we use heuristics.
This section presents four heuristics.
Each iteratively selects vertices to add to the cover set $S$ until all edges in $K$ are covered.

\textbf{Max-Degree} is a standard MVC approximation that selects high-conflict \lbs first. 
At each step, it selects the vertex with the highest degree in the conflict graph (i.e., the \lb with the most conflicts). Removes it and its incident edges from the $G_k$. 

\textbf{Top-Z}
sorts $V_k$ in descending order of Z-coordinate. At each step, it selects the highest \lb that is an incident to an uncovered edge. The intuition is that downwash interferences propagate downwards, prioritizing the selection of the topmost \lbs resolves cascades of conflicts.

\textbf{Bottom-Z} is the counterpart to Select:Top-Z. It selects the lowest \lb involved in a conflict at each step. This approach addresses downwash from the impacted \lb.% receiver's perspective.

\textbf{Random} chooses a vertex randomly from the set of conflicting \lbs at each step. It serves as a baseline.

We compare these alternatives in Section \ref{sec:ss_eval}.
Obtained results show Select:Max-Degree performs the same as brute-force in all our experiments. A key difference is that it is significantly faster.

\subsection{Resolve}
\label{sec:resolve}

\begin{figure}
    \centering
    \includegraphics[width=\linewidth]{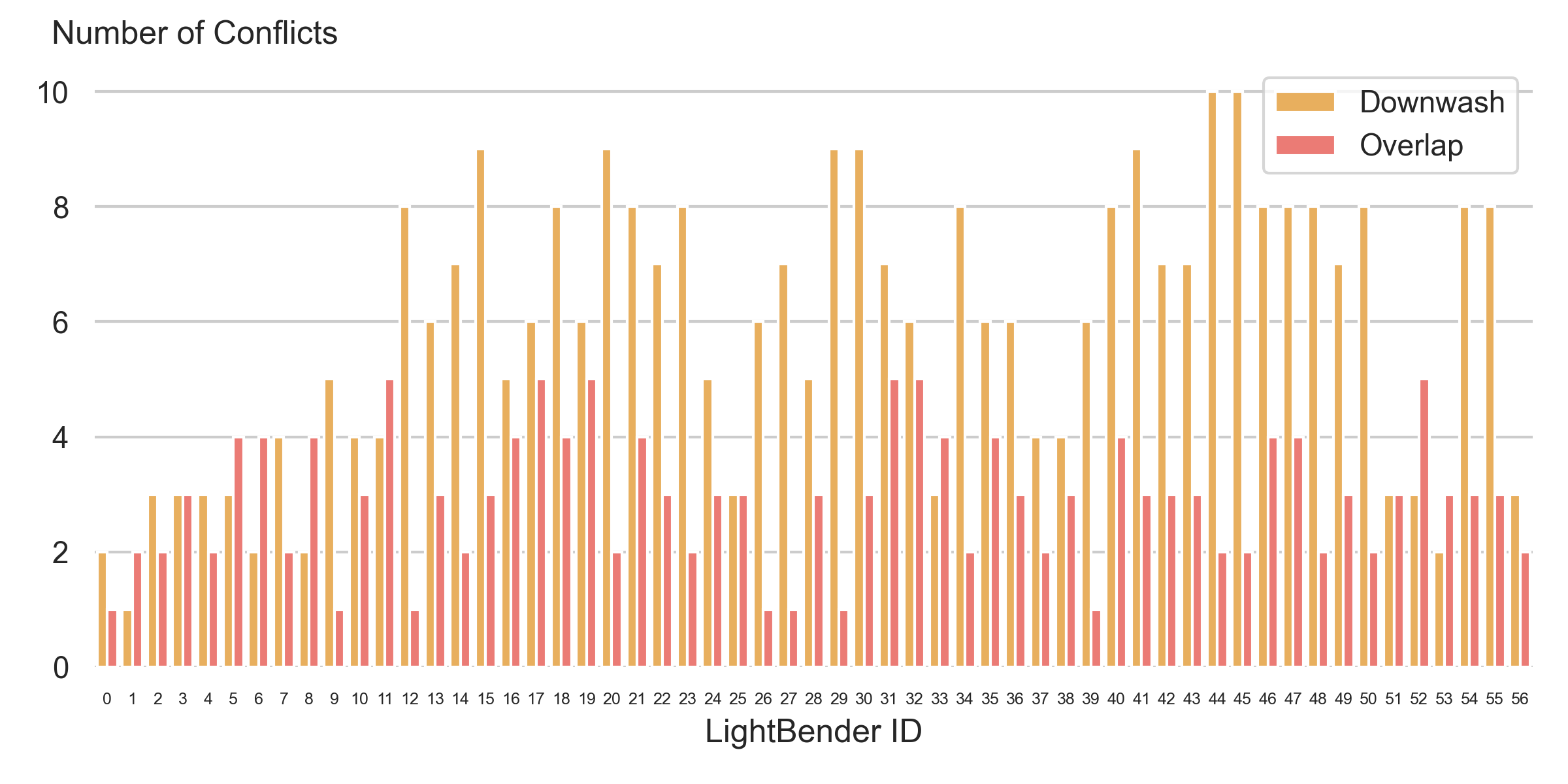}
    \caption{Downwash and overlap conflict with the skyline graphic, VFG placement, and one \lb type.}
    \label{fig:dtla_conflicts}
\end{figure}

\begin{figure}
    \centering
    \includegraphics[width=\linewidth]{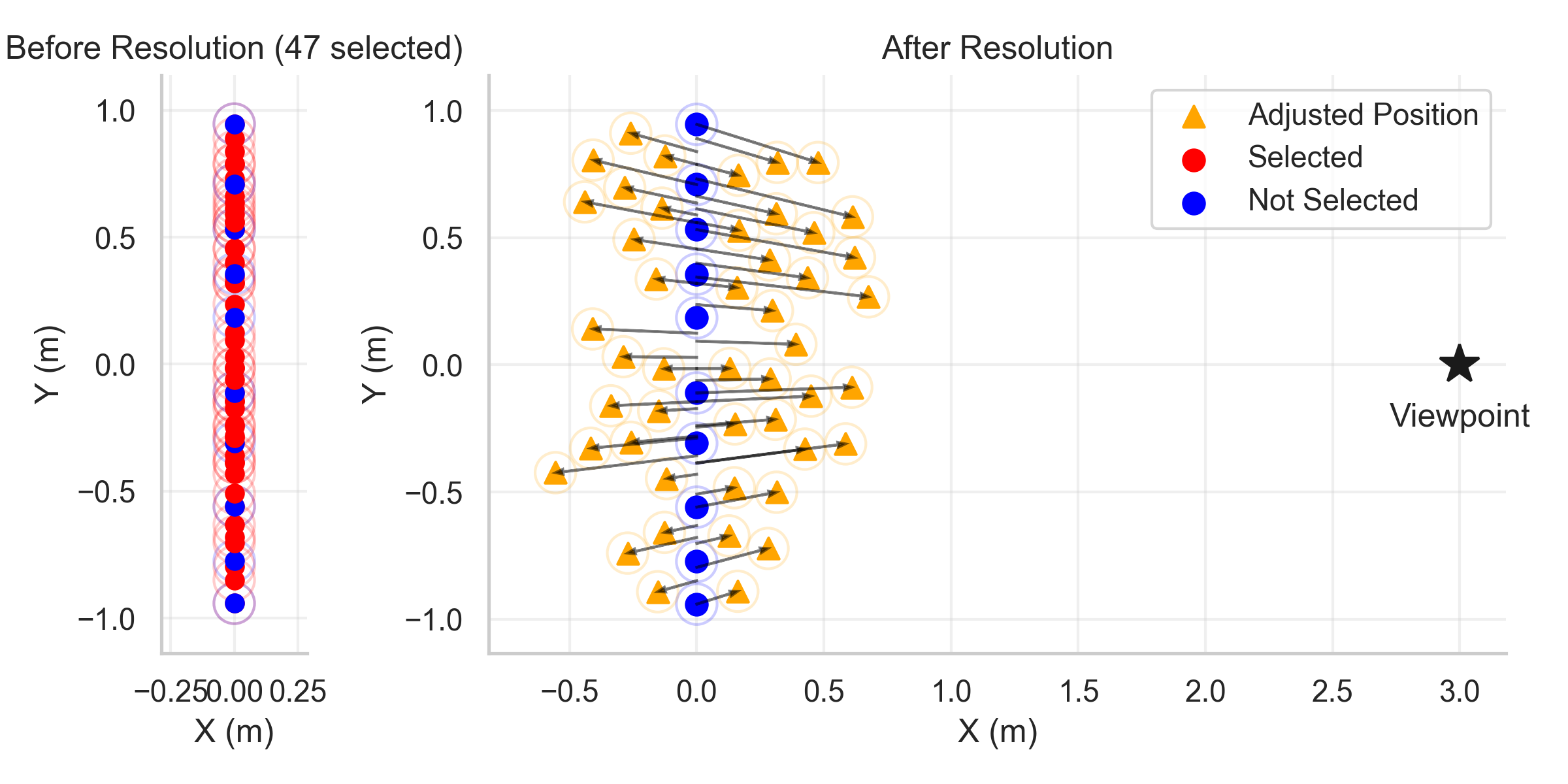}
    \caption{{\small Top view of \lbs before and after resolution with LoS and VFG placement and 1 type of \lb.}}
    \label{fig:dtla_stagger}
\end{figure}

Resolve adjusts the position of the \lbs in the cover set $S$ to eliminate conflicts.
Its objective is to preserve the QoI~\cite{tomm2025,reliability2024} from the viewpoint of a single user.
It decides the {\em order} in which \lbs move to resolve conflicts, the {\em trajectory} of movement for each \lb, its {\em distance} of movement, and adjustment to the lit LEDs of an illuminated line segment when it is {\em displaced closer or away} from the user's viewpoint.
We describe each in turn.
%It uses four policies: ordering, trajectory, distance, and perspective correction.
%, with the objective of preserving the QoI from the user's viewpoint.

% [x] replace camera with user
\subsubsection{Ordering}
\label{sec:ordering}
Resolving a conflict should not introduce a new one.
The order in which conflicts are resolved affects the final layout of \lbs.
To determine this ordering, we use the heuristics of Section~\ref{sec:ss_heuristics}: Max-Degree, Top-Z, Bottom-Z, and Random. 

\subsubsection{Trajectory}
\label{sec:trajectory}
When repositioning a \lb, Resolve selects a displacement trajectory to preserve the 2D projection of the line drawing from the user's viewpoint.
It may compute this trajectory for each \lb individually or globally once using the swarm's centroid.
These two alternatives are termed Line-of-Sight (LoS) and Global, respectively.
We describe them in turn.

LoS defines the trajectory for each \lb $i \in S$ as 
%its individual line of sight: 
the unit vector $\hat{v}_i$ originating at the user's gaze 
and passing through the centroid of the \lb $i$.
See Figure~\ref{fig:dtla_stagger}.
Displacing along $\hat{v}_i$ leaves the 2D projection of \lb $i$ invariant with respect to depth, because any point on a line of sight maps to the same 2D position under perspective projection regardless of its distance from the viewer.

\begin{figure}
    \centering
    \includegraphics[width=.9\linewidth]{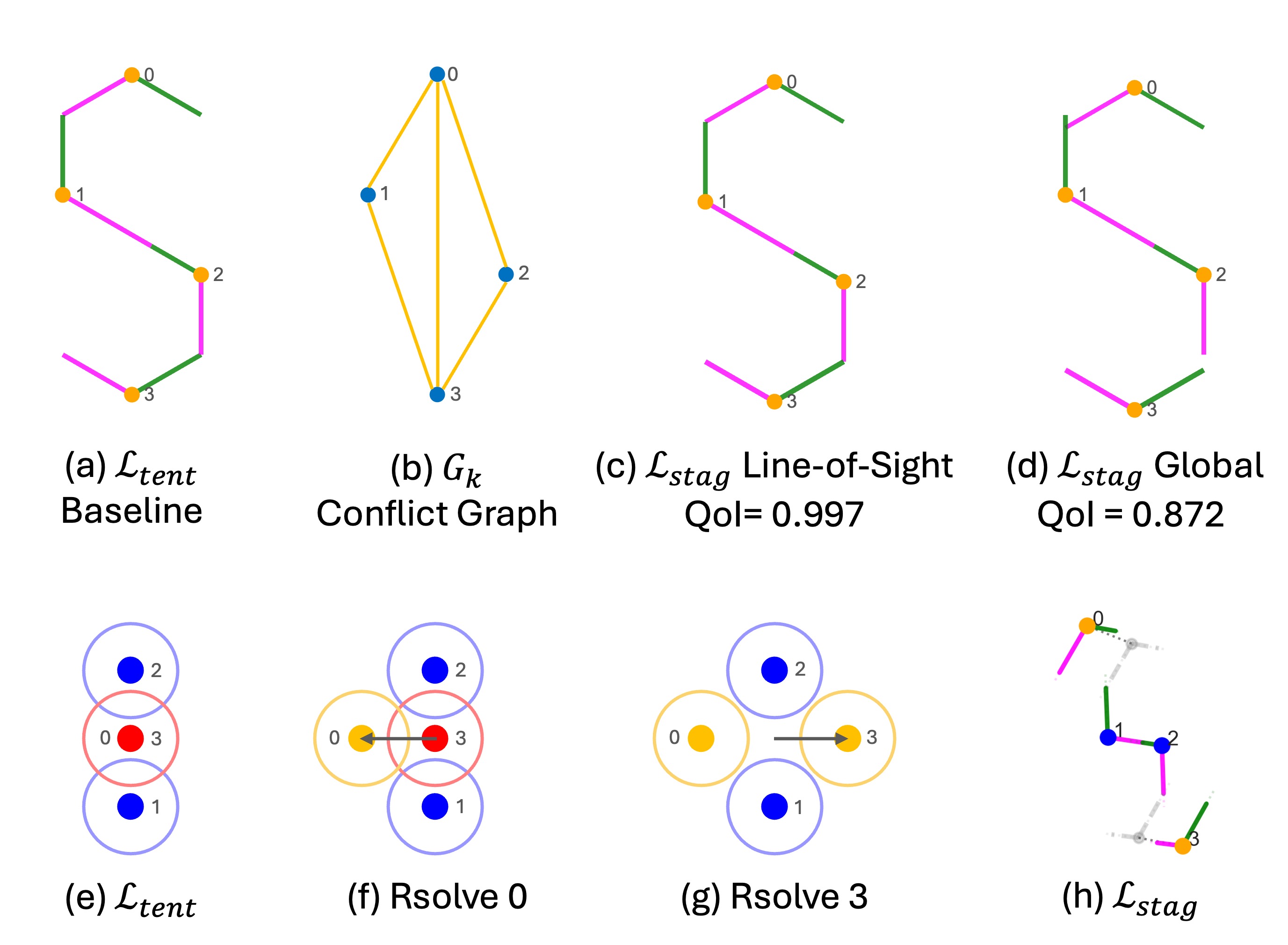}
    \caption{A trajectory towards the swarm's centroid, Global, results in visible gaps and misalignments as shown in d.  LoS prevents these gaps as shown in c from the user's viewpoint.  The side view of c is shown in h.}
    \label{fig:s_stagger}
\end{figure}

%An alternative that adjusts along a global direction, 
%Adjusting instead along a global direction, such as 
Global uses 
the vector toward the swarm's centroid.
It introduces parallax errors in the form of adjusted \lb projecting to a different 2D position than intended, resulting in visible gaps and misalignment in the perceived line drawing as shown in Figure~\ref{fig:s_stagger}d.

\subsubsection{Displacement distance}\label{sec:dist}
For each \lb $i \in S$, the Resolve step computes the minimum displacement $\delta_i$ along $\hat{v}_i$ such that \lb $i$ exits all conflict regions. 
\begin{equation}
    \delta_i = \min \{ \delta \in \mathbb{R} \mid 
    \text{no conflict exists at position } 
    \mathbf{p}_i + \delta \cdot \hat{v}_i \}
    \label{eq:delta}
\end{equation}
where $\mathbf{p}_i$ is the current position of \lb $i$.
The search considers both directions along $\hat{v}_i$, toward and away from the user, and selects the direction that yields the smaller $|\delta_i|$. This minimizes the displacement and thus the introduced QoI degradation. 

We solve Equation~\ref{eq:delta} by stepping along $\hat{v}_i$ with step size $\sigma$. By starting from $\delta = 0$ and incrementing (decrementing) it, we search away from (toward) the user until the conflict is resolved. A low $\sigma$ value yields a solution closer to the true minimum at the cost of additional iterations. A large $\sigma$ value terminates faster but may overshoot and produce a displacement larger than necessary\footnote{It is possible to use hybrids using large $\sigma$ values to identify an interval that is subsequently searched using smaller $\sigma$ values.}. We consider searching away from the user, toward the user, or both as \textit{Away}, \textit{Toward}, and \textit{Hybrid} policies, respectively.

% We quantify the degradation of the Quality of Illumination (QoI) as a function of the displacement magnitude. Therefore, the system calculates the minimum distance $\delta$ required to move a \lb along the user's line of sight until it exits the conflict region. 
% [ ] we want to minimze the distance between the new and the original position.
% [ ] equations.

% [x] add post processing section to describe the scaling of lines.

% \paragraph{Perspective Correction}\label{sec:pcorr}
Displacing a \lb 
%by $\delta_i$ 
changes its distance from the user and thus its apparent size. To maintain the intended length of the line, the displayed line length by the \lb (the number of lit LEDs) must be adjusted.

{\bf Towards} the user ($\delta_i < 0$): The \lb appears larger.
%The line must be shortened by turning off the LEDs at the segment tips to restore the intended length. 
Its LEDs at the segment tips are turned off to restore the intended length.
This is always feasible since turning LEDs off is unconstrained.
% \textbf{Moving Away from the User:} The line length is increased. If the required length exceeds the hardware limit ($l_{max}$), the move is flagged as invalid, and the system reports an error.

{\bf Away} from the user ($\delta_i > 0$): The \lb appears smaller, requiring the intended illuminated line to become longer. This is bounded by the maximum segment limit, $l_{max}$, of the \lbs. The required physical length is:
\begin{equation}
    l_{req} = l_{initial} \cdot 
    \frac{d_{user} + \delta_i}{d_{user}}
    \leq l_{max}
    \label{eq:perspective}
\end{equation}
where $d_{user}$ is the initial distance from the user to the \lb $i$. 
If $l_{req}$ > $l_{max}$, the LEDs of the segment are exhausted.
%the away-from-user direction is infeasible for this \lb. In this case, 
Resolve uses new \lb to elongate the existing line segment.
%that allows longer line segments.
%to be displayed by two \lbs instead of one. 
The original and new align to illuminate one 
%assumes the responsibility of one 
of the original segments.
%in the original placement.  
% An alternative is to replace the \lb with one with a larger segment length (if available) to display a longer line.

% [ ] Evaluation section

\begin{example}
With S in Figure~\ref{fig:s_stagger}a, \lb 0 has a downwash conflict with \lbs 1, 2, and 3. And, \lbs 1 and 2 have a downwash conflict with 3. The resulting conflict graph has four vertices and five edges, as shown in Figure~\ref{fig:s_stagger}b. Vertices 0 and 3 have degree three; vertices 1 and 2 have degree two. 

To find a minimum vertex cover $\Omega$, Select's Max-Degree heuristic selects either vertex 0 or 3 and removes it and its three incident edges from $G_k$.
Assuming vertex 0 is selected, $\Omega=\{0\}$.
The remaining graph contains vertex 3 with degree two and vertices 1 and 2 with degree one. Select's Max-Degree next selects vertex 3, whose removal eliminates all remaining conflict edges, yielding $\Omega=\{0,3\}$.

With LoS and the Hybrid displacement heuristic, Resolve first repositions \lb 0 away from the viewpoint until it no longer overlaps the conflict regions of the fixed \lbs (1 and 2). See Figures~\ref{fig:s_stagger}e-f. Subsequently, it adjusts \lb 3. Because the region away from the viewpoint is now occupied by \lb 0, displacing \lb 3 in that direction would require a larger adjustment than moving it toward the viewpoint. Thus, \lb 3 is displaced toward the viewpoint. This yields the staggered configuration shown in Figure~\ref{fig:s_stagger}h.~$\blacksquare$
\end{example}

% {\color{blue}

\subsection{Comparison and Ablation Study}\label{sec:ss_eval}
We compare the alternative Select and Resolve heuristics using the skyline graphic\footnote{VFG placement with 1 \lb type.}. Its tentative layout, $\mathcal{L}_{tent}$, produced by the placement technique (see Figure~\ref{fig:pipeline} and Section~\ref{sec:placement}) contains 167 downwash and 83 \ov conflicts, see Figure~\ref{fig:dtla_conflicts}. 
This is a complex scenario in which every \lb conflicts with at least three others.
Figure~\ref{fig:dtla_stagger} shows the staggered layout, $\mathcal{L}_{stag}$, after resolving these conflicts.
%Since every \lb conflicts with at least 3 other \lbs, this is a stress test for the Stagger step.

Below, we define QoI as a metric to quantify the impact of staggering on $\mathcal{L}_{tent}$, comparing it with $\mathcal{L}_{stag}$.
Subsequently, we present an ablation study to quantify the benefits of alternative techniques in Sections~\ref{sec:staggerselect}-\ref{sec:resolve}.
A key finding is that Resolve's LoS trajectory technique (Section~\ref{sec:trajectory}) is fundamental, providing the highest QoI.
Other heuristics improve QoI modestly at best.
%However, their improvement is not that significant.

\subsubsection{Quality of Illumination, QoI}
\label{sec:qoi_metric}
We quantify QoI from the user's viewpoint, which is modeled as a camera.
This camera points towards the swarm centroid, see the star labeled ``Viewpoint'' in Figure~\ref{fig:dtla_stagger}.
Stagger's movement of an illuminated line segment impacts its visual properties in two ways.
First, its width increases (decreases) when the line is moved closer (away) from the viewpoint.
Second, the line's endpoints may change.
With the first, we simply quantify its impact on QoI. 
With the second, we are able to turn dark (lit) LEDs on (off) to compensate, see discussions of Equation~\ref{eq:perspective}.
Below, we quantify these two in turn.
Subsequently, we define QoI.

%To evaluate QoI, we project both $\mathcal{L}_{tent}$ and $\mathcal{L}_{stag}$ into 2D image space using a perspective camera model. Each projection results in an SVG with $n$ line segments. Stagger affects two visual properties of each segment. 
%First, the line width changes because \lbs closer to the viewpoint appear larger.
Line widths change because \lbs moved closer to (farther from) the viewpoint appear thicker and longer (slimmer and shorter). 
The width after displacement is: $w_{new} = \frac{d_{old}}{d_{new}} \cdot w_{old}$ where $d_{old}$ and $d_{new}$ are distances of a \lb from viewpoint before and after displacement, and $w_{old}$ is the original line width.
The average width error across all $n$ line segments is:
\begin{equation}
    E_{width} = \frac{1}{n} \sum_{j=1}^{n} |w_j - w'_j|
    \label{eq:width_error}
\end{equation}
% The 2D endpoint positions shift for any line displaced off its exact line of sight.
% %Each line has two endpoints in each of $\mathcal{L}_{tent}$ and $\mathcal{L}_{stag}$.
% The following quantifies the difference across all $n$ line segments:
Furthermore, displacing a line off its exact line of sight shifts its 2D endpoint positions. We quantify this positional error across all $n$ line segments as:
%We quantify both across all $n$ line segments:
\begin{equation}
    E_{pos} = \frac{1}{2n} \sum_{j=1}^{n} 
    \left( \|\mathbf{a}_j - \mathbf{a}'_j\| + 
           \|\mathbf{b}_j - \mathbf{b}'_j\| \right)
    \label{eq:pos_error}
\end{equation}
where $\mathbf{a}_j$ and $\mathbf{b}_j$ are the endpoints of a line segment in $\mathcal{L}_{tent}$ and their prime, $a'_j$ and $b'_j$, are the endpoints of the same line in $\mathcal{L}_{stag}$. 

We combine these two metrics to quantify QoI using exponential decay:
\begin{equation}
    %QoI = \exp\!\left(-\alpha \cdot 
    %\frac{E_{pos} + E_{width}}{D}\right), %\quad \alpha = 200
    QoI = e^{\!\left(-\alpha \cdot 
    \frac{E_{pos} + E_{width}}{D}\right)} \quad \alpha = \{1, 100, 200\}
    \label{eq:qoi}
\end{equation}
where $D$ is the image diagonal $D = \sqrt{W^2 + H^2}$ ($D = 2202.91$~px). 
The constant $\alpha$ magnifies small differences to differentiate heuristics.
A larger $\alpha$ amplifies the value for small errors.
See Table~\ref{tab:ablation} and its discussion.

QoI equals 1 when $\mathcal{L}_{tent}=\mathcal{L}_{stag}$.
More generally, a QoI score close to 1 indicates the staggered projection is identical to the tentative projection from the camera viewpoint.

% Calibrate alpha using the quality ratings of the user study.

\subsubsection{Comparison}
With skyline, we evaluated all 192 combinations of Select heuristics (Max-Degree, Top-Z, Bottom-Z, Random), Resolve ordering (Max-Degree, Top-Z, Bottom-Z, Random), trajectory (Line-of-Sight, Global), and displacement (Away, Towards, Hybrid) with two $\sigma$ values, $\sigma=\{10,200\}$~mm. A combination resolved all conflicts of Figure~\ref{fig:dtla_conflicts}.

Obtained results highlight the following lessons:
\begin{enumerate}
    \item Trajectory of Section~\ref{sec:trajectory} is the dominant factor and LoS is superior to Global.
    \item The Hybrid displacement technique of Section~\ref{sec:dist} achieves the shortest mean displacement and higher QoI relative to its alternatives.
    % \item Smaller $\sigma$ in Section~\ref{sec:dist} yields higher QoI.
    % \item Select:Max-Degree of Section~\ref{sec:ss_heuristics} selects fewest \lbs.
    % \item Select heuristics of Section~\ref{sec:ss_heuristics} do not impact the QoI.
    %\item QoI is not impacted by the heuristics of Section~\ref{sec:ss_heuristics} that select \lbs to stagger.  If the objective is to stagger the minimum number of \lbs then Select:Max-Degree is superior to its alternatives.
    \item Resolve ordering techniques of Section~\ref{sec:ordering} have a negligible impact on QoI.
    
    %Searching both toward-user and away-from-user (Hybrid) is necessary for full conflict resolution.
    
\end{enumerate}
The highest QoI is consistently achieved by combining LoS with the Hybrid displacement, the max-degree heuristic for both Select and Resolve steps, and small $\sigma$ values.
To better explain the contribution of each component and the sensitivity to parameter selection, we present an ablation study below.

\begin{table*}[htbp]
\centering
\caption{Ablation study.  First two rows use Max-Degree for Select and Resolve phases, Hybrid displacement, and $\sigma$=10 mm.}
\label{tab:ablation}
\begin{tabular}{ccccccccccccc}
\toprule
\toprule
\multirow{2}{*}{\textbf{}} & & \multirow{2}{*}{\textbf{Heuristic}}
 & \textbf{Moved (added)} & \multicolumn{3}{c}{{\textbf{QoI}}}
 & \multicolumn{3}{c}{\textbf{Distance Moved (mm)}} & \multirow{2}{*}{\textbf{$E_{pos}$}} & \multirow{2}{*}{\textbf{$E_{width}$}} \\

 & & & \textbf{\lbs} & \textbf{$\alpha=1$} & \textbf{$\alpha=100$} & \textbf{$\alpha=200$} & \textbf{Min} & \textbf{Avg} & \textbf{Max} &  &  \\
\midrule
\midrule

\multirow{4}{*}{\rotatebox[origin=c]{90}{\small Trajectory}} & \multirow{4}{*}{\rotatebox[origin=c]{90}{\small \S~\ref{sec:trajectory}}} & \multirow{2}{*}{LoS} & \multirow{2}{*}{47} & \multirow{2}{*}{0.99996} & \multirow{2}{*}{0.986} & \multirow{2}{*}{0.971} & \multirow{2}{*}{120} & \multirow{2}{*}{310} & \multirow{2}{*}{680} & \multirow{2}{*}{0.0447} & \multirow{2}{*}{0.276}  \\

 & &  & & &  &  &  &  &  &   &  \\
 
 & & \multirow{2}{*}{Global} & \multirow{2}{*}{47} & \multirow{2}{*}{0.994}  & \multirow{2}{*}{0.525} & \multirow{2}{*}{0.275} & \multirow{2}{*}{110} & \multirow{2}{*}{320} & \multirow{2}{*}{620} & \multirow{2}{*}{13.925} & \multirow{2}{*}{0.282} \\

  & &  &  &  &  &  &  &  &  &  \\
   
\bottomrule

\multirow{3}{*}{\rotatebox[origin=c]{90}{\small Select}} & \multirow{3}{*}{\rotatebox[origin=c]{90}{\small \S~\ref{sec:ss_heuristics}}} & Bottom-Z & 48  & 0.9996 & 0.962 & 0.967 & 200 & 450 & 770 & 0.0480 & 0.317  \\

 & &  Top-Z & 48 & 0.9996 & 0.960 & 0.968 & 40 & 327 & 810 & 0.0573 & 0.307 \\

 & &  Random & 49 & 0.9996 & 0.964 & 0.965 & 90 & 338 & 880 & 0.0784 & 0.317 \\
\bottomrule

\multirow{3}{*}{\rotatebox[origin=c]{90}{\small Resolve}} & \multirow{3}{*}{\rotatebox[origin=c]{90}{\small \S~\ref{sec:ordering}}} & Bottom-Z & 47  & 0.9996 & 0.962 & 0.966 & 80 & 335 & 800 & 0.0781 & 0.302 \\
							
 & &  Top-Z & 47  & 0.9996 & 0.961 & 0.967 & 130 & 347 & 840 & 0.0476 & 0.320 \\

  & &  Random & 47  & 0.9996 & 0.957 & 0.971 & 130 & 347 & 840 & 0.0129 & 0.316 \\

\bottomrule

\multirow{4}{*}{\rotatebox[origin=c]{90}{\small Displacem.}} & \multirow{4}{*}{\rotatebox[origin=c]{90}{\small \S~\ref{sec:dist}}} &  Towards & 47 & 0.9996 & 0.963 & 0.928 & 130 & 130 & 140 & 0.1496 & 0.679 \\

  & &  Away & 47 (+29) & 0.9998 & 0.978 & 0.956 & 70 & 639 & 1410 & 0.0638 & 0.427 \\

%\bottomrule
\cdashline{3-12}

 &  & \multirow{2}{*}{$\sigma$=200 mm} & \multirow{2}{*}{47} & \multirow{2}{*}{0.9998} & \multirow{2}{*}{0.980} & \multirow{2}{*}{0.959} & \multirow{2}{*}{200} & \multirow{2}{*}{472} & \multirow{2}{*}{1000} & \multirow{2}{*}{0.0190} & \multirow{2}{*}{0.438} \\

%\multirow{2}{*}{\rotatebox[origin=c]{90}{\small Dist}} & \multirow{2}{*}{\rotatebox[origin=c]{90}{\small \S~\ref{sec:dist}}} & \multirow{2}{*}{$\sigma$=200 mm} & \multirow{2}{*}{47} & \multirow{2}{*}{0.9998} & \multirow{2}{*}{0.980} & \multirow{2}{*}{0.959} & \multirow{2}{*}{200} & \multirow{2}{*}{472} & \multirow{2}{*}{1000} & \multirow{2}{*}{0.0190} & \multirow{2}{*}{0.438} \\

 &  &  & &  &  &  &  &  &  &  &  \\
		
\bottomrule
\bottomrule
\end{tabular}
\end{table*}

\begin{figure}[h]
    \centering
    \begin{subfigure}{\columnwidth}
        \centering
        \includegraphics[width=\linewidth]{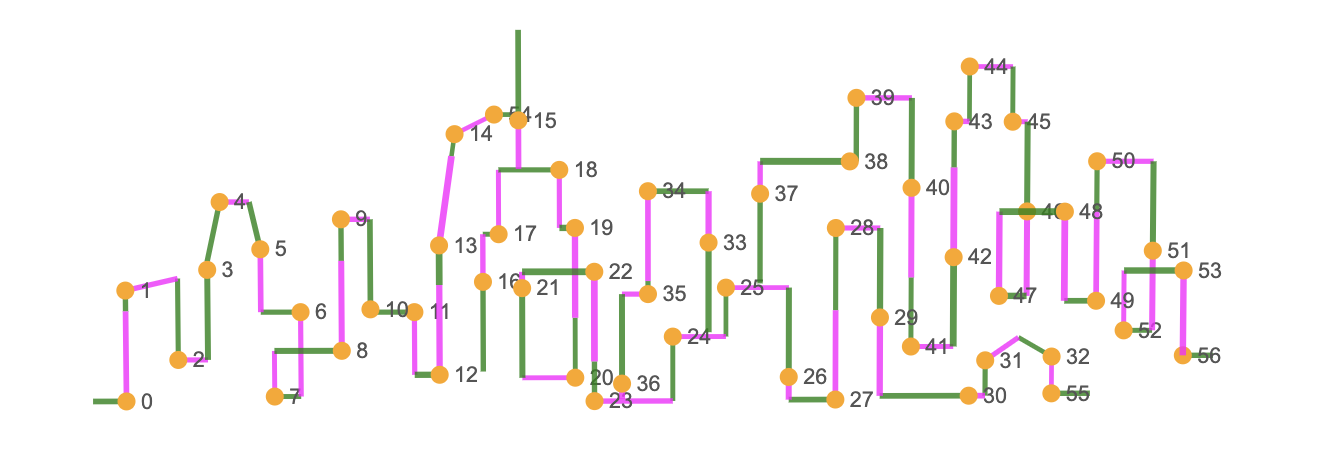}
        \caption{LoS staggering of Skyline graphic, Row 1 of Table~\ref{tab:ablation}}\label{fig:skyline_los}
    \end{subfigure}

    \begin{subfigure}{\columnwidth}
        \centering
        \includegraphics[width=\linewidth]{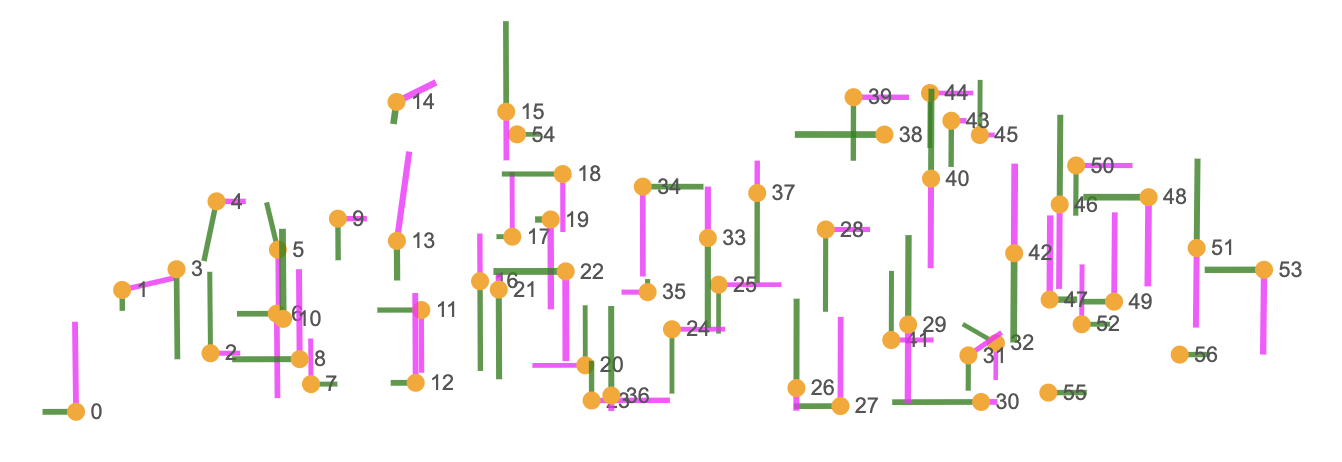}
        \caption{Global staggering of Skyline graphics, Row 2 of Table~\ref{tab:ablation}.}\label{fig:skyline_global}
    \end{subfigure}
    
    \caption{QoI differences impact illuminations significantly.}
    \label{fig:staggered-skyline}
\end{figure}

\subsubsection{Ablation study}
Table~\ref{tab:ablation} presents an ablation study of Stagger’s techniques and heuristics.
The first two rows compare the two trajectory techniques ( \S~\ref{sec:trajectory}) using Max-Degree for both Select (\S~\ref{sec:ss_heuristics}) and Resolve (\S~\ref{sec:ordering}), Hybrid displacement (\S~\ref{sec:dist}), and $\sigma$=10 mm for displacement steps (\S~\ref{sec:dist}).
They show Line-of-Sight, LoS, provides a significantly (3.5x) higher QoI than Global.
Both move the same number of \lbs with the same overall average displacement.
Hence, their $E_{width}$ (Eq.~\ref{eq:width_error}) is comparable.
However, LoS moves each \lb relative to the center of its rod and the viewpoint.  This introduces a small error in $E_{pos}$ for each \lb (Eq.~\ref{eq:pos_error}).
Global moves each \lb relative to the center of the swarm.
This increases $E_{pos}$ error by more than two orders of magnitude, diminishing QoI and resulting in a distorted organization of \lbs.
Compare 
Figure~\ref{fig:skyline_los} with~\ref{fig:skyline_global} (and Figure~\ref{fig:s_stagger}c with~\ref{fig:s_stagger}d for letterform S).  
The remaining rows, Rows 3-11, of Table~\ref{tab:ablation} assume the LoS trajectory.

Table~\ref{tab:ablation} shows QoI with three different $\alpha$ values:
1, 100, and 200. 
This value magnifies the small differences. 
With LoS, Row 1, $\alpha$=1 shows the impact of staggering is insignificant, producing a QoI with four nines.
Global has two nines.
The differences are magnified using $\alpha$=100 and 200.
We assume $\alpha$=200 for the remainder of this section.

With the Select phase of Stagger (\S~\ref{sec:ss_heuristics}), Max-Degree is superior to its alternatives, providing the highest QoI while moving the fewest number of \lbs.  Compare first row with Rows 3-5.
However, with the Resolve phase, Max-Degree provides comparable QoI to its alternatives when ordering \lbs (\S~\ref{sec:ordering}).
Compare Row 1 with Rows 6-8.

With displacement (\S~\ref{sec:dist}), Hybrid is superior to its alternatives that move the conflicting \lbs either away or towards the viewpoint always.
Compare Row 1 with Rows 9-10.
Moving conflicting \lbs away causes stagger to introduce 29 additional \lbs to compensate for lines becoming shorter, see discussions of Eq.~\ref{eq:perspective}.
This is a 52\% increase in the original swarm size of 57.
It reduces LED utilization to 40\%.

Finally, the last row of Table~\ref{tab:ablation} increases the Resolve displacement step from $\sigma = 10$ mm to $\sigma = 200$ mm.
A larger $\sigma$ induces greater \lb displacement,
increasing error in the width of illuminated lines.
This is reflected in the higher $E_{width}$ observed in Row~1 compared to the final row.

%Both the Stagger technique and its alternative heuristics execute in less than a millisecond. The Blender add-on allows users to revert $\mathcal{L}_{stagger}$ to $\mathcal{L}_{tent}$ and subsequently apply a different heuristic to compute an alternative $\mathcal{L}_{stagger}$ for evaluation.
%may execute multiple heuristics to provide a user with choices prior to generating an SFL file.  
%They highlight the superiority of LoS with the Hybrid perspective correction.  The other heuristics provide modest gains. Max-Degree is required to minimize the number of moved \lbs.

\section{Illumination using a Swarm of LightBenders}\label{sec:illuminate}

%Our authoring pipeline generates SFL files to control the execution of $L$ \lbs is an autonomous mode.
%Autonomous and Leader-Follower. An SFL file defines {\em setpoints} for each \lb, which specifies its position, actuation angle, and lighting color over time. A \lb processes these setpoints by issuing commands to its flight controller, servo, and lighting modules to transition smoothly to the next state. 

Stagger, used by LB-Author and the Blender add-on, outputs an SFL file to illuminate a line drawing or animation.
%Drawings are produced by LB-Author and Blender add-on.
%Animations are produced by the Blender add-on only.
The SFL file defines {\em setpoints} for each \lb, which specifies its position, actuation angle, and lighting color over time. A \lb processes these setpoints by issuing commands to its flight controller, servo, and lighting modules to transition smoothly to the next state. 

The {\em Orchestrator} of Figure~\ref{fig:pipeline} initiates the illumination process. It prompts each \lb to download the SFL file and wait in a ready state.  
Once all \lbs confirm readiness, the Orchestrator broadcasts a start to initiate the illumination after a synchronized delay of $\Delta = \max(\text{RTT}, t)$ time units.
Here RTT is the maximum round-trip time between the Orchestrator and the \lbs, and
$t$ is the granularity of the RPi clock, i.e., the shortest reliably measurable time interval\footnote{We set $t = 10$ ms in our experiments of Section~\ref{sec:eval}.
RTT is sub-millisecond.}.

Each \lb executes its portion of the SFL file independently using its local clock~\cite{integrate2025} and the position data of a motion capture system, Vicon~\cite{preiss2017whitewash}.
The RPi receives the Vicon's broadcast containing the position of all $L$ \lbs, filters its own, and provides it to the flight controller. It concurrently issues commands to the servo, LED lighting, and flight controller based on the SFL specifications.
Failure of one or more \lbs does not cascade to the rest of the swarm, allowing
the functioning \lbs to continue rendering their portion of illumination.

\subsection{An Evaluation}\label{sec:eval}
We evaluated the execution of different SFL files by a swarm of \lbs to quantify how closely the resulting formations corresponded to the target ground truth.

To quantify accuracy, let $t$ denote the time index. For an illumination with $L$ \lbs, each with $E$ LEDs, let $P_{i,e,t}^{GT}$ and $P_{i,e,t}^{Act}$ be the ground truth and actual 3D positions of the $e$-th LED on \lb $i$ at time $t$, respectively. 
The Root Mean Square Error (RMSE) for \lb $i$ at time $t$ is calculated across all its $E$ LEDs as: 
$RMSE_{i,t} = \sqrt{ \frac{1}{E} \sum_{e=1}^E \lVert P_{i,e,t}^{GT} - P_{i,e,t}^{Act} \rVert^2 }$
The average across all $L$ \lbs at time $t$ is:
$\overline{ RMSE_t }=\sqrt{ \frac{1}{ L } \sum_{i=1}^L (RMSE_{i,t})^2 }$
The error observed across all $L$ \lbs for the entire $T$ time steps is defined as a single value:
$RMSE_{Illumination} = \sqrt{ \frac{1}{T} \sum_{t=1}^T (\overline{ RMSE_t })^2 }$.

We consider absolute and relative RMSE.
Relative RMSE subtracts the centroid of LEDs illuminating a shape at time $t$,
$\mu_t$, from the actual and ground truth position of each LED:
$\hat{P}_{i,e,t}^{Act}$= $P_{i,e,t}^{Act}-\mu_t^{Act}$ and
$\hat{P}_{i,e,t}^{GT}$= $P_{i,e,t}^{GT}-\mu_t^{GT}$.
It uses these to quantify the RMSE metrics.

\begin{table}[htbp]
\centering
\caption{Relative and Absolute RMSE (mm) of Shapes.}
\label{tab:rmse_transposed}
\resizebox{\columnwidth}{!}{
\begin{tabular}{lccccccc}
\toprule
\textbf{Metric} & \textbf{Arrow} & \textbf{S} & \textbf{Blue Emoji} & \textbf{Yellow Emoji} & \textbf{X} & \textbf{ACM} & \textbf{NSF} \\
\midrule
\textbf{Relative RMSE} & 6.6 & 7.4 & 9.1  & 14.1 & 9.0  & 7.6 & 7.5 \\
\textbf{Absolute RMSE} & 13.2 & 9.2 & 11.5 & 16.5 & 11.2 & 8.9 & 8.6 \\
\bottomrule
\end{tabular}
}
\end{table}

Table~\ref{tab:rmse_transposed} presents the relative and absolute RMSE for different shapes and letterforms.
It highlights two key observations.
First, RMSE remains comparable across illuminations with different numbers of \lbs.
Second, mid-flight actuation does not adversely impact RMSE.
These observations are supported by comparisons across formations with different sizes and motion characteristics.
The arrow consists of two moving \lbs that adjust the actuation and lighting of their rods during flight, whereas the letterforms are stationary and may be subject to downwash effects. Similarly, the ACM and NSF formations each contain seven \lbs.
This is more than twice as many as the arrow.
Yet the resulting illuminations achieve comparable relative RMSE values.
Together, these results suggest that neither formation size nor mid-flight actuation significantly impacts illumination accuracy.
To further illustrate these findings, we present detailed evaluation results for two representative cases: the letterform S and the arrow.

\begin{figure*}
    \centering
    \begin{subfigure}{\columnwidth}
        \centering
        \includegraphics[width=\linewidth]{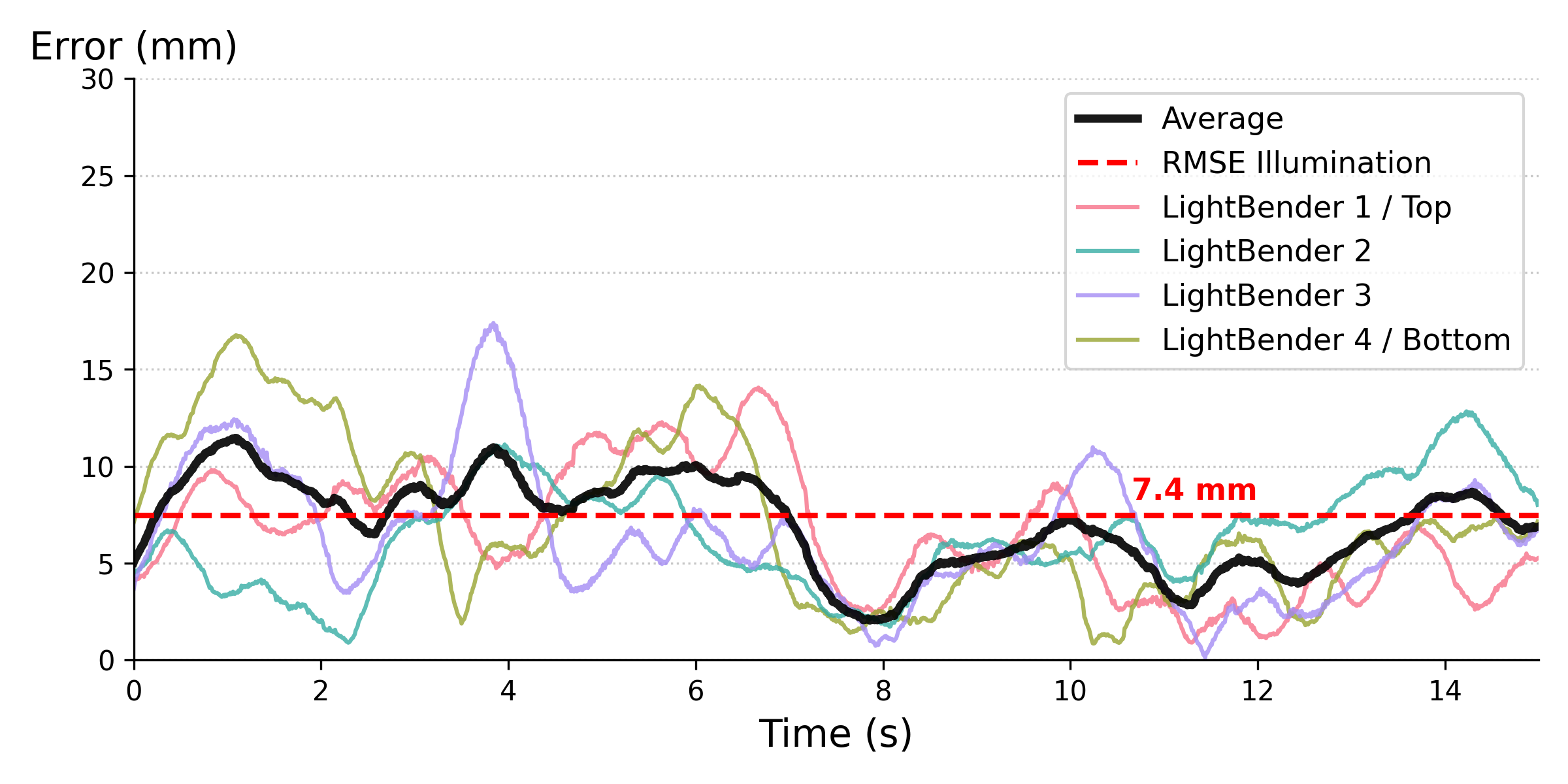}
        \caption{Relative RMSE.}
        \label{fig:rel_s}
    \end{subfigure}
    \begin{subfigure}{\columnwidth}
        \centering
        \includegraphics[width=\linewidth]{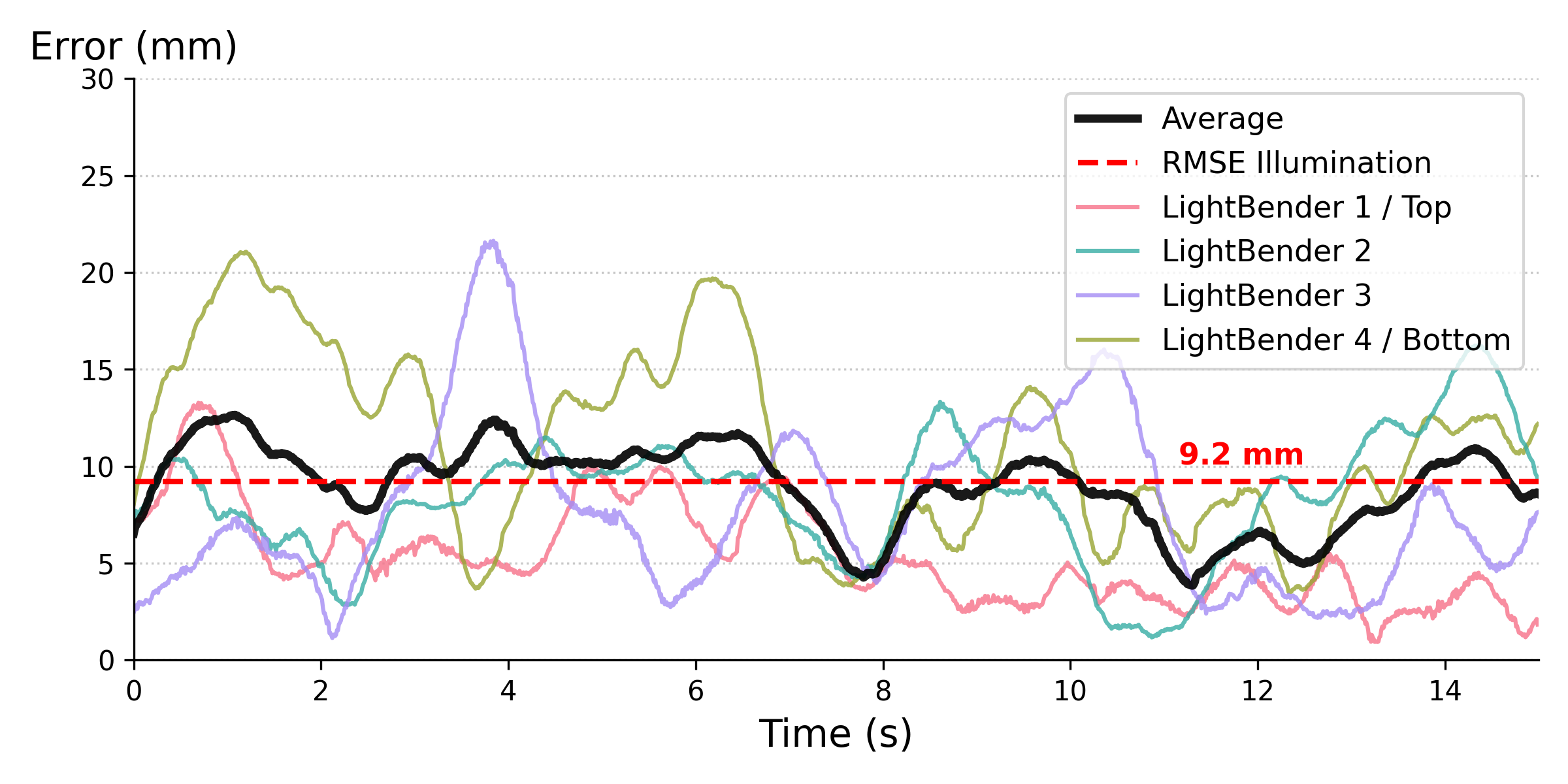}
        \caption{Absolute RMSE}\label{fig:abs_s}
    \end{subfigure}
    \caption{Four \lbs illuminating S.}\label{fig:rmse_s}
\end{figure*}

%The RMSE is lower with Autonomous when compared with Leader-Follower. 
\subsubsection{Letterform S}
`S' consists of four \lbs that are staggered to compensate for downwash.
Figure~\ref{fig:rmse_s} shows the observed RMSE in millimeters (y-axis) as a function of the total illumination time at 10 ms granularity (x-axis) for each \lb.
The thick black line shows the average error across all four \lbs as a function of time.
The relative and absolute
$RMSE_{Illumination}$ is 7.4 mm and 10 mm, respectively.
%The y-axis is the observed error in millimeters.
The relative RMSE is lower because the error in \lbs relative to the center of the illumination is lower.

% Autonomous is superior to Leader-Follower in the presence of drift, see Section~\ref{sec:drift}.
% With Leader-Follower, the leader's drift causes all following LightBenders to deviate from the ground truth.
% They compute their destination using an offset with the erroneous position of the leader due to drift.

Repeated illumination of the same SFL file demonstrate high predictability, with the absolute $RMSE_{Illumination}$ bounded between 9.2 mm and 10.1 mm. 
Furthermore, Figure~\ref{fig:rmse_s} shows that the bottom \lb incurs the highest individual error, primarily due to downwash generated by the three \lbs above it. 

% \begin{figure*}[h]
%     \centering
%     \includegraphics[width=\linewidth]{fig/arrow_temporal_fix_rmse.png} 
%     \caption{Root mean square error of 2 moving LightBenders that animate a moving arrow with (a) Autonomous and (b) Leader-Follower techniques.}
%     \label{fig:eval-arrow-temporal-fix}
% \end{figure*}

\begin{figure*}
    \centering
    \begin{subfigure}{\columnwidth}
        \centering
        \includegraphics[width=\linewidth]{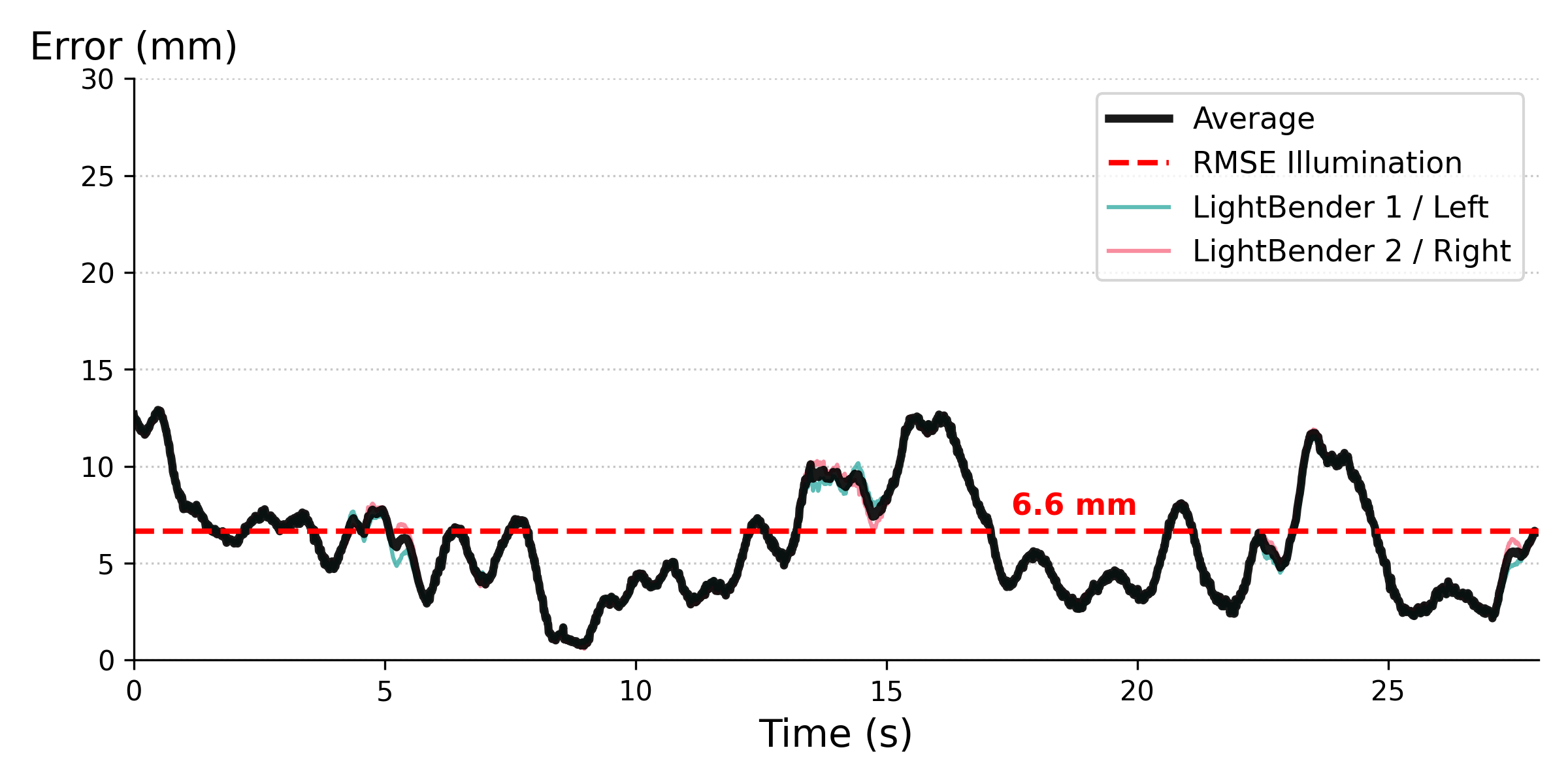}
        \caption{Relative RMSE.}
        \label{fig:rel_arrow}
    \end{subfigure}
    \begin{subfigure}{\columnwidth}
        \centering
        \includegraphics[width=\linewidth]{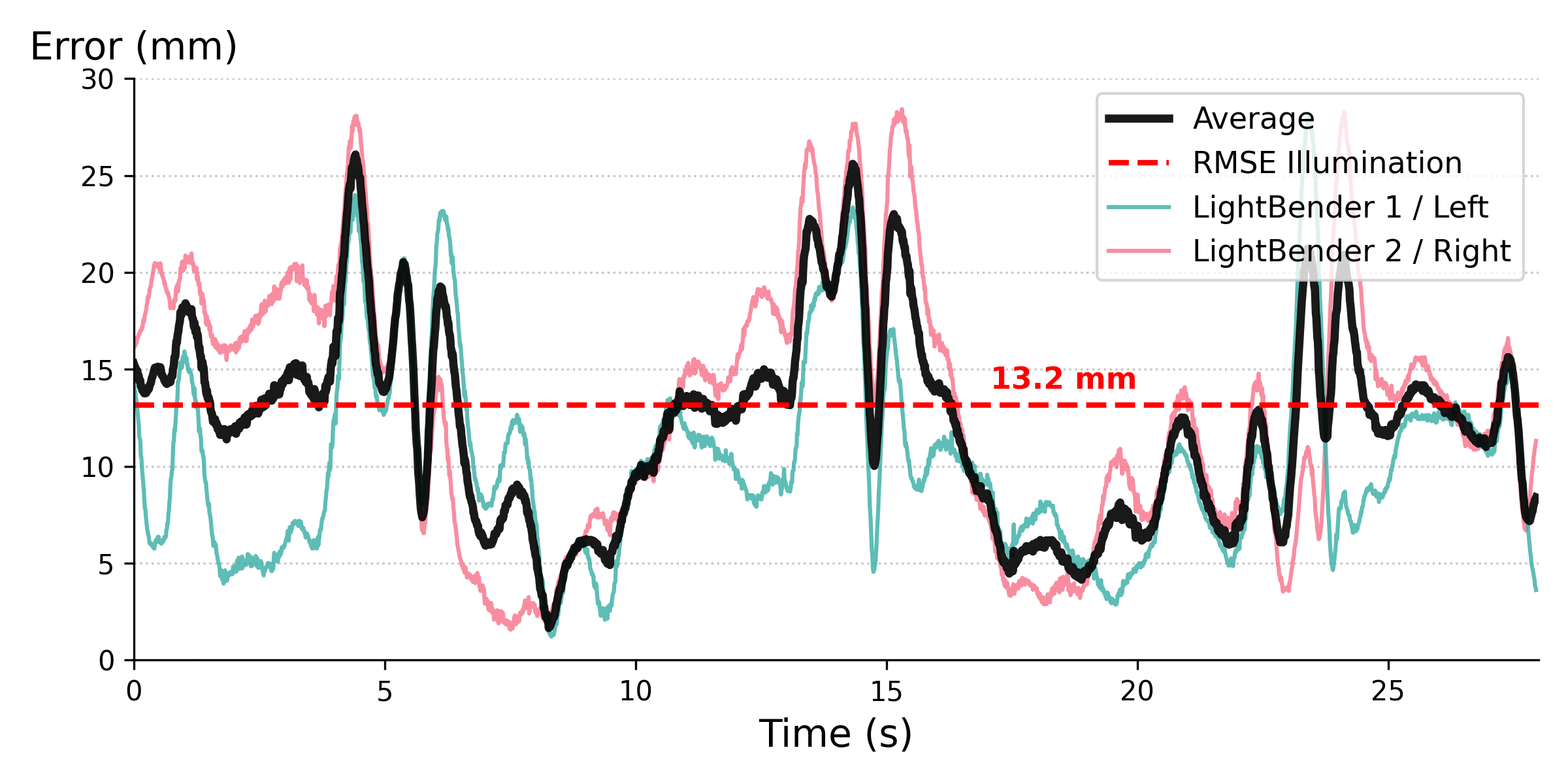}
        \caption{Absolute RMSE}\label{fig:abs_arrow}
    \end{subfigure}
    \caption{Two \lbs illuminating a moving arrow.}\label{fig:rmse_arrow}
\end{figure*}

\subsubsection{Moving Arrow}
The arrow is animated using two \lbs. One \lb illuminates the arrowhead while the other illuminates the shaft. The two \lbs move from left to right while decelerating, then actuate to switch roles and travel in the opposite direction. The second row of Figure~\ref{fig:teaser} illustrates this behavior. The illumination sequence repeats three times.
%The arrow consists of two \lbs and is animated. One \lb illuminates the head of the arrow and the other illuminates the body of the arrow. These two \lbs move from the left to right, decelerating, actuating to switch roles, and travel in the reverse direction. See the second row of Figure~\ref{fig:teaser}. The illumination shows three repetitions.
%The switch requires mid-flight actuation.
%The moving arrow evaluates an animated illumination consisting of two \lbs, one for the head and the other for the body.After moving from left to right, the \lbs decelerate, switch roles and reverse direction, see the second row of Figure~\ref{fig:teaser}.The switch requires mid-flight actuation. 
%In the illumination, the arrow reverses direction three times.
% They repeat this several times, see the second row of Figure~\ref{fig:teaser}. 
Its $RMSE_{Illumination}$ is consistent across multiple repetitions of the illumination ranging in value from 11.5 to 13.2.
Figure~\ref{fig:rmse_arrow} shows the relative and absolute RMSE 
for the illumination with the highest $RMSE_{Illumination}$.
%The results show a low absolute $RMSE_{Illumination}$ of 13.2 mm.
The average RMSE spikes because of deceleration, mid-flight actuation, and subsequent acceleration when the \lbs switch roles.
%It reveals spikes in the average error when the \lbs begin to decelerate, actuate their rods to show the reversed arrow, and start to move in the opposite direction. 
These spikes are attributed to physical inertia, causing the absolute positions of the \lbs to lag slightly behind the ground truth position.
When considering the relative RMSE, both $RMSE_{Illumination}$ and the average RMSE are significantly reduced.  
Similar to `S', the error in \lbs relative to the center of the illumination is lower.
Note that the relative RMSE of individual \lbs is approximately the same.

% \footnote{After removing the discrepancy in the initial acceleration of LightBenders.}.
%The error is high because the ground truth does not consider the acceleration of the LightBenders. In reality, the LightBenders must accelerate to reach the speed specified by the animation.  
%We compensate for this discrepancy by considering the region where the speed of the LightBenders matches the speed required by the ground truth.  
%After this compensation, the errors are reduced dramatically and shown in Figure~\ref{}.
% The obtained results show that Autonomous has a low error, less than 2 cm.  
% Leader-Follower has the same error for the Leader.
% However, the follower's error is significantly higher.  
% The follower is $\Delta t$=20 milliseconds behind the leader, see Section~\ref{sec:swarm} and Figure~\ref{fig:arrow-comp}.
% The motion capture system transmits location data at 100 Hz.
% A LightBender also transmits the destination to its Flight Controller at 100 Hz. 
% Their sum is 20 milliseconds.
% Mid-flight actuation does not impact the error.

%Similar trends are observed for other illumination patterns. Overall, the Autonomous technique is typically about 50\% more accurate than the Leader–Follower technique.

%With repeated illumination of the same SFL file, absolute $RMSE_{Illumination}$ is bounded between 6.6 and 6.7 mm.

\section{Human Subject Study}\label{sec:human}
\begin{figure}
    \centering
    \includegraphics[width=\linewidth]{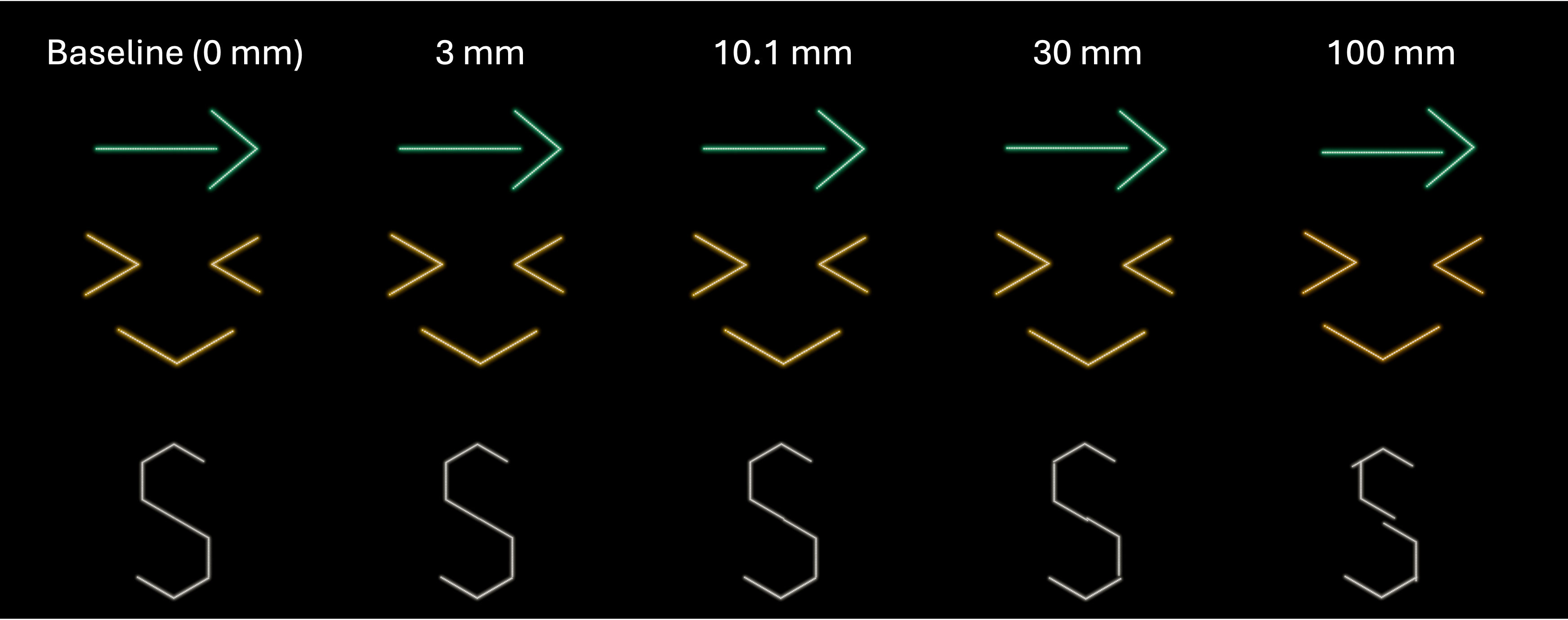}
    \caption{Example renderings of Arrow, Emoji, and S with five misalignments.}
    \label{fig:stimuli}
\end{figure}

RMSE of Section~\ref{sec:eval} highlights misalignment between \lbs.
We conducted an IRB approved human subject study UP-25-00102 to evaluate the impact of this misalignment on the human perceived quality of illuminations. 
Here the quality of illumination is subjective, referring to the structural integrity, stability of a shape, and the degree to which the shape is recognizable.
The main research questions that we strive to answer are:
\begin{enumerate}
    \item 
    %Is a 10.1 mm misalignment of \lbs illuminating a shape in Section~\ref{sec:illuminate} acceptable to users?
    Is a misalignment of a few millimeters in a swarm of \lbs illuminating a shape acceptable to subjects? 
    \item Do subjects perceive a small amount of misalignment (say 3 mm) compared with a baseline of no misalignment? 
    \item At what misalignment does the perceived quality of the illumination become noticeable while still remaining usable?
    \item Does the geometry of a shape (e.g., arrow compared with `S') impact the perceived misalignment?
    %Is misalignment perceived differently depending of the geometry of the shape (e.g., an arrow compared to a letter like such as 'S')
    % The same question with an illumination that involves real-time actuation such as the emoji?
\end{enumerate}
%See Table \ref{tab:results_summary} for results summary.
%Table~\ref{tab:results_summary} summarizes the answers to these questions.
To answer these questions, we used a behavioral Likert scale~\cite{sullivan2013likert} for subjects to rate the quality of an illumination with no (0 mm) misalignment and across four different misalignments:
3, 10.1, 30, and 100 mm.
See Figure \ref{fig:stimuli}.
% of 0, 3, 10.1, 30, and 100 millimeters. 
%On each trial, five videos were presented in randomized order, one for each misalignment.

\begin{table}[ht]
\centering
\caption{Absolute and relative $RMSE_{Illumination}$ (mm) per misalignment.}
\label{tab:misalignment}
\begin{tabular}{l rr rr rr rr}
\toprule
& \multicolumn{2}{c}{3 mm}
& \multicolumn{2}{c}{10.1 mm}
& \multicolumn{2}{c}{30 mm}
& \multicolumn{2}{c}{100 mm} \\
\cmidrule(lr){2-3}\cmidrule(lr){4-5}
\cmidrule(lr){6-7}\cmidrule(lr){8-9}
Shape & Abs & Rel & Abs & Rel & Abs & Rel & Abs & Rel \\
\midrule
S     & 2.8 &  2.6 &  9.6 &  9.0 & 28.1 & 25.8 &  97.4 & 84.7 \\
Arrow & 2.2 &  1.8 &  8.1 &  5.1 & 28.2 & 19.7 & 113.7 & 90.0 \\
Emoji & 2.9 &  2.2 & 10.7 &  8.0 & 26.7 & 21.2 &  92.3 & 67.2 \\
\bottomrule
\end{tabular}
\end{table}

We generated controlled video stimuli of S, Arrow, and Emoji using Blender to precisely manipulate the misalignment. 
The same is difficult (if not impossible) with illuminations.
This is because, in practice, drift is stochastic and not directly controllable\footnote{If this was possible then we would have minimized drift close to zero with a constant variation.}.  
%The use of synthetic misalignment is necessary as it allows controlled variation of drift to enable systematic observation. In practice, drift is difficult to hold constant across different illuminations as it is stochastic and not directly controllable.
Misalignment was introduced as an added positional error to the position of \lbs using a Blender noise modifier. For each shape, the misalignment may be randomized. Table~\ref{tab:misalignment} shows RMSE values for the generated videos.
We used this observation to generate 3 videos for each misalignment, resulting in 12 videos.
With one video for no misalignment, 0 mm, there is a total of 13 videos for each illumination.

%\subsection{Apparatus and stimuli}

%This produced a total of 13 videos\footnote{13 because 0 mm has only one video.} per shape. 

%Each video is 10 seconds in duration. We used a 32-inch 4K monitor to display the stimuli, with participants seated in a chair behind a desk at a fixed viewing distance (\~1m).

\begin{figure}
    \centering
    \includegraphics[width=\linewidth]{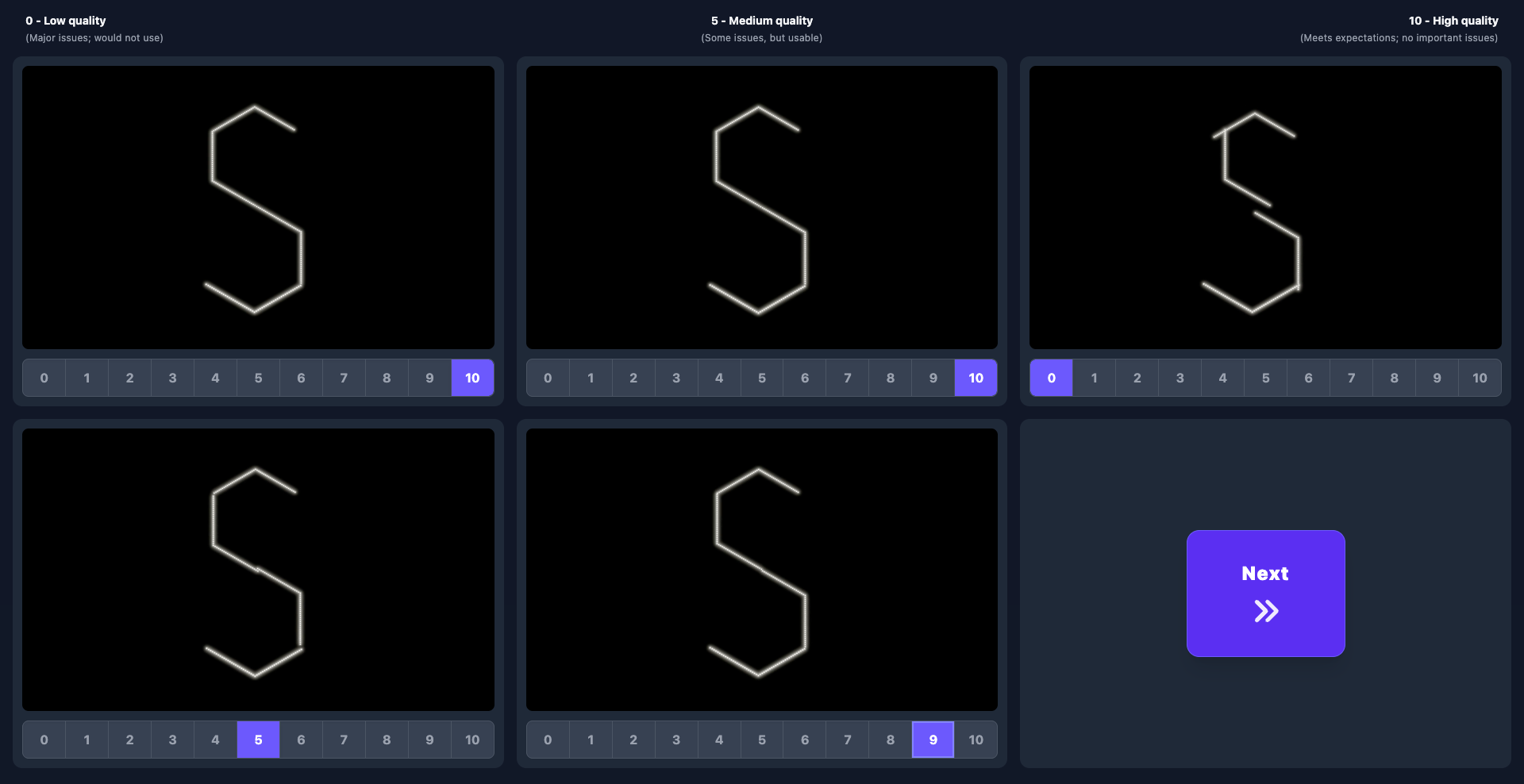}
    \caption{Survey interface showing S with five misalignments. The subject rates the quality of each misalignment using the scale shown below each illumination.}
    \label{fig:ui}
\end{figure}

%\subsection{Experimental Design}
We employed a within-subjects experimental design \cite{kim2010withinsubjects}. In each trial, participants were simultaneously shown five videos of the same shape at different levels of misalignment.
See Figure \ref{fig:ui}.
%, representing the five misalignments (0, 10.1, 30, and 100 mm). 
The videos were arranged side-by-side in a randomized order on the screen.
Each is 10 seconds in duration and loops infinitely while a subject rates its quality
%Participants rated the quality of each video 
using an 11-point Likert scale (0 to 10) with three behavioral anchors:  
\begin{itemize}
    \item[0] (Low quality): The shape has major issues and would not be usable.
    \item[5] (Medium quality): The shape has some noticeable issues but is still usable.
    \item[10] (High quality): The shape meets expectations with no important issues.
\end{itemize}
Participants were explicitly instructed that they could use any value on the 0–10 scale to reflect perceived differences in quality.
%Participants were explicitly instructed that they may use any value across the full range of the scale to capture differences in quality (e.g., between 0 and 5, or 5 and 10).

%each with a different misalignment level. 
%Participants rated the overall quality of each video on a scale of 0 to 10.
Twenty-one participants rated the overall quality of each video
displayed on a 32-inch 4K monitor.
They were seated in a chair behind a desk at a fixed viewing distance of approximately 1 meter.
%\subsection{Participants}
%21 participants completed the study. 
%All completed a demographic and a post-study feedback survey. 10 were Female and 11 were Male. The average age was 25.4 years (min: 18, max: 34).
All completed a demographic questionnaire and a post-study feedback survey. Ten were female and eleven were male. The average age was 25.4 years (range: 18–34).
Our findings are summarized in Table~\ref{tab:results_summary}.

\subsection{Protocol} 
Each participant completed one tutorial, one warmup, and nine experimental trials (three trials per shape). 
During the tutorial, participants were instructed on how to rate the videos. They were asked to focus on whether a shape appeared distorted, using reference examples of letter “X” with 0, 10.1, 30, and 100 mm misalignment. Subsequently, participants completed a warm-up trial to become familiar with the interface and rating process. Data from these initial trials were excluded from the analysis.

For the nine trials, the presentation order of the shapes was counterbalanced across participants using three different start sequences.
Each participant rated the quality of five videos shown in each trial.
After completing all trials, participants answered an open-ended post-study feedback question: "Did you notice any differences in the quality of the symbols? If yes, what kinds of differences did you notice?”

\subsection{Results}

\begin{figure}
    \centering
    \includegraphics[width=\linewidth]{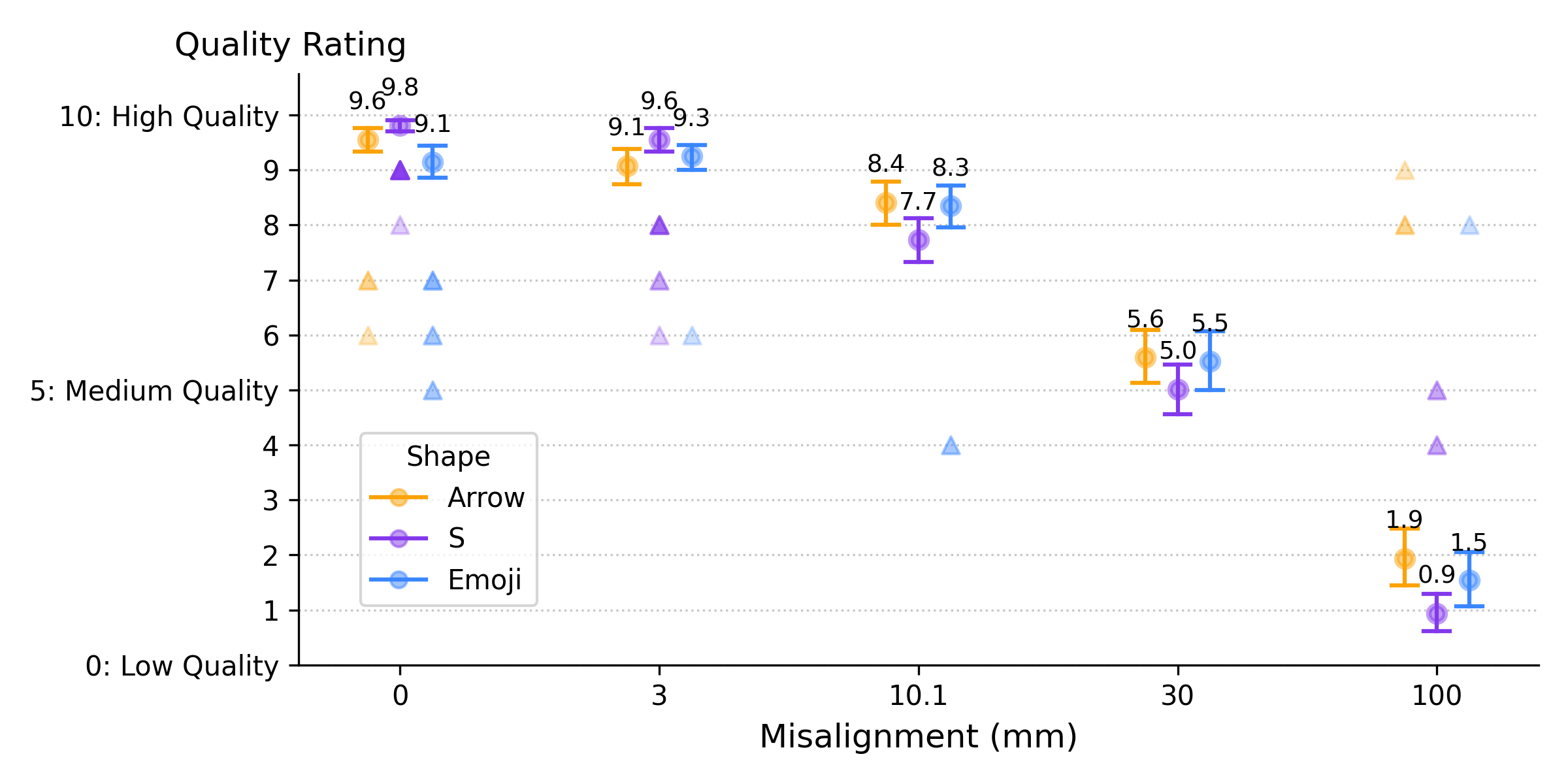}
    \caption{Results of experiment. The circle markers represent the average quality across all participants.
    Bars show 95\% confidence.}
    \label{fig:qbox}
\end{figure}
Figure~\ref{fig:qbox} shows the box plot of quality rating (y-axis) as a function of misalignment (x-axis) for each shape.
In general, participants' perceived quality of an illumination decreases as we increased misalignment from 0 to 100 mm.
The perceived decrease with a low misalignment of 3 mm is shape dependent.
The subjects do not notice it with the Emoji.
However, they report a slight decrease in quality with S and Arrow.
%Except for the Emoji, even at a low misalignment of 3 mm, users perceived a drop in quality with Arrow and S.
%Even at a low misalignment of 3 mm, users perceived a drop in quality, except for the Emoji. 
At a 10.1 mm misalignment, quality ratings is relatively high (mean = 8.16, median = 8), indicating the illuminations are usable. 
At 30 mm, ratings dropped to medium quality (mean = 5.38, median = 5), denoting noticeable issues with the illumination while still remaining usable. However, at the extreme 100 mm misalignment, perceived quality diminished severely; ratings for the S is less than 1 (completely unusable), while the Arrow and Emoji drops below 2.

A Mann-Whitney U test~\cite{10.1214/aoms/1177730491} found no significant difference in quality ratings between male and female participants (U = 61.0, p = .698). This lack of gender effect held true across all shapes and all misalignment.

Because our Likert data is ordinal, we evaluated statistical significance using non-parametric Friedman tests \cite{Friedman01121937} for simple effects, followed by pairwise Wilcoxon signed-rank tests \cite{c4091bd3-d888-3152-8886-c284bf66a93a} with Bonferroni corrections for pairwise comparisons. We conducted these test using custom Python scripts and SciPy package \cite{2020SciPy-NMeth}.

\textbf{Effect of misalignment:} We found a significant effect of misalignment on quality ratings within all three shapes: Arrow ($\chi^2(4) = 76.44, p < .001$), Emoji ($\chi^2(4) = 73.86, p < .001$), and Letter S ($\chi^2(4) = 79.49, p < .001$).

Post-hoc pairwise comparisons revealed that, despite quality ratings dropping at higher misalignments, users struggle to differentiate between the lowest misalignments. Specifically, for the Arrow, differences between 0 and 3 mm ($p = .166$) and 3 and 10.1 mm ($p = .069$) were not significant. For the Emoji, users did not significantly distinguish between 0 and 3 mm ($p = 1.000$) or between 0 and 10.1 mm ($p = .089$). For the 'S', only the 0 vs. 3 mm comparison was non-significant ($p = .748$). This indicates that for multi-part shapes like the Emoji and Arrow, illuminations with small misalignments ($\leq 10.1$) look almost identical to users, resulting in inconsistent ratings within repeated trials. In contrast, users could more easily detect misalignments beyond 3 mm in the S.

\textbf{Effect of Shape:} We also analyzed how the shape affected quality perception at fixed misalignments. Fridman tests revealed significant differences between the shapes at every error level: 0 mm ($p < .001$), 3 mm ($p = .004$), 10.1 mm ($p = .031$), 30 mm ($p = .014$), and 100 mm ($p < .001$).

Pairwise comparisons show a key insight about shapes. At the 0 mm baseline, the S was rated significantly higher than the Emoji ($p = .004$). This means users easily identified the baseline S as the highest quality, whereas they inconsistently rated the baseline Emoji (not always as the highest quality). At the 100 mm misalignment,  the S was rated significantly lower than both the arrow ($p = 0.004$) and the Emoji ($p = 0.011$). These findings suggest that the continuous geometry of the S makes both its perfect and completely broken states highly obvious to users. In contrast, the multi-part structure of the Arrow and Emoji masks the extreme misalignments, making them harder to consistently judge.

\textbf{Qualitative Feedback:} Participants' open-ended responses supported our quantitative findings regarding shape geometry. A dominant theme was the distinction between continuous and multi-part shapes.
Multiple participants noted that the distortions were more obvious in shapes intended to be single connected component, S, because the misalignment destroyed the continuity (e.g., P1, P11, P19). For the shapes composed of multiple pieces, Emoji and Arrow, one participant reported a "higher tolerance" (P1) for misalignment. Finally, participants attributed visual stability with higher quality and noted that "wavering" and "vibration" degraded the quality (P5, P13, P21). 

\begin{table}
\centering
    \begin{threeparttable}
        \caption{Summary of findings from the human subject study.}
        \label{tab:results_summary}
        \begin{tabular}{@{} p{0.35\columnwidth} p{0.64\columnwidth} @{}}
            \toprule
            \textbf{Research Question} & \textbf{Summary of Finding} \\
            \midrule
            \textbf{RQ1: Is 10.1 mm acceptable?} & Subjects found the 10.1 mm misalignment acceptable (mean = 8.16, median = 8). \\
            \addlinespace
            \textbf{RQ2: Do subjects perceive small misalignment compared with no misalignment?} & When compared with no misalignment, the perceived decrease in quality at a low misalignment (3 mm) is shape-dependent. Subjects reported a slight decrease with the `S' and Arrow but not with the Emoji; this observation was not statistically significant
            \tnote{$\dagger$}.
            % \footnotemark
            % \tablefootnote{This observation is encouraging, as it suggests that minor misalignments do not substantially degrade perceived quality.}.
            \\
            \addlinespace
            \textbf{RQ3: What misalignment has noticeable issues but is usable?} & Illuminations begin to show noticeable visual issues at 30 mm, but are still usable (mean = 5.38, median = 5). \\
            \addlinespace
            \textbf{RQ4: Does the shape geometry impact the perceived misalignment?} & At both low (e.g., 3 mm) and extreme (e.g., 100 mm) misalignments, the perceived quality is shape-dependent.
            Shapes with disconnected lines mask misalignments better than those with connected lines. \\
            \bottomrule
        \end{tabular}
        \begin{tablenotes}
            \item[$\dagger$] This observation is encouraging, as it suggests that minor misalignments do not substantially degrade perceived quality.
        \end{tablenotes}
    \end{threeparttable}
\end{table}
% \footnotetext{This observation is encouraging, as it suggests that minor misalignments do not substantially degrade perceived quality.}
% \vspace{-.1in}

\balance

\section{Conclusions and Future Research}\label{sec:future}
%\lbs are an exciting and emerging area of multimedia research.
A \lb is an exciting and emerging area of multimedia research.  It is a drone with an actuated LED rod as its lighting primitive. 
This paper introduced the hardware and software architecture of Figure~\ref{fig:pipeline} to author and render illuminations using a swarm of \lbs. 
We are extending both the hardware and software in several ways.
With the hardware, we are extending the \lb's rod to consist of multiple rows of smaller LEDs instead of one row of medium sized LEDs.
This will enable us to adjust for the width of the lines when we stagger \lbs, see discussion of Eq.~\ref{eq:width_error}.
With the software, we are designing techniques to compute medial axis skeleton~\cite{au2008skeleton,rodrigues2018part,lin2020seg,lu2025survey} for 3D volumes.
We envision an iterative optimization algorithm to adjust the position and the angles of the \lbs' rod segments to maximize similarity between the LightBender illumination and the 3D volume.
This will include support for animations.  

In the longer term, we will investigate techniques that support simultaneous viewing by multiple users from different viewpoints using spatial staggering. We will also explore FLSs~\cite{shahram2021} equipped with three-dimensional lighting primitives, such as cubes and pyramids, whose volumetric illuminations complement the line-rendering capabilities of \lbs.

%This enables a ``what"-oriented Blender add-on that reduces artist provided volumetric meshes into line drawings that use the placement technique of Section~\ref{sec:placement}. These complement one another because we can extend the Blender add-on to use the volumetric variant in addition to \lbs. This will enable animated line drawings and volumetric meshes.

We will extend our human subject study to focus on the user experience.
A key research question is how usable and intuitive is our Blender add-on? 
This will include participants with and without previous background in the use of Blender, comparing the experiences of the two groups with one another.

We will extend our human-subject study to investigate the user experience of our Blender add-on. In particular, we seek to evaluate its usability and intuitiveness. The study will include participants with and without prior Blender experience, allowing us to compare the experiences of the two groups.

%Human subject studies are another future research direction that investigates creator experience and how an artist perceives the \lb illuminations. Studying creator experience involves evaluating authoring usability in studies with participants with or without a relevant artistic or technical background. The human perception study presented in this work can be extended to evaluate the impact of spatial staggering and other aspects of \lbs on perceived quality under different conditions like lighting, distance, and viewing angle. The quantitative results of such studies enable us to calibrate QoI metrics.

\begin{acks}
This research is supported by the NSF grant CMMI-2425754.
We gratefully acknowledge CloudBank~\cite{cloudbank2021} and CloudLab~\cite{emulab} for the use of their resources.
\end{acks}

%%
%% The next two lines define the bibliography style to be used, and
%% the bibliography file.
\bibliographystyle{ACM-Reference-Format}
\bibliography{refs}

\appendix
\section{Blender add-on and LED Color Expressions}\label{sec:pointer}
\begin{figure}[h]
\includegraphics[width=1\linewidth]{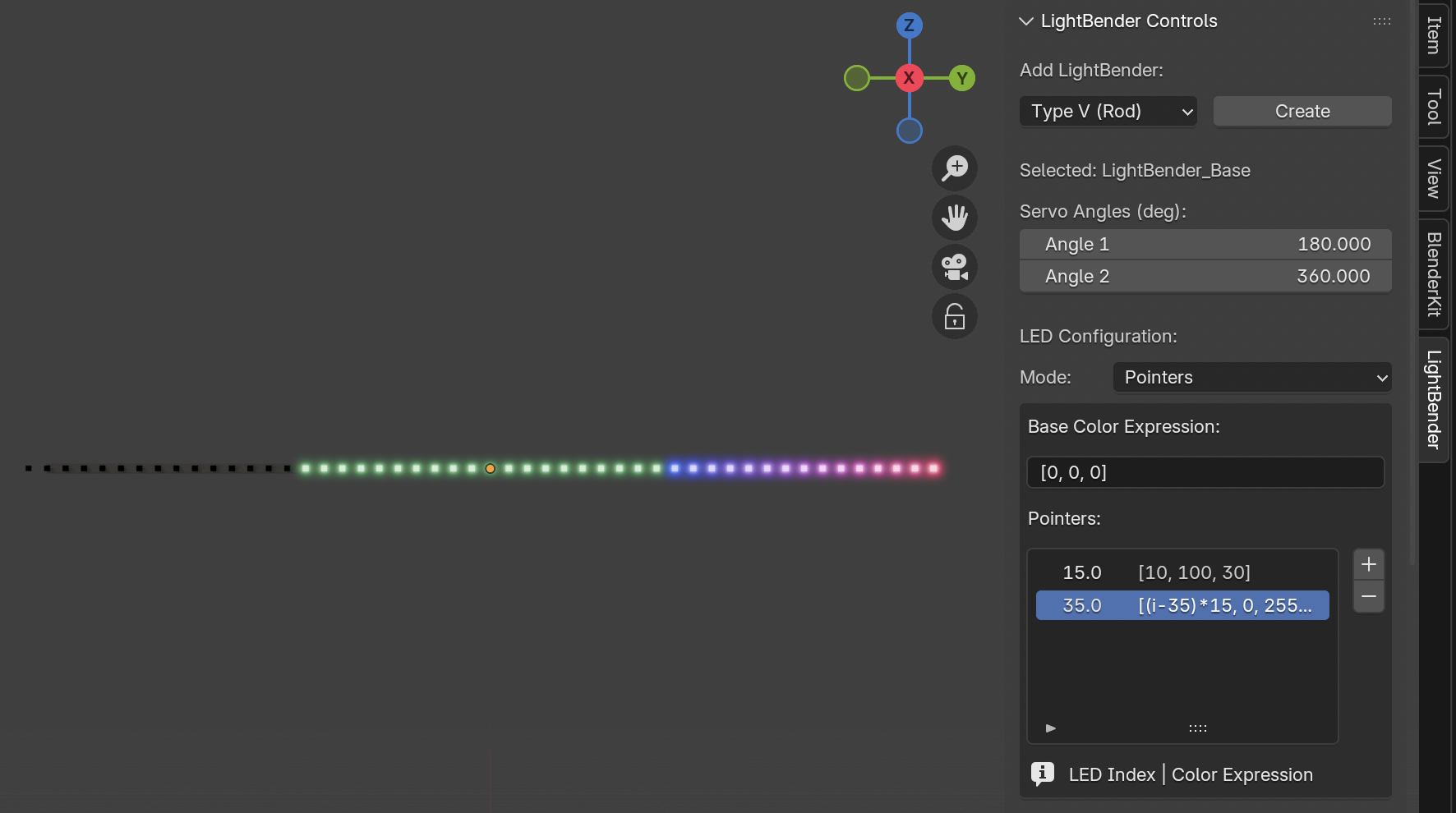} 
\caption{Pointers and expressions to control LED colors.}
\label{fig:pointer}
\end{figure}

Once an artist places a \lb in Blender’s 3D viewport using our add-on, they may specify LED illumination using expressions over the two segments of its actuated rod.
A pointer consists of an index and a color expression.
For the 160 mm \lb, LEDs are indexed from 0 to 49.
An expression is evaluated for a range of LEDs and may produce a color dependent on time.
The range is specified using a pair of consecutive pointers.
A single LED at index $i$ is identified by the range [$i$,$i+1$).
%This expression is a mathematical formula that is evaluated per LED to produce a color, and may depend on time and the LED index.
Figure~\ref{fig:pointer} shows three examples. 
First, bottom right hand side of the figure shows an expression to turn the range of LEDs [0,15) off by setting their color values to black, RGB $(0,0,0)$.
Second, an expression may assign the same color to all LEDs in a range to display a solid color.
The expression with RGB $(10,100,30)$ and range [15, 35) results in solid green in the middle of rod, see Figure~\ref{fig:pointer}.
It is used to illuminate the uniform green of the arrow in Figure~\ref{fig:teaser}.
Third, an expression may define a color gradient for an LED range.
An example is range [35, 50) with $(i-35)*15$ in Figure~\ref{fig:pointer}. In this case, the LEDs transition smoothly between two colors throughout the range of the 15 LEDs. Blue to red in Figure~\ref{fig:pointer}. The color of each LED is interpolated linearly from the start color to the end color on the basis of its position. This expression type is used to create smooth color transitions along a rod segment.

\end{document}